\documentclass[journal]{IEEEtran}

\ifCLASSINFOpdf
\else
\fi

\usepackage{graphicx}
\usepackage{amsmath}
\usepackage{amsfonts}
\usepackage{cite}
\usepackage{algorithm}
\usepackage{algpseudocode}
\usepackage{booktabs}
\usepackage{color}
\usepackage{array}

\begin{document}

\title{Symmetric Bilinear Regression for Signal Subgraph Estimation}

\author{Lu~Wang,~
	Zhengwu~Zhang~
	and~David~Dunson
  \thanks{Lu Wang is with the Department of Statistics, Central South University, Changsha, China (e-mail: wanglu\_stat@csu.edu.cn)}%
   \thanks{Zhengwu Zhang is with the Department of Biostatistics and Computational Biology, University of Rochester, Rochester, NY 14604 USA (e-mail: Zhengwu\_Zhang@URMC.Rochester.edu)}%
   \thanks{David Dunson is with the Department of Statistical Science, Duke University, Durham, NC 27708 USA (e-mail: dunson@duke.edu)}%
  \thanks{Matlab code associated with this article can be found at https://doi.org/10.24433/CO.69bf9b86-3276-40fa-a979-b68a6ff1e562}
}

\maketitle

\begin{abstract}
There is increasing interest in learning a set of small outcome-relevant subgraphs in network-predictor regression. The extracted signal subgraphs can greatly improve the interpretation of the association between the network predictor and the response. In brain connectomics, the brain network for an individual corresponds to a set of interconnections among brain regions and there is a strong interest in linking the brain connectome to human cognitive traits. Modern neuroimaging technology allows a very fine segmentation of the brain, producing very large structural brain networks. Therefore, accurate and efficient methods for identifying a set of small predictive subgraphs become crucial, leading to discovery of key interconnected brain regions related to the trait and important insights on the mechanism of variation in human cognitive traits. We propose a symmetric bilinear model with $L_1$ penalty to search for small clique subgraphs that contain useful information about the response. A coordinate descent algorithm is developed to estimate the model where we derive analytical solutions for a sequence of conditional convex optimizations. Application of this method on human connectome and language comprehension data shows interesting discovery of relevant interconnections among several small sets of brain regions and better predictive performance than competitors.
\end{abstract}


\IEEEpeerreviewmaketitle

\section{Introduction}

\IEEEPARstart{I}{n} this article, we study methods for predicting an outcome variable $y_i$ from a network-valued variable $W_i$, measured on $n$ subjects, where $W_i$ is a $V\times V$ symmetric matrix. In the typical scenario, the number of free elements of $W_i$, $V(V-1)/2$, is much larger than $n.$ In our motivating example, $W_i$ is the weighted adjacency matrix of an individual's brain structural network, where the brain is segmented into $V$ regions and each entry in $W_i$ denotes the connectivity strength of neural fibers between a pair of regions. The outcome $y_i$ is a cognitive trait of an individual which is a continuous variable. The goal is to select neurologically interpretable subgraphs in the brain connectome, corresponding to a subset of neural connections, that are relevant to the outcome $y_i$.

One typical approach to this large $p$ small $n$ problem would be a linear regression with some regularization, such as lasso \cite{tibshirani1996regression}, elastic-net regression \cite{zou2005elastic-net} and SCAD \cite{fan2001variable}. These approaches require first flattening out each adjacency matrix into a long vector, which could induce ultra high dimensionality for huge networks \cite{da2015flashgraph}. In addition, for large signal subgraphs with small sample size $n$, lasso cannot recover the truth because it cannot select more than $n$ variables (edges). The most serious problem for these methods is that the selected connections generally do not have any structure in brain connectivity, making the results hard to interpret.

Graphical learning methods with sparsity regularization such as graphical lasso \cite{friedman2008sparse} aim to learn the conditional independence structure among multiple variables, which are usually assumed to have a multivariate Gaussian distribution and the focus is on estimating a sparse inverse covariance matrix for the variables. It may be possible to jointly model the outcome $y_i$ and all the connection strengths in the network $W_i$ as a multivariate Gaussian. But this would involve estimating an $O(V^2) \times O(V^2)$ inverse covariance matrix, which may not be appealing in practice. Also the interpretation would be a big issue as the selected connections relevant to $y_i$ may not have any structure as with lasso.

Existing feature extraction approaches \cite{beckmann2005investigations, kolda2009tensor, varoquaux2011multi, wang2017common, zhang2018relationships} typically employ a two-stage procedure where some latent representations of the networks are first learnt and a prediction model is trained on the low-dimensional representations. 
For example, tensor network principal components analysis (TN-PCA)  \cite{zhang2018relationships} is an unsupervised dimension reduction method, which approximates a semi-symmetric 3-way tensor $\mathcal{W}$ by a sum of rank-one tensors:
\begin{equation}
\mathcal{W}\approx\sum_{k=1}^{K}d_{k}\boldsymbol{v}_{k}\circ\boldsymbol{v}_{k}\circ\boldsymbol{u}_{k},
\label{eq:semi_sym_tensor_decomp}
\end{equation}
where $\mathcal{W}$ is a concatenation of symmetric (demeaned) adjacency matrices $\{W_{i}\}_{i=1}^{n}$, $d_k$ is a positive scaling parameter, $\circ$ denotes the outer product, $\boldsymbol{v}_{k}$ is a $V \times 1$ vector of unit length that stores the PC score for each node in component $k$,  and $\boldsymbol{u}_{k}$ is a $n \times 1$ vector of unit length that stores the PC score for each network in component $k$. \cite{zhang2018relationships} places orthogonality constraints on the component vectors $\boldsymbol{v}_{k}$'s but leaves the vectors $\boldsymbol{u}_{k}$'s unconstrained.
The TN-PCA \eqref{eq:semi_sym_tensor_decomp} embeds the $V \times V$ undirected networks $\{W_{i}\}_{i=1}^{n}$ into a low dimensional $n \times K$ matrix $U=(\boldsymbol{u}_{1},\dots,\boldsymbol{u}_{K})$, where each row $i$ represents a $1 \times K$ embedded vector for network $i$. When $K < n$, we can study the relationship between the network $W_i$ and an outcome $y_i$ via a simple linear regression on the low dimensional embeddings $U$. The set of rank-one matrices $\{\boldsymbol{v}_{k}\boldsymbol{v}_{k}^{\top}\}_{k=1}^K$ can be viewed as basis networks and the ones corresponding to the significant components in the regression of $y$ are selected as signal sub-networks.
However, such an unsupervised approach has the disadvantage that the low-dimensional structure $\{\boldsymbol{v}_{k}, \boldsymbol{u}_{k}\}_{k=1}^{K}$ is extracted to minimize the reconstruction error in network approximation \eqref{eq:semi_sym_tensor_decomp}, which may not produce network features that are particularly predictive of the response $y$.

Another related method is the low-rank sensing model, which considers the problem of recovering a low-rank matrix from affine equations. That is, 
\begin{align}
\mbox{minimize } & \mbox{rank}(B) \nonumber \\
\mbox{subject to } & y_{i}=\mbox{trace}(W_{i}^{\top}B)=\left\langle W_{i},B\right\rangle, \label{low-rank-sensing} \\ 
 & \; i=1,\dots,n. \nonumber
\end{align}
\cite{recht2010guaranteed} proves that under a restricted isometry property (RIP), minimizing the nuclear norm, or the sum of the singular values of $B$, over the affine subset, is guaranteed to produce the minimum-rank solution. \cite{jain2013low} later studies the performance of alternating minimization for matrix sensing and matrix completion problems. However, without any sparse regularization, a low-rank solution for $B$ could be a dense matrix where the nonzero entries correspond to almost all the edges in the network. 

Tensor regression models \cite{zhou2013tensor, zhou2014regularized, hoff2015multilinear, li2016sparse} provide a promising tool for estimating outcome-relevant subgraphs in this situation. Initially proposed for neuroimaging analysis, tensor regression methods can effectively exploit the array-valued covariates to identify regions of interest in brains that are relevant to a clinical response \cite{zhou2013tensor}. Considering a rank-$K$ tensor regression of the response on the matrix-valued network predictor, 
\begin{equation}
E(y_{i}\mid W_{i}) = \alpha+\sum_{k=1}^{K}\boldsymbol{\beta}_{1}^{(k)\top}W_{i}\boldsymbol{\beta}_{2}^{(k)}, \label{eq:rank-k-tensor-reg}
\end{equation}
where $\boldsymbol{\beta}_{d}^{(k)}\in\mathbb{R}^{V}, d=1,2; k=1,\dots,K$. 
The set of rank-1 coefficient component matrices $\{\boldsymbol{\beta}_{1}^{(k)}\boldsymbol{\beta}_{2}^{(k)\top}\}_{k=1}^K$ in the bilinear form \eqref{eq:rank-k-tensor-reg} naturally selects a collection of subgraphs where the nonzero edges are predictive of the response. However, the symmetric matrix predictor $W_i$ does not necessarily lead to a symmetric coefficient matrix estimate for 
\[
B = \sum_{k=1}^K \boldsymbol{\beta}_{1}^{(k)}\boldsymbol{\beta}_{2}^{(k)\top}
\]
in model \eqref{eq:rank-k-tensor-reg}, which makes the interpretation difficult.

We propose to use a symmetric bilinear model with $L_1$ penalty to estimate a set of small signal subgraphs. The model puts symmetry constraints on the coefficient matrix of tensor regression due to the symmetry in predictors - the adjacency matrices of undirected networks are symmetric. In this case, the block relaxation algorithm \cite{zhou2013tensor} of tensor regression cannot be applied. As far as we know, there is no available algorithm for estimating $L_1$-penalized symmetric bilinear regression in the literature. We therefore develop an effective algorithm based on the idea of the efficient coordinate descent algorithm \cite{friedman2010regularization} of lasso, which involves solving a sequence of conditional convex optimizations.

The rest of the paper is organized as follows. We describe the symmetric bilinear model and the special format of $L_1$ regularization in the next section. A coordinate descent algorithm for estimation of this model is introduced in Section \ref{estimate}. Section \ref{simulate} contains a simulation study demonstrating the good performance of our algorithm in recovering true signal clique subgraphs in high and low signal-to-noise ratio. We apply the method on brain connectome and cognitive traits data in Section \ref{application} to search for sub-structure in the brain that is relevant to certain cognitive ability. Section \ref{conclude} concludes.

\section{Symmetric Bilinear Regression with $L_1$ Regularization} 
\label{model}

The notations and symbols used in this paper are summarized in Table \ref{table_notation}. The classical linear model relates a vector-valued covariate $\boldsymbol{x}\in\mathbb{R}^{p}$ to the conditional expectation of the response $y$ via $E(y\mid\boldsymbol{x})=\alpha+\boldsymbol{\beta}^{\top}\boldsymbol{x}$. For a matrix-valued covariate $W\in\mathbb{R}^{V\times V}$, one can choose a coefficient matrix $B$ of the same size to capture the effect of each element. Then the linear model has the following form
\begin{equation}
E(y\mid W)=\alpha+\left\langle B,W\right\rangle, \label{eq:linear regression}
\end{equation}
where $\left\langle B,W\right\rangle =\mbox{trace} (B^{\top}W) =\mbox{vec}(B)^{\top}\mbox{vec}(W)$. If $W$ is symmetric, the coefficient matrix $B$ should also be symmetric. In this case, $B$ has the same number of parameters, $V(V-1)/2$, as $W$, which grows quadratically with $V$ and can quickly exceed the sample size $n$ when $V$ is large. For example, typical structural brain networks of size $68\times68$ require $68\times67/2=2278$ regression parameters. Hence, the goal is to approximate $B$ with fewer parameters. If $B$ admits a rank-$1$ decomposition
\[
B=\lambda\boldsymbol{\beta}\boldsymbol{\beta}^{\top}
\]
where $\boldsymbol{\beta}\in\mathbb{R}^{V}$, the linear part in \eqref{eq:linear regression} has the symmetric bilinear form
\[
E(y\mid W)=\alpha+\lambda\boldsymbol{\beta}^{\top}W\boldsymbol{\beta}.
\]

\begin{table}[hbt]
\caption{Notations and symbols used in this paper.}
\label{table_notation}
\centering
\begin{tabular}{l>{\raggedright}p{7cm}}
\toprule 
Symbols & Description\tabularnewline
\midrule
$y_{i}$ & scalar response of observation $i$\tabularnewline
$W_{i}$ & a $V\times V$ symmetric matrix predictor of observation $i$ \\with zero diagonal entries\tabularnewline
$W_{i[u\cdot]}$ & the $u$-th row of $W_{i}$\tabularnewline
$W_{i[\cdot u]}$ & the $u$-th column of $W_{i}$\tabularnewline
$W_{i[uv]}$ & the $(u,v)$ entry of $W_{i}$\tabularnewline
$W_{i}^{(u)}$ & $W_{i}$ with $u$-th row and $u$-th column set to zero\tabularnewline
$B$ & a $V\times V$ symmetric coefficient matrix\tabularnewline
$\alpha$ & intercept of regression\tabularnewline
$\lambda_{h}$ & scalar of component $h$ in decomposition \eqref{eq: decomposition_B}\tabularnewline
$\boldsymbol{\beta}_{h}$ & the $V\times 1$ vector of component $h$ in decomposition \eqref{eq: decomposition_B}\tabularnewline
$\beta_{hu}$ & the $u$-th entry of $\boldsymbol{\beta}_{h}$\tabularnewline
$\boldsymbol{\beta}_{d}^{(k)}$ & the $d$-th $V\times 1$ vector of component $k$ in tensor regression \eqref{eq:rank-k-tensor-reg} \tabularnewline
$K$ & the rank of decomposition \eqref{eq: decomposition_B}\tabularnewline
$\gamma$ & penalty factor\tabularnewline
$e_{i}^{(h)}$ & the partial residual of subject $i$ excluding the fitting \\from component $h$, $e_{i}^{(h)}=y_{i}-\alpha-\sum_{k\neq h}\lambda_{k}\boldsymbol{\beta}_{k}^{\top}W_{i}\boldsymbol{\beta}_{k}$\tabularnewline
$M_{u}$ & intermediate matrix, $M_{u}=\sum_{i=1}^{n}W_{i[\cdot u]}W_{i[u\cdot]}$\tabularnewline
$a_{hu}$ & intermediate scalar, \\$a_{hu}=2\lambda_{h}/n\cdot\sum_{i=1}^{n}(e_{i}^{(h)}-\lambda_{h}\boldsymbol{\beta}_{h}^{\top}W_{i}^{(u)}\boldsymbol{\beta}_{h})W_{i[u\cdot]}\boldsymbol{\beta}_{h}$\tabularnewline
$d_{hu}$ & intermediate scalar, $d_{hu}=4\lambda_{h}^{2}/n\cdot\boldsymbol{\beta}_{h}^{\top}M_{u}\boldsymbol{\beta}_{h}$\tabularnewline
$c_{h}$ & intermediate scalar, $c_{h}=\sum_{i=1}^{n}\boldsymbol{\beta}_{h}^{\top}W_{i}\boldsymbol{\beta}_{h}e_{i}^{(h)}/n$\tabularnewline
$b_{h}$ & intermediate scalar, $b_{h}=\sum_{i=1}^{n}(\boldsymbol{\beta}_{h}^{\top}W_{i}\boldsymbol{\beta}_{h})^{2}/n$\tabularnewline
\bottomrule
\end{tabular}
\end{table}

A more flexible symmetric bilinear model would be a rank-$K$ approximation to the general coefficient matrix $B$. Specifically, suppose $B$ admits a rank-$K$ decomposition 
\begin{equation}
B=\sum_{h=1}^{K}\lambda_{h}\boldsymbol{\beta}_{h}\boldsymbol{\beta}_{h}^{\top}, \label{eq: decomposition_B}
\end{equation}
where $\boldsymbol{\beta}_h\in\mathbb{R}^{V}$, $\lambda_h \in \mathbb{R}$, $h=1,\dots,K$. We do not constrain $\{\boldsymbol{\beta}_h\}_{h=1}^K$ to be orthogonal or linearly independent, because we want the component matrices $\{\lambda_{h}\boldsymbol{\beta}_{h}\boldsymbol{\beta}_{h}^{\top}\}_{h=1}^{K}$ to be sparse, while such constraints discourage sparsity and do not provide interpretable results in practice. Therefore the rank $K$ in \eqref{eq: decomposition_B} refers to the number of component matrices $\{\lambda_{h}\boldsymbol{\beta}_{h}\boldsymbol{\beta}_{h}^{\top}\}_{h=1}^{K}$ instead of rank($B$). Note that $\{\lambda_h\}_{h=1}^K$ is necessary in the decomposition \eqref{eq: decomposition_B} as we don't want to constrain $B$ to be positive semi-definite.

The decomposition \eqref{eq: decomposition_B} leads to a rank-$K$ symmetric bilinear regression model
\begin{eqnarray}
E(y\mid W) & = & \alpha+\left\langle \sum_{h=1}^{K}\lambda_{h}\boldsymbol{\beta}_{h}\boldsymbol{\beta}_{h}^{\top},W\right\rangle \nonumber \\
 & = & \alpha+\sum_{h=1}^{K}\lambda_{h}\boldsymbol{\beta}_{h}^{\top}W\boldsymbol{\beta}_{h}.\label{eq:rank-K regression}
\end{eqnarray}
The decomposition \eqref{eq: decomposition_B} may not be unique even up to permutation and scaling \cite{sidiropoulos2000uniqueness, liu2001cramer, de2006link}. Hence, we introduce an $L_1$ penalty on the entries of component matrices $\{\lambda_{h}\boldsymbol{\beta}_{h}\boldsymbol{\beta}_{h}^{\top}\}_{h=1}^{K}$ to ensure both the identifiability of the model and the sparsity of the coefficient components $\{\lambda_{h}\boldsymbol{\beta}_{h}\boldsymbol{\beta}_{h}^{\top}\}_{h=1}^{K}$. 
The loss function of model \eqref{eq:rank-K regression} under $L_1$ regularization is given by
\begin{eqnarray}
\dfrac{1}{2n}\sum_{i=1}^{n}\left(y_{i}-\alpha-\sum_{h=1}^{K}\lambda_{h}\boldsymbol{\beta}_{h}^{\top}W_{i}\boldsymbol{\beta}_{h}\right)^{2} \nonumber \\
\qquad+\gamma\sum_{h=1}^{K}\left|\lambda_{h}\right|\sum_{u=1}^{V}\sum_{v<u}\left|\beta_{hu}\beta_{hv}\right|
\label{objective_function}
\end{eqnarray}
where $\gamma$ is a penalty factor that can be optimized via test data or cross validation in practice.
Here we choose to penalize the sum of absolute values of the lower-triangular entries in the matrices $\{\lambda_{h}\boldsymbol{\beta}_{h}\boldsymbol{\beta}_{h}^{\top}\}_{h=1}^{K}$ instead of the $L_1$ norms of the vectors $\{\boldsymbol{\beta}_{h}\}_{h=1}^{K}$ for two reasons: (i) this form achieves an adaptive penalty on each $\beta_{hu}$ (the $u$-th entry of $\boldsymbol{\beta}_{h}$) given others; (ii) this form avoids scaling problems between $\lambda_h$ and $\boldsymbol{\beta}_{h}$. 

Regarding (i), by ``adaptive penalty" we mean that the penalty factor for $\beta_{hu}$ in \eqref{objective_function} given all the other parameters tends to be high with many nonzero entries in $\boldsymbol{\beta}_{h}$ and low with few nonzero entries. Refer to Section \ref{update_beta} for technical details on this property. Overall, this conditional adaptive $L_1$ penalty will lead to sparser matrix estimates for $\{\lambda_{h}\boldsymbol{\beta}_{h}\boldsymbol{\beta}_{h}^{\top}\}_{h=1}^{K}$ than simply penalizing the $L_1$ norms of $\{\boldsymbol{\beta}_{h}\}_{h=1}^{K}$. 

Regarding (ii), note that our main interest is in the nonzero entries in the coefficient matrices $\{\lambda_{h}\boldsymbol{\beta}_{h}\boldsymbol{\beta}_{h}^{\top}\}_{h=1}^{K}$ instead of $\{\lambda_{h}\}_{h=1}^{K}$ and $\{\boldsymbol{\beta}_{h}\}_{h=1}^{K}$ separately. Therefore we want to ensure that each component matrix $\lambda_{h}\boldsymbol{\beta}_{h}\boldsymbol{\beta}_{h}^{\top}$ is identifiable when minimizing the loss function comprising two parts: the mean squared error (MSE) and the $L_1$ regularization term as in \eqref{objective_function}. If we only penalized the $L_1$ norms of $\{\boldsymbol{\beta}_{h}\}_{h=1}^{K}$ as in the regularized tensor regressions \cite{zhou2013tensor}, the loss function would be reduced by simply manipulating the scales of $\lambda_h$ and $\boldsymbol{\beta}_{h}$ simultaneously. For example, if we shrink $\boldsymbol{\beta}_{h}$ to be $0.1\boldsymbol{\beta}_{h}$ and grow $\lambda_h$ to be $100\lambda_h$ so that the matrix $\lambda_{h}\boldsymbol{\beta}_{h}\boldsymbol{\beta}_{h}^{\top}$ remains unchanged, the MSE would stay the same but the $L_1$ regularization term would decline, making the loss function decrease. Therefore the component matrix $\lambda_{h}\boldsymbol{\beta}_{h}\boldsymbol{\beta}_{h}^{\top}$ is non-identifiable under such regularization form. However, if we use the $L_1$ regularization form in \eqref{objective_function}, the loss function \eqref{objective_function} will not be affected when changing the scales for both $\lambda_h$ and $\boldsymbol{\beta}_{h}$ while leaving the matrix $\lambda_{h}\boldsymbol{\beta}_{h}\boldsymbol{\beta}_{h}^{\top}$ unchanged. This ensures the identifiability of the matrix $\lambda_{h}\boldsymbol{\beta}_{h}\boldsymbol{\beta}_{h}^{\top}$ when minimizing \eqref{objective_function}. The $L_1$ regularization form in \eqref{objective_function} also saves us from putting unit length constraints on $\{\boldsymbol{\beta}_{h}\}_{h=1}^{K}$, as often done in CP decomposition \cite{allen2012sparse}, while such constraints would make the optimization more difficult.

The symmetric bilinear model achieves the goal of reducing parameters while maintaining flexibility. Model \eqref{eq:rank-K regression} only has $(1+K+KV)$ parameters, which is much smaller than the number of parameters, $(1+V(V-1)/2)$, in the unstructured linear model \eqref{eq:linear regression} when $V$ is large and $K \ll V$. According to \cite{zhou2013tensor}, such a massive reduction in dimensionality provides a reasonable approximation to many low-rank signals. If the true signal edges in the undirected network form several clique subgraphs, the symmetric bilinear model \eqref{eq:rank-K regression} will be much more efficient in requiring many fewer parameters to capture the structure. If this is not the case, model \eqref{eq:rank-K regression} is still flexible at capturing any structure of signal edges in the network with $K$ being large. For example, if we set $K = V(V-1)/2$ and choose $\{\boldsymbol{\beta}_{h}\}_{h=1}^{K}=\{\boldsymbol{e}_{u}+\boldsymbol{e}_{v}\}_{u<v}$ where $\{\boldsymbol{e}_{u}\}_{u=1}^{V}$ is the standard basis for $\mathbb{R}^{V}$, then the symmetric bilinear model \eqref{eq:rank-K regression} becomes unstructured linear regression \eqref{eq:linear regression} and equivalent to usual lasso. 

The interpretation of the symmetric bilinear model \eqref{eq:rank-K regression} is very appealing in the context of networks. The nonzero entries in each coefficient component matrix $\lambda_{h}\boldsymbol{\beta}_{h}\boldsymbol{\beta}_{h}^{\top}$ locate a clique subgraph where the edge weight between any two nodes is relevant to the response, and the number of nodes equals the number of nonzero entries in $\boldsymbol{\beta}_{h}$.

\section{Estimation Algorithm}
\label{estimate}

The parameters of the symmetric bilinear model \eqref{eq:rank-K regression} are estimated by minimizing the loss function \eqref{objective_function}
\begin{align}
\underset{\alpha,\{\lambda_{h}\},\{\boldsymbol{\beta}_{h}\}}{\min}\  & \dfrac{1}{2n}\sum_{i=1}^{n}\left(y_{i}-\alpha-\sum_{h=1}^{K}\lambda_{h}\boldsymbol{\beta}_{h}^{\top}W_{i}\boldsymbol{\beta}_{h}\right)^{2} \nonumber \\
& \quad+\gamma\sum_{h=1}^{K}\left|\lambda_{h}\right|\sum_{u=1}^{V}\sum_{v<u}\left|\beta_{hu}\beta_{hv}\right|. \label{eq:SBL_optimization}
\end{align}
Note that $K$ is fixed in our model \eqref{eq:SBL_optimization} and the selection of $K$ in practice is discussed in Section \ref{remarks}.

We consider a coordinate descent step \cite{schmidt2005least, ruszczynski2006nonlinear} for solving \eqref{eq:SBL_optimization}. Note that the objective function in \eqref{eq:SBL_optimization} is a fourth order with $\boldsymbol{\beta}_{h}$. Therefore the block relaxation algorithm \cite{zhou2013tensor}, which alternatively updates each component vector, is not efficient for \eqref{eq:SBL_optimization}, because partially optimizing $\boldsymbol{\beta}_{h}$ when fixing the other parameters is not a convex problem and there is no closed form solution. However, since the undirected networks of interest do not have self loops, the diagonal of each adjacency matrix $W_i$ can be set to zero. In this case, the objective function in \eqref{eq:SBL_optimization} is indeed a partial convex function of each entry $\beta_{hu}$ in $\boldsymbol{\beta}_{h}$ and has an analytical form solution, which makes coordinate descent very appealing in solving \eqref{eq:SBL_optimization}. The challenge then lies in deriving the closed form update for each parameter due to the nonsmoothness of the objective function in \eqref{eq:SBL_optimization} and the technical details are discussed below. 

\subsection{Updates for entries in $\{\boldsymbol{\beta}_{h}\}_{h=1}^{K}$}
\label{update_beta}

Suppose we want to optimize with respect to $\beta_{hu}$, the $u$-th entry in $\boldsymbol{\beta}_{h}$, given all the other parameters. The problem becomes 
\begin{align}
\underset{\beta_{hu}}{\min} & \ L_{\beta,h}(\beta_{hu}) = f_{h}(\lambda_h, \boldsymbol{\beta}_{h})+\left(\gamma\left|\lambda_{h}\right|\sum_{v\neq u}\left|\beta_{hv}\right| \right)\left|\beta_{hu}\right|, 
\label{eq:obj_beta_hu}
\end{align}
where 
\begin{equation}
f_{h}(\lambda_h, \boldsymbol{\beta}_{h})=\dfrac{1}{2n}\sum_{i=1}^{n}(e_{i}^{(h)}-\lambda_{h}\boldsymbol{\beta}_{h}^{\top}W_{i}\boldsymbol{\beta}_{h})^{2}, \label{eq:RSS}
\end{equation}
and $e_{i}^{(h)}$ is the partial residual of subject $i$ excluding the fitting from component $h$, 
\[e_{i}^{(h)}=y_{i}-\alpha-\sum_{k\neq h}\lambda_{k}\boldsymbol{\beta}_{k}^{\top}W_{i}\boldsymbol{\beta}_{k}.\]
An important remark on \eqref{eq:obj_beta_hu} is that the penalty factor for $\left|\beta_{hu}\right|$, $\gamma\left|\lambda_{h}\right|\sum_{v\neq u}\left|\beta_{hv}\right|$, is related to the nonzero entries in $\boldsymbol{\beta}_{h}$ excluding $\beta_{hu}$. Hence $\beta_{hu}$ is more likely to be shrunk to zero if the current number of nonzero entries in $\boldsymbol{\beta}_{h}$ is large. This adaptive penalty will lead to a set of sparse vectors $\{\boldsymbol{\beta}_{h}\}_{h=1}^{K}$ and hence a set of small signal subgraphs.

Since the diagonal elements of each $W_i$ are all equal to zero, $f_{h}(\lambda_h, \boldsymbol{\beta}_{h})$ is actually a partial quadratic function of $\beta_{hu}$ given $\{\beta_{hv}\}_{v\neq u}$ and hence a partial convex function of $\beta_{hu}$ with
\begin{eqnarray}
\dfrac{\partial f_{h}}{\partial\beta_{hu}} & = & -\dfrac{2\lambda_{h}}{n}\sum_{i=1}^{n}\left(e_{i}^{(h)}-\lambda_{h}\boldsymbol{\beta}_{h}^{\top}W_{i}\boldsymbol{\beta}_{h}\right)W_{i[u\cdot]}\boldsymbol{\beta}_{h}  \label{eq:1st_deriv_beta_hu}\\
\dfrac{\partial^{2}f_{h}}{\partial\beta_{hu}^{2}} & = & \dfrac{4\lambda_{h}^{2}}{n}\sum_{i=1}^{n}\left(W_{i[u\cdot]}\boldsymbol{\beta}_{h}\right)^{2}\geq0  \label{eq:2nd_deriv_beta_hu}
\end{eqnarray}
where $W_{i[u\cdot]}$ is the $u$-th row of $W_i$ and $W_{i[\cdot u]}$ is the $u$-th column of $W_i$ below. To find the optimal $\beta_{hu}$, we write \eqref{eq:1st_deriv_beta_hu} as
\begin{align}
\dfrac{\partial f_{h}}{\partial\beta_{hu}} = & -\dfrac{2\lambda_{h}}{n}\sum_{i=1}^{n} \left(e_{i}^{(h)}-\lambda_{h}\boldsymbol{\beta}_{h}^{\top}W_{i}^{(u)}\boldsymbol{\beta}_{h} \right)W_{i[u\cdot]}\boldsymbol{\beta}_{h} \nonumber \\
& \quad +\dfrac{4\lambda_{h}^{2}}{n}\left(\boldsymbol{\beta}_{h}^{\top}M_{u}\boldsymbol{\beta}_{h}\right)\beta_{hu}, \label{eq:1st_deriv_beta_hu_explicit}
\end{align} 
where $W_{i}^{(u)}$ is $W_i$ with $u$-th row and $u$-th column set to zero, and $M_{u}=\sum_{i=1}^{n}W_{i[\cdot u]}W_{i[u\cdot]}$. Let $a_{hu} = 2\lambda_{h} / n \cdot \sum_{i=1}^{n}(e_{i}^{(h)}-\lambda_{h}\boldsymbol{\beta}_{h}^{\top}W_{i}^{(u)}\boldsymbol{\beta}_{h})W_{i[u\cdot]}\boldsymbol{\beta}_{h}$ and $d_{hu} = 4\lambda_{h}^{2} / n \cdot \boldsymbol{\beta}_{h}^{\top}M_{u}\boldsymbol{\beta}_{h}$. Note that $W_{i[uu]}=0$, so $a_{hu}$ and $d_{hu}$ do not depend on $\beta_{hu}$. Therefore the first derivative $\partial f_{h}/\partial\beta_{hu}$ is a linear function of $\beta_{hu}$.

The derivative of the second term in the objective function of \eqref{eq:obj_beta_hu} with respect to 
$\beta_{hu}$ only exists if $\beta_{hu} \neq 0$.  Hence
\begin{align}
\dfrac{\partial L_{\beta,h}}{\partial\beta_{hu}} & =\begin{cases}
- a_{hu} + d_{hu} \beta_{hu}+\gamma\left|\lambda_{h}\right|\sum_{v\neq u}\left|\beta_{hv}\right|, \ \mbox{if }\beta_{hu}>0\\
- a_{hu} + d_{hu} \beta_{hu}-\gamma\left|\lambda_{h}\right|\sum_{v\neq u}\left|\beta_{hv}\right|, \ \mbox{if }\beta_{hu}<0
\end{cases} \label{eq:deriv_loss_beta}
\end{align}
Simple calculus \cite{friedman2007pathwise} shows that the solution to \eqref{eq:obj_beta_hu} has the soft-thresholding form
\begin{eqnarray}
\hat{\beta}_{hu}=\dfrac{1}{d_{hu}}\mbox{sign}(a_{hu})\left(\left|a_{hu}\right|-\gamma\left|\lambda_{h}\right|\sum_{v\neq u}\left|\beta_{hv}\right|\right)_{+}. \label{eq:optim_beta_hu} 
\end{eqnarray}
Thus \eqref{eq:optim_beta_hu} gives the analytical form for coordinate-wise update for $\{\beta_{hu}:h=1,\dots,K;u=1,\dots,V\}$. The computational complexity of updating each entry $\beta_{hu}$ is $O(nV^2)$ and hence that of updating $\{\boldsymbol{\beta}_{h}\}_{h=1}^{K}$ is $O(nKV^3)$. This step requires storing a $V\times V$intermediate matrix $M_u$ for each $u=1,\dots,V$ and a $V\times K$ matrix for $\{\boldsymbol{\beta}_{h}\}_{h=1}^{K}$, and therefore the memory complexity is $O(V^3 + VK)$. 

\subsection{Updates for $\{\lambda_{h}\}_{h=1}^{K}$ }

Partial optimization with respect to each $\lambda_{h}$ while fixing other parameters, solves the following convex optimization
\begin{align}
\underset{\lambda_{h}}{\min} & \ L_{\lambda,h}(\lambda_{h})=f_{h}(\lambda_{h},\boldsymbol{\beta}_{h}) + \left(\gamma\sum_{u=1}^{V}\sum_{v<u}\left|\beta_{hu}\beta_{hv}\right| \right)\left|\lambda_{h}\right|. \label{eq:obj_lambda}
\end{align}
The derivative of $L_{\lambda,h}$ only exists if $\lambda_{h}\neq0$ and has a similar form to \eqref{eq:deriv_loss_beta} as
\begin{equation}
\dfrac{\partial L_{\lambda,h}}{\partial\lambda_{h}}=\begin{cases}
-c_{h}+b_{h}\lambda_{h}+\gamma\sum_{u=1}^{V}\sum_{v<u}\left|\beta_{hu}\beta_{hv}\right|,\  \mbox{if }\lambda_{h}>0\\
-c_{h}+b_{h}\lambda_{h}-\gamma\sum_{u=1}^{V}\sum_{v<u}\left|\beta_{hu}\beta_{hv}\right|,\  \mbox{if }\lambda_{h}<0
\end{cases}\label{eq:deriv_loss_lambda}
\end{equation}
where $c_{h}=\sum_{i=1}^{n}\boldsymbol{\beta}_{h}^{\top}W_{i}\boldsymbol{\beta}_{h}e_{i}^{(h)}/n$ and $b_{h}=\sum_{i=1}^{n}(\boldsymbol{\beta}_{h}^{\top}W_{i}\boldsymbol{\beta}_{h})^{2}/n$. The coordinate-wise update for each $\lambda_{h}$ has the form
\begin{equation}
\hat{\lambda}_{h}=\dfrac{1}{b_{h}}\mbox{sign}(c_{h})\left(\left|c_{h}\right|-\gamma\sum_{u=1}^{V}\sum_{v<u}\left|\beta_{hu}\beta_{hv}\right|\right)_{+}, \ h=1,\dots,K. \label{eq:optim_lambda_h}
\end{equation}
The computational complexity for updating $\{\lambda_{h}\}_{h=1}^{K}$ is $O(nKV^2)$. This step requires storing the intermediate results $\{\boldsymbol{\beta}_{h}^{\top}W_{i}\boldsymbol{\beta}_{h}: h=1,\dots,K; i=1,\dots,n\}$, which uses $O(nK)$ memory.

\subsection{Update for $\alpha$}

Given other parameters, the optimal $\alpha$ is
\begin{equation}
\hat{\alpha}=\dfrac{1}{n}\sum_{i=1}^{n} \left(y_{i}-\sum_{h=1}^{K}\lambda_{h}\boldsymbol{\beta}_{h}^{\top}W_{i}\boldsymbol{\beta}_{h} \right). \label{eq:optim_alpha}
\end{equation}
The computational and memory complexity of this step is $O(nK)$ and $O(1)$ respectively.

\subsection{Other details}
\label{remarks}

The above procedure is cycled through all the parameters until convergence, where the diagonal of each adjacency matrix $W_i$ is set to zero. This coordinate descent algorithm ensures the loss function to converge to a local minimum as each update always decreases the objective function in \eqref{eq:SBL_optimization} \cite{bezdek2002some}. In general, the algorithm should be run from multiple initializations to locate a good local minimum. One important remark is that although the entries in $\{\boldsymbol{\beta}_{h}\}_{h=1}^{K}$ and $\{\lambda_{h}\}_{h=1}^{K}$ have closed form solution of 0 under sufficiently large penalty factor $\gamma$, we cannot initialize them at zero as the results will get stuck at zero. Update form \eqref{eq:optim_beta_hu} and \eqref{eq:optim_lambda_h} imply that given others being zero, the optimal $\beta_{hu}$ or $\lambda_{h}$ will also be zero. In fact, we recommend to initialize all the parameters to be nonzero in case some components unexpectedly degenerate at the beginning. In practice, we initialize each $\beta_{hu}\sim U(-1,1)$ and initialize $\alpha$ and $\{\lambda_{h}\}_{h=1}^{K}$ by a least-square regression of $y_{i}$ on $\{\boldsymbol{\beta}_{h}^{\top}W_{i}\boldsymbol{\beta}_{h}\}_{h=1}^{K}$. 

Another remark relates to the invariance of loss function \eqref{objective_function} under rescaling between $\lambda_{h}$ and $\boldsymbol{\beta}_{h}$. The estimated component matrices $\{\lambda_{h}\boldsymbol{\beta}_{h}\boldsymbol{\beta}_{h}^{\top}\}_{h=1}^{K}$ from our algorithm do not depend on the magnitude of initial values for $\{\lambda_{h}\}_{h=1}^{K}$ and $\boldsymbol{\beta}_{h}$ as long as the initial matrices of $\{\lambda_{h}\boldsymbol{\beta}_{h}\boldsymbol{\beta}_{h}^{\top}\}_{h=1}^{K}$ remain unchanged.

Our proposed model \eqref{eq:rank-K regression} assumes a known rank $K$. In practice, we choose an upper bound for the rank, and then allow the $L_1$ penalty to discard unnecessary components, leading to a data-driven estimate of the rank. This has the distinct advantage of avoiding the introduction of an additional tuning parameter. That is, if we followed the usual model selection criteria to choose an optimal rank, such as BIC, AIC or cross validation \cite{zhou2013tensor}, this would incur heavy computational burden since we have to tune the $L_1$ penalty factor under each rank. We assess the performance of our procedure and verify its lack of sensitivity to the chosen upper bound in simulation studies of Section \ref{sensitivity K}. 

Considerable speedup is obtained by organizing the iterations around the nonzero parameters -- active set, as recommended in \cite{friedman2010regularization}. After a few complete cycles through all the parameters, we iterate on only the active set till convergence. The general procedure of the coordinate descent algorithm is summarized in Algorithm \ref{CD-SBL}.

\begin{algorithm}[htb]
\caption{Coordinate descent for $L_1$-penalized symmetric bilinear model \eqref{eq:SBL_optimization} \label{CD-SBL}}

\begin{algorithmic}[1]

\State \textbf{Input:} Adjacency matrices $W_{i}$ of size $V\times V$, outcome $y_{i}$, $i=1,\dots,n$; rank $K$, penalty factor $\gamma$, tolerance $\epsilon\in\mathbb{R}_{+}$.

\State \textbf{Output:} Estimates of $\alpha$, $\{\lambda_{h}\}_{h=1}^{K}$, $\{\boldsymbol{\beta}_{h}\}_{h=1}^K$.

\State Initialize $\{\boldsymbol{\beta}_{h}\}_{h=1}^K$ at nonzero random vectors;  initialize $\alpha$ and $\{\lambda_{h}\}_{h=1}^{K}$ by a least-square regression of $y_{i}$ on $\{\boldsymbol{\beta}_{h}^{\top}W_{i}\boldsymbol{\beta}_{h}\}_{h=1}^{K}$.  
\Repeat
	\For{$h=1:K$}
		\For{$u=1:V$}
			\State Update $\beta_{hu}$ by \eqref{eq:optim_beta_hu}
		\EndFor
	\EndFor
	\For{$h=1:K$}
		\State Update $\lambda_{h}$ by \eqref{eq:optim_lambda_h}
	\EndFor
	\State Update $\alpha$ by \eqref{eq:optim_alpha}
\Until{relative change of objective function \eqref{eq:SBL_optimization} $<\epsilon$}

\end{algorithmic}
\end{algorithm}

\section{Simulation Study}
\label{simulate}

In this section, we first conduct a number of simulation experiments to study the empirical  computational and memory complexity of Algorithm \ref{CD-SBL}. We then compare the inference results to several competitors.

\subsection{Computational and memory complexity}
\begin{figure}[htb]
\centering
\includegraphics[width=0.23\textwidth]{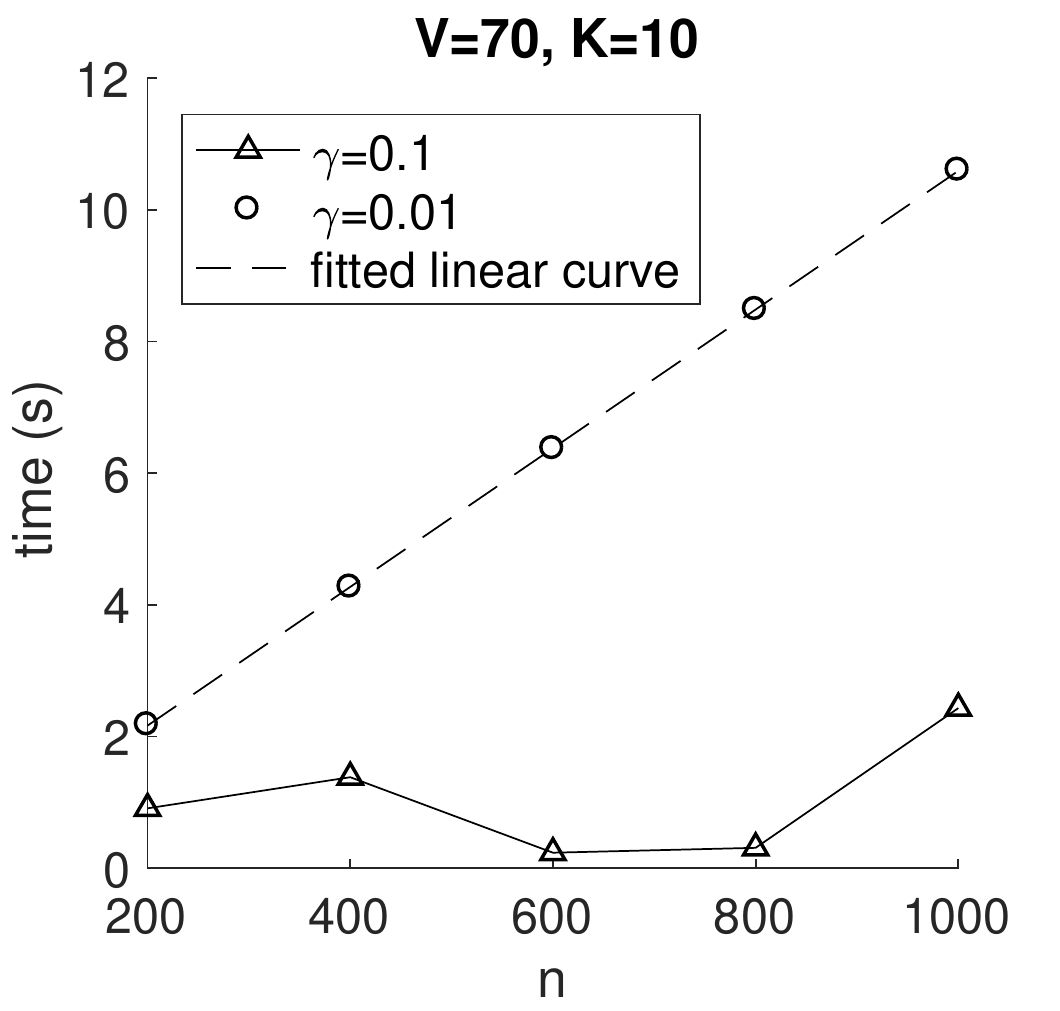}
\includegraphics[width=0.23\textwidth]{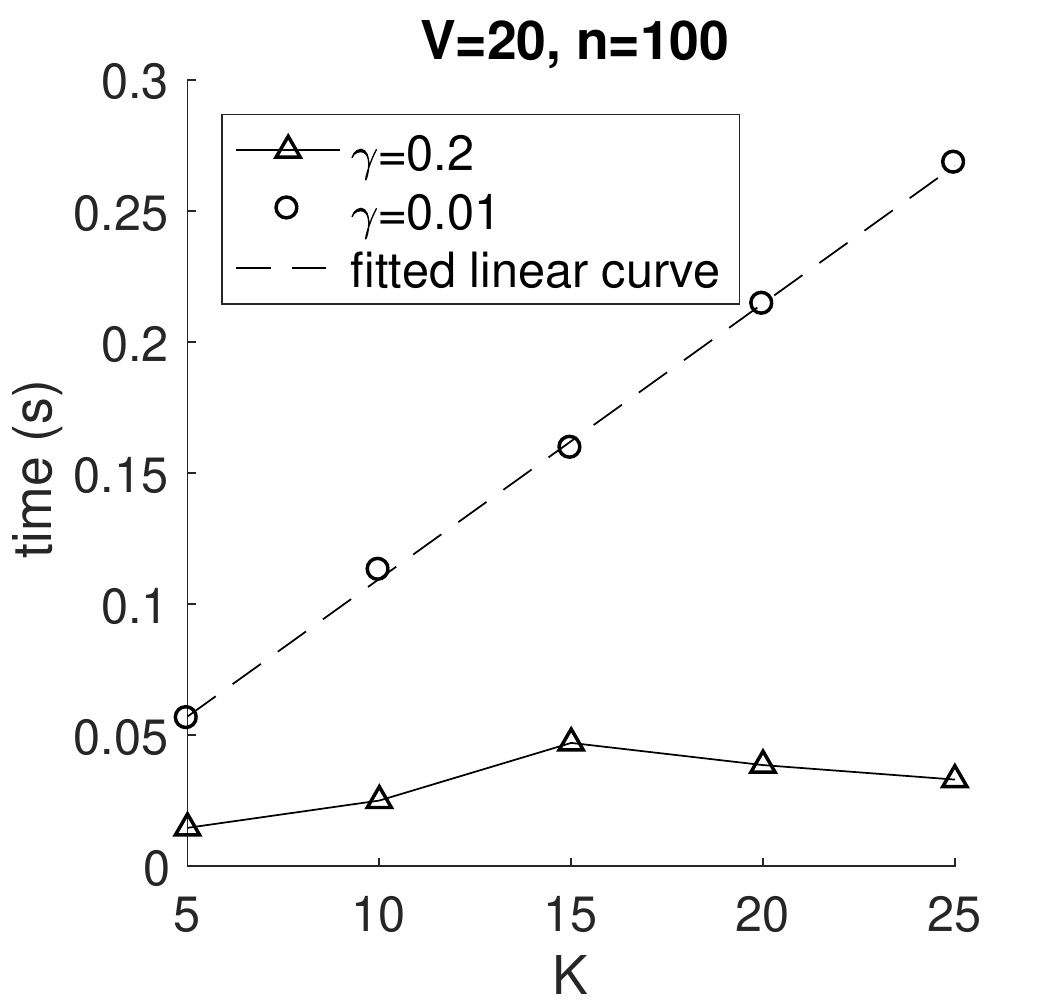}
\caption{Average computation time (in seconds) per iteration of Algorithm \ref{CD-SBL} for 30 runs versus the number of graphs n (left) and rank K (right). \label{fig:comp_complex_nK}}
\end{figure}

\begin{figure}[htb]
\centering
\includegraphics[width=0.25\textwidth]{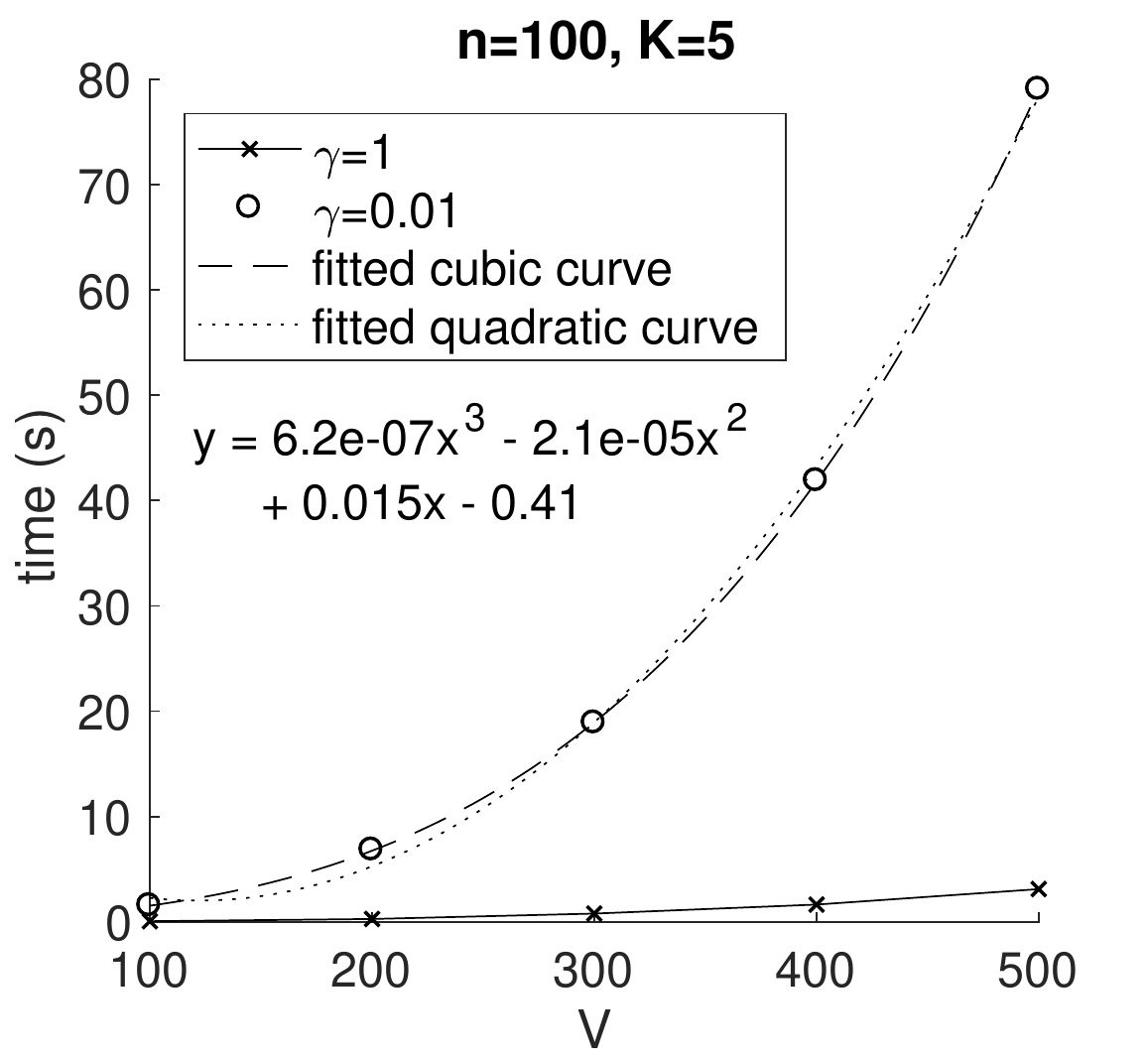}\includegraphics[width=0.25\textwidth]{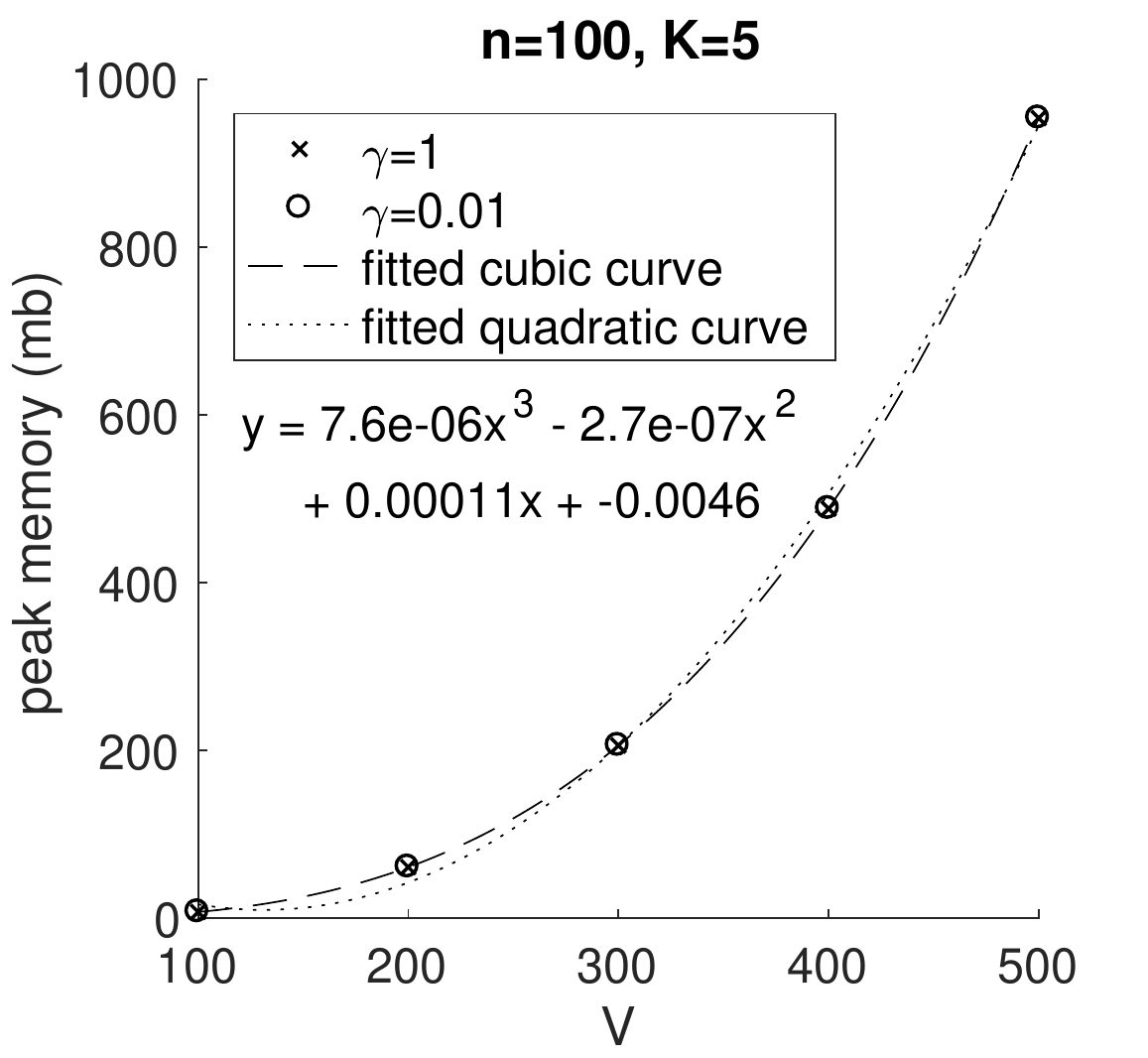}
\caption{Average computation time (in seconds) per iteration (left) and average peak memory (in mb) in use (right) during the execution of Algorithm \ref{CD-SBL} for 30 runs versus the number of nodes V. The equation of the fitted cubic curve is shown on either plot. \label{fig:comp_mem_V}}
\end{figure}

Algorithm \ref{CD-SBL} is implemented in Matlab (R2017a) and all the numerical experiments are conducted in a machine with one Intel Core i5 2.7 GHz processor and 8 GB of RAM. We simulated different number $n$ of observation pairs $\{(W_i, y_i): W_{i[uv]} = W_{i[vu]} \sim N(0,1), y_i \sim N(0,1) \}_{i=1}^n$ for different number of nodes $V$ (each $W_i$ is a $V\times V$ symmetric matrix with zero diagonal entries), and then assess how the execution time and peak memory (maximum amount of memory in use) increase with the problem size. In practice, the computational time of Algorithm \ref{CD-SBL} also depends on the penalty factor $\gamma$. When a small $\gamma$, e.g. $\gamma=0.01$, is applied so that most of the estimated parameters are nonzero, the runtime per iteration is a linear order with $n$ and $K$, and a cubic order with $V$ as shown in Figure \ref{fig:comp_complex_nK} and the left plot of Figure \ref{fig:comp_mem_V}. This is in accordance with the theoretical analysis of the computational complexity per iteration of Algorithm \ref{CD-SBL} in Section \ref{estimate}, which is $O(nKV^3)$ in the worst case scenario. But the computational time declines considerably when a large penalty $\gamma$, e.g. $\gamma=0.1, 0.2$ or $1$, is applied, which increases sparsity in the parameters. The reason is that some computation cost can be saved in the sparsity scenario even though we run complete cycles through all the parameters per iteration. For example, if some $\lambda_h$ becomes $0$ at a certain step, changing the entries of $\boldsymbol{\beta}_{h}$ will not affect the loss function \eqref{objective_function} and hence we could set $\boldsymbol{\beta}_{h}=\boldsymbol{0}$ later on (the component degenerates). 

The right plot in Figure \ref{fig:comp_mem_V} shows that the peak memory during the execution of Algorithm \ref{CD-SBL} is a cubic order with $V$ no matter what penalty factor is used. This is in accordance with the theoretical memory complexity of Algorithm \ref{CD-SBL}, $O(V^3+VK+nK)$, in Section \ref{estimate}. We do not show the peak memory of Algorithm \ref{CD-SBL} versus the number of observations $n$ or the rank $K$ here because the peak memory is dominated by the cubic term of $V$ and does not vary much with $n$ or $K$ in these cases.

Algorithm \ref{CD-SBL} was coded in the Matlab (R2017a) programming environment using no \texttt{C} or \texttt{FORTRAN} code. It is likely that the computational time of Algorithm \ref{CD-SBL} would improve relative to lasso or tensor regression if such code were used, as each iteration of Algorithm \ref{CD-SBL} involves for-loops over the elements of component vectors $\{\boldsymbol{\beta}_{h}\}_{h=1}^K$ which are particularly slow in Matlab.

\subsection{Inference on signal subgraphs}
In this experiment, we compare the performance of recovering true signal subgraphs among lasso,  TN-PCA (see model \eqref{eq:semi_sym_tensor_decomp}), tensor regression with $L_1$ regularization, low-rank sensing (LRS) model \eqref{low-rank-sensing} and symmetric bilinear regression with $L_1$ penalty (SBL). 

For tensor regression (TR), we consider a linear TR model \eqref{eq:rank-k-tensor-reg} with the same rank $K$ as in SBL. The penalty function in TR model has the form of $\rho\sum_{k=1}^{K}\sum_{d=1}^{2}\left\Vert \boldsymbol{\beta}_{d}^{(k)}\right\Vert _{1}$, where $\rho$ is the tuning parameter. Considering the symmetric property of the matrix predictor in this case, a naive method is to symmetrize the TR estimator $\hat{B}$ by $(\hat{B}+\hat{B}^{\top})/2$. Then the symmetrized component matrices 
\[
\left\{\left(\hat{\boldsymbol{\beta}}_{1}^{(k)}\hat{\boldsymbol{\beta}}_{2}^{(k)\top}+\hat{\boldsymbol{\beta}}_{2}^{(k)}\hat{\boldsymbol{\beta}}_{1}^{(k)\top}\right)/2\right\}_{k=1}^{K}
\]
assumably locate the signal subgraphs. We refer to this method as \emph{naive TR} later on.

\begin{figure}[htb]
    \centering
    \includegraphics[width=0.3\textwidth]{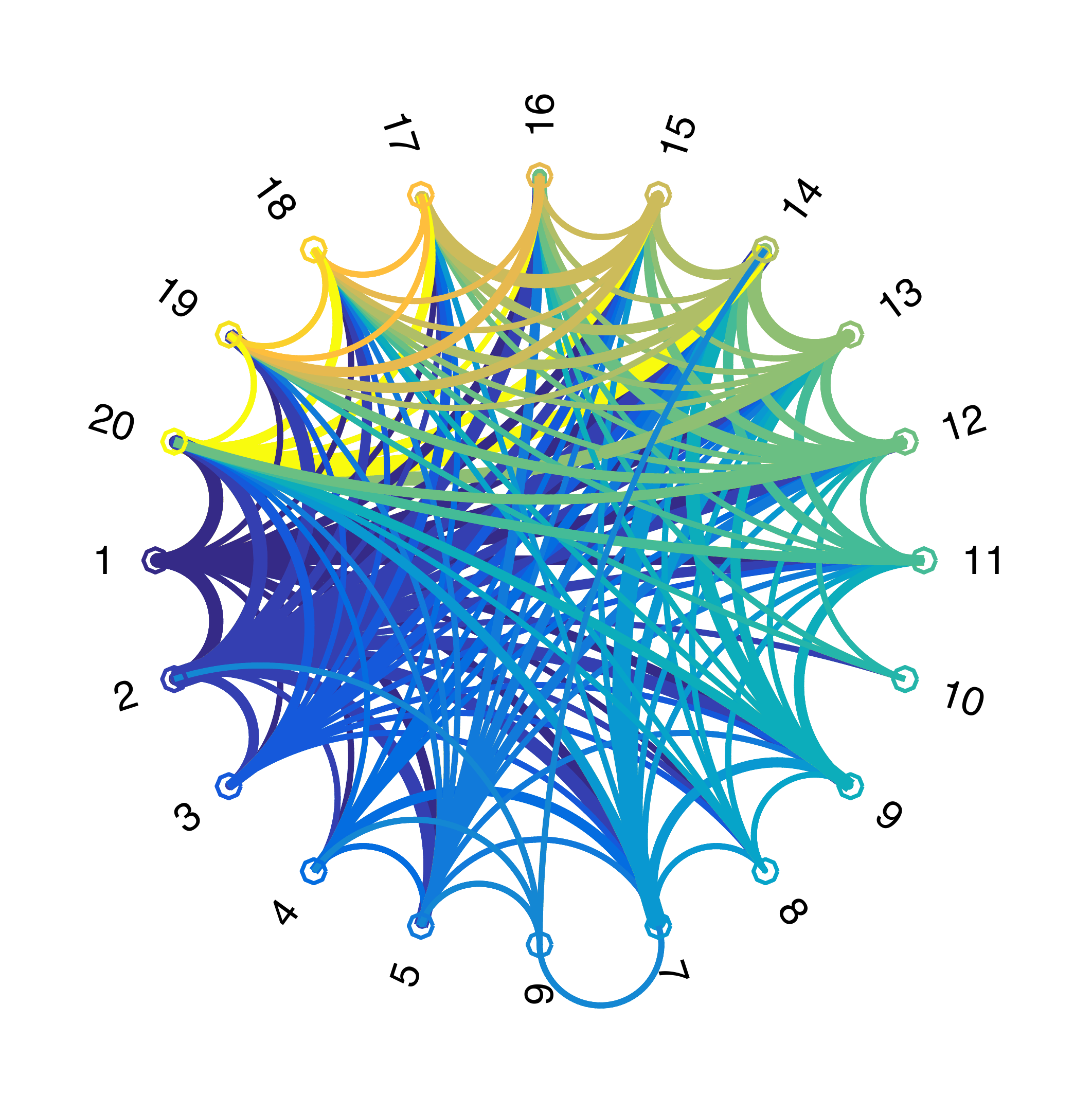} 
    \caption{Overlay of 10 basis subgraphs corresponding to $\{\boldsymbol{q}_{h}\boldsymbol{q}_{h}^{\top} \}_{h=1}^{10}$.  \label{fig:basis_subgraph_overlay}}
\end{figure}

We simulate a synthetic dataset consisting of 100 pairs of observations $\{(W_{i},y_{i}): i=1,\dots,100\}$ as follows. Each pair consists of a $20\times20$ adjacency matrix $W_{i}$ and a scalar $y_{i}\in\mathbb{R}$. Specifically, each network $W_{i}$ is generated from a set of basis subgraphs with an individual loading vector as
\begin{eqnarray}
W_{i} & = & \sum_{h=1}^{10}\lambda_{ih}\boldsymbol{q}_{h}\boldsymbol{q}_{h}^{\top}+\Delta_{i}, \label{eq:simu_dgp}
\end{eqnarray}
where $\boldsymbol{q}_{h}\in\{0,1\}^{20}$ is a random binary vector with $ \left\Vert \boldsymbol{q}_{h}\right\Vert _{0}=h+1$, $h=1,\dots,10$. 

The loadings $\{\lambda_{ih}\}$ in \eqref{eq:simu_dgp} are generated independently from $N(0,1)$. $\Delta_{i}$ is a symmetric $20\times20$ noise matrix with each entry $\Delta_{i[uv]}\sim N(0,0.1^{2})$, $u>v$. This generating process produces dense networks with complex structure. Figure \ref{fig:basis_subgraph_overlay} visualizes the 10 basis subgraphs superimposed together. 

\begin{figure}[htb]
\centering
\includegraphics[width=0.4\textwidth]{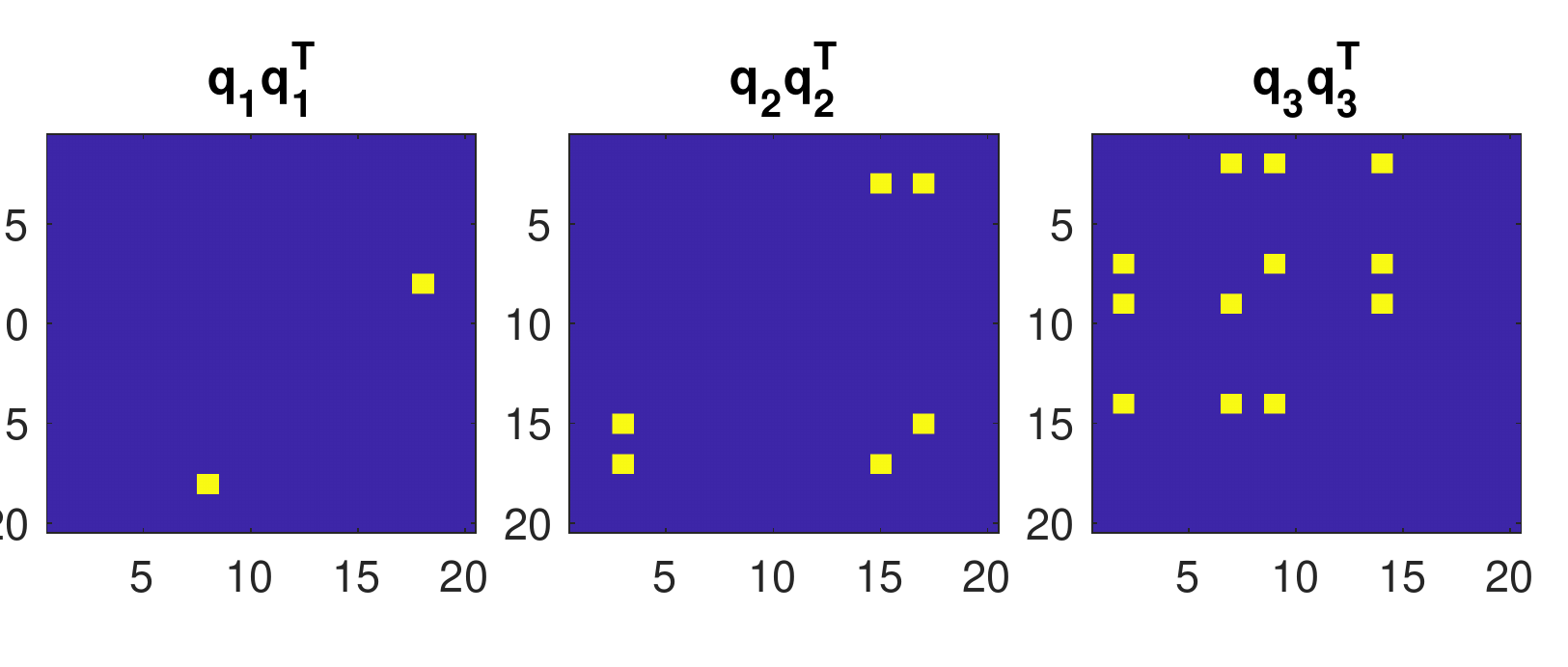} \\
\includegraphics[width=0.1\textwidth]{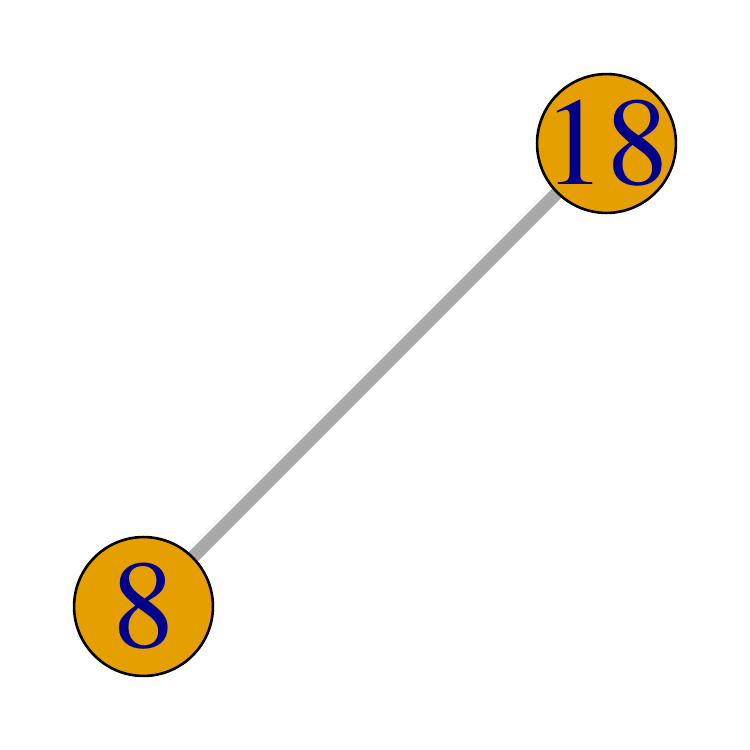} \includegraphics[width=0.11\textwidth]{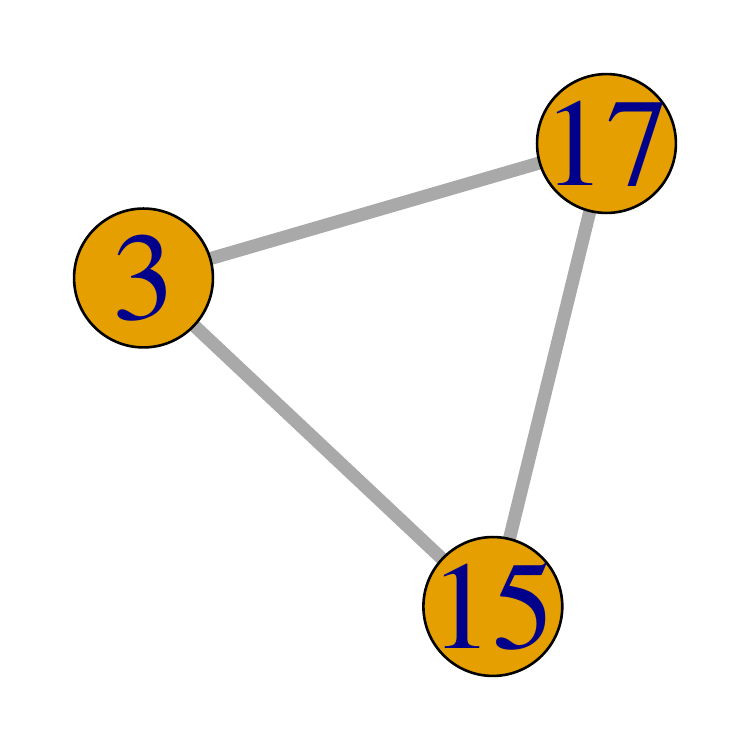} \includegraphics[width=0.12\textwidth]{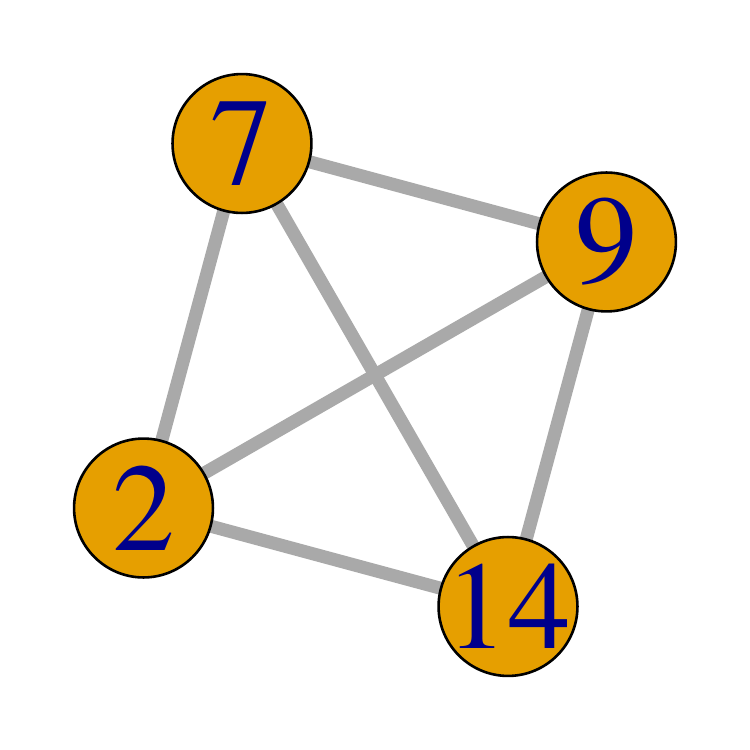} 
\caption{True signal subgraphs in simulation: $\{\boldsymbol{q}_{h}\boldsymbol{q}_{h}^{\top}:h=1,2,3\}$ (upper panel) and the corresponding clique subgraphs (lower panel). \label{fig:true_signal}}
\end{figure}

The response $y_{i}$ is generated by
\begin{equation}
y_{i}=\boldsymbol{q}_{1}^{\top}W_{i}\boldsymbol{q}_{1}+\boldsymbol{q}_{2}^{\top}W_{i}\boldsymbol{q}_{2}+\boldsymbol{q}_{3}^{\top}W_{i}\boldsymbol{q}_{3}+\varepsilon_{i}, \label{eq:simu_y}
\end{equation}
where $\varepsilon_{i}\sim N(0,\sigma^{2})$. We consider two noise levels: $\sigma=10\%$ and $100\%$ of the standard deviation of the conditional mean $E(y_i \mid W_i)$. The generating process \eqref{eq:simu_y} indicates that the true signal subgraphs relevant to $y_i$ correspond to the first three basis subgraphs $\{\boldsymbol{q}_{h}\boldsymbol{q}_{h}^{\top}:h=1,2,3\}$ as displayed in Figure \ref{fig:true_signal}, so that the true signal subgraphs have nontrivial variations across observations, as is often the case in practice.

\subsubsection{High signal-to-noise ratio}
\label{low nsr}

In this case, we set the noise level $\sigma=10\%$ of the standard deviation of the conditional mean $E(y_i \mid W_i)$ in the generating process \eqref{eq:simu_y}.

The input parameters of Algorithm \ref{CD-SBL} for SBL are set as follows. $K$ is set at 5 and the tolerance $\epsilon=10^{-5}$ in this simulation study. It is easy to find a \textit{roughly} smallest value $\gamma_{\max}$ for which $\{\boldsymbol{\beta}_{h}\}_{h=1}^{K}$ and $\{\lambda_{h}\}_{h=1}^{K}$ become zero. We set $\gamma_{\min}=0.01\gamma_{\max}$ and choose a sequence of 50 equally spaced $\gamma$ values on the logarithmic scale. 

\begin{figure}[hbt]
\centering
\includegraphics[width=.25\textwidth]{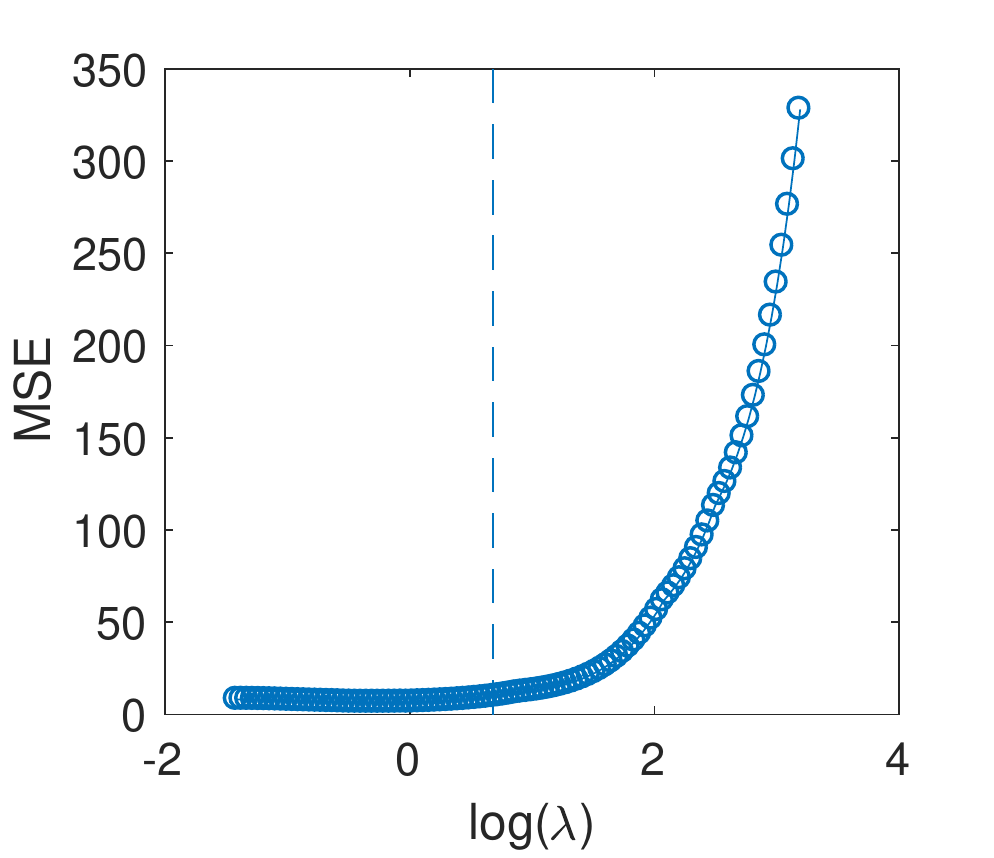}\includegraphics[width=0.25\textwidth, trim={0 0.1cm 0.8cm 0.8cm}, clip]{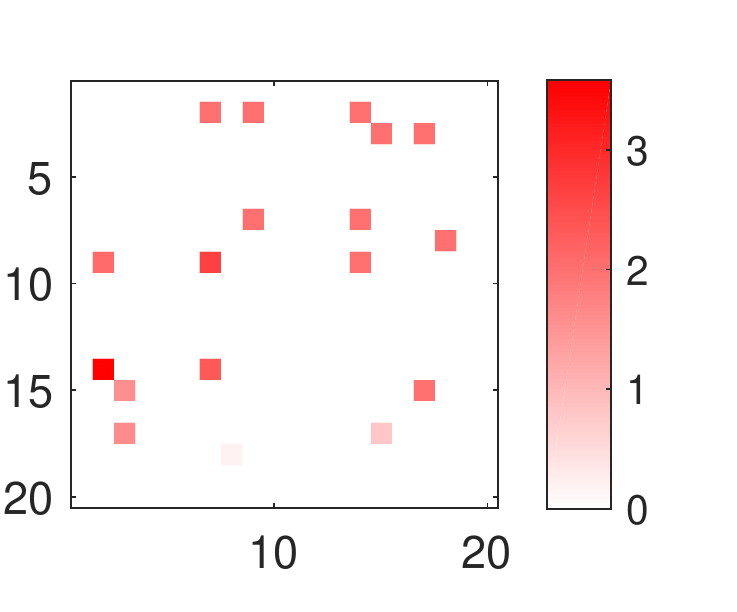}
\caption{Left: out-of-sample MSE from lasso under high signal-to-noise ratio. Right: estimated coefficients from lasso (lower-triangular) where the $L_1$ penalty factor is set corresponding to the vertical line on the left plot; the true coefficients for each edge of the network are shown in the upper triangle. \label{fig:sim_lasso}}
\end{figure}

\begin{figure}[htb]
    \centering
    \includegraphics[width=.25\textwidth]{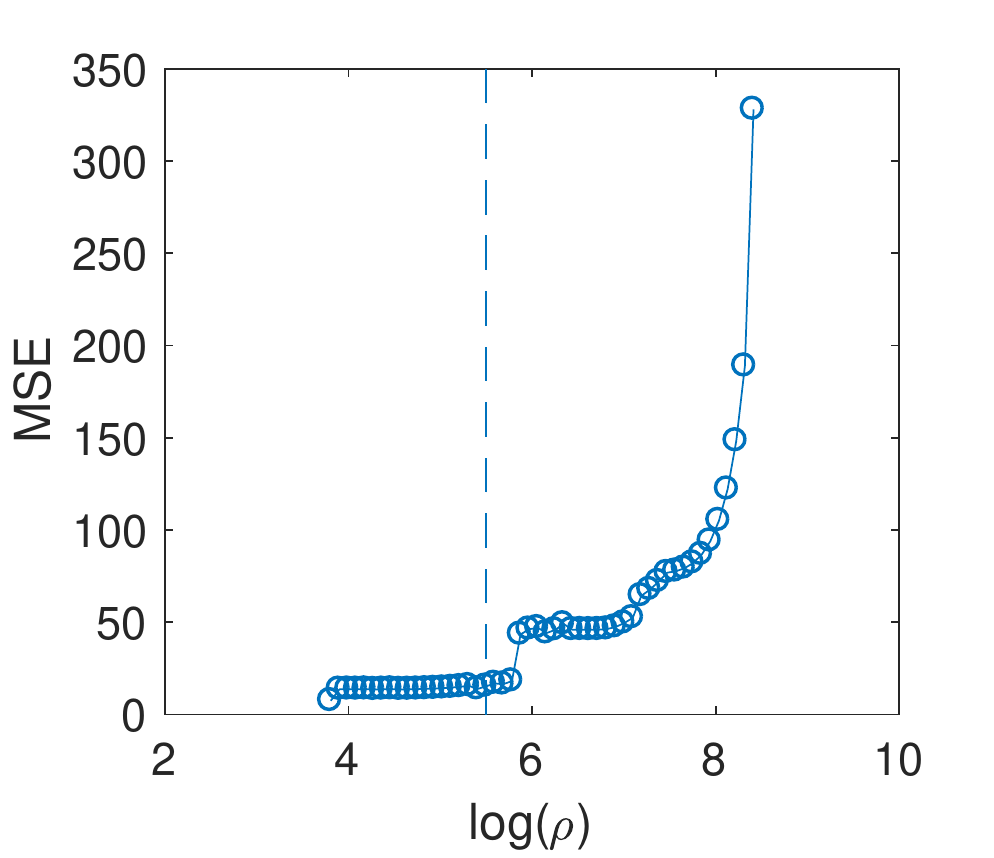}\includegraphics[width=.25\textwidth]{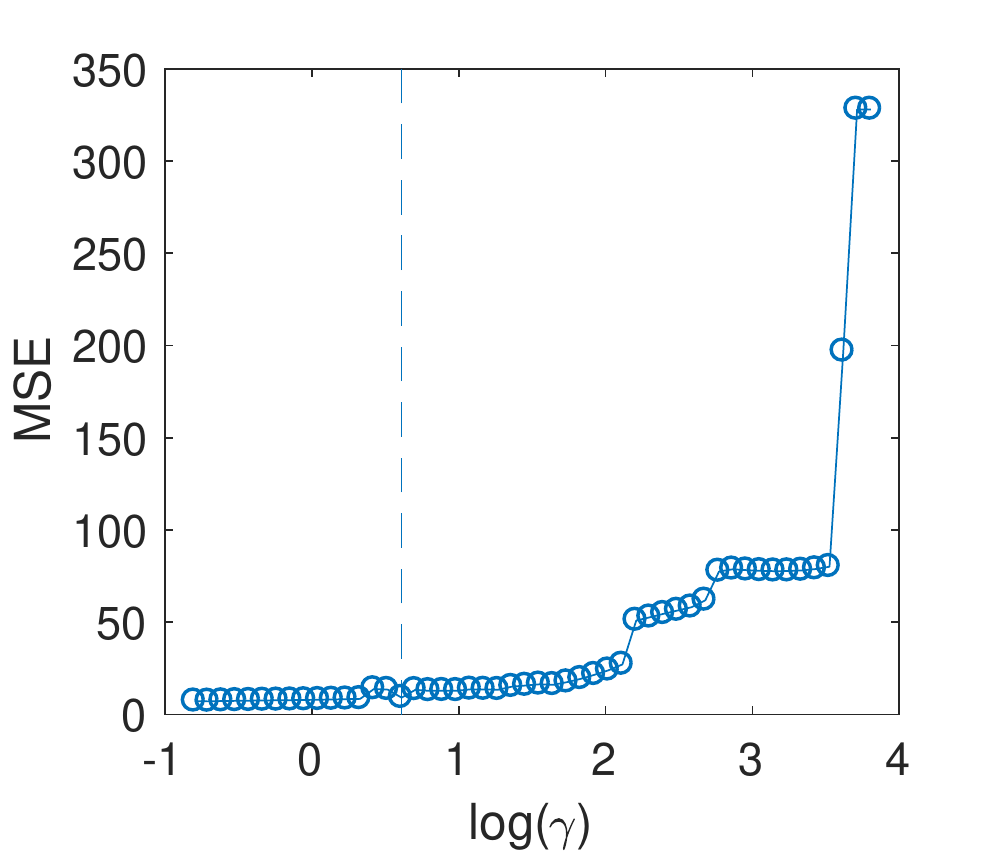}
    \caption{Out-of-sample MSE from naive TR (left) and SBL (right) under high signal-to-noise ratio. The vertical line in either plot indicates the selected value of the $L_1$ penalty factor in coefficient estimation. \label{fig:MSE_SBL_TR}}
\end{figure}

The dataset is split into a training set and a test set with each consisting of 50 observations, for tuning the $L_1$ penalty factor. Figure \ref{fig:sim_lasso} and \ref{fig:MSE_SBL_TR} display the mean squared error (MSE) on test data across different values of the $L_1$ penalty factor for lasso, naive TR and SBL respectively. As can be seen, the out-of-sample MSE does not vary much with small values of the penalty factor for each method. Therefore we set the optimal $L_1$ penalty factor at the largest possible value that produces small MSE (e.g. less than $3\%$ of the maximum MSE when all the parameters are zero in this case) for all models as indicated in Figure \ref{fig:sim_lasso} and \ref{fig:MSE_SBL_TR}. 

The estimated coefficients from lasso are displayed in the lower-triangular matrix in the right plot of  Figure \ref{fig:sim_lasso} with the true coefficients in the upper-triangular. As can be seen, lasso misses some true signal edges and it is not straightforward to identify meaningful structure among the selected edges. 

For the linear regression based on TN-PCA, we set the rank $K=20$ in \eqref{eq:semi_sym_tensor_decomp}, which explains approximately 100\% of the variation in the networks. The MSE on test data from TN-PCA is $19.46$, higher than the MSE at the optimal $L_1$ penalty factor, 15.32 for naive TR, 9.67 for lasso and 9.17 for SBL. The linear regression on the network PC scores shows that all the 20 components are significant at the 5\% significance level, which is noninformative of the subgraphs relevant to $y$ since all the basis networks $\{\boldsymbol{v}_{k}\boldsymbol{v}_{k}^{\top}\}_{k=1}^{20}$ are dense.

For the low-rank sensing (LRS) model, we solve the optimization \eqref{low-rank-sensing} by minimizing the nuclear norm \cite{recht2010guaranteed} with the CVX toolbox in matlab. The solution for the coefficient matrix $B$ does not have low rank but actually full rank in this case. This is probably due to the randomness in the generating process for $y$, which is closer to the reality in neuroimaging studies, while model \eqref{low-rank-sensing} does not contain any randomness. In addition, the estimated $B$ is a dense matrix with all the entries nonzero, and hence selects all the edges in the network. The MSE on test data from LRS is 13.46.

\begin{figure}[htb]
\centering
\includegraphics[width=0.5\textwidth]{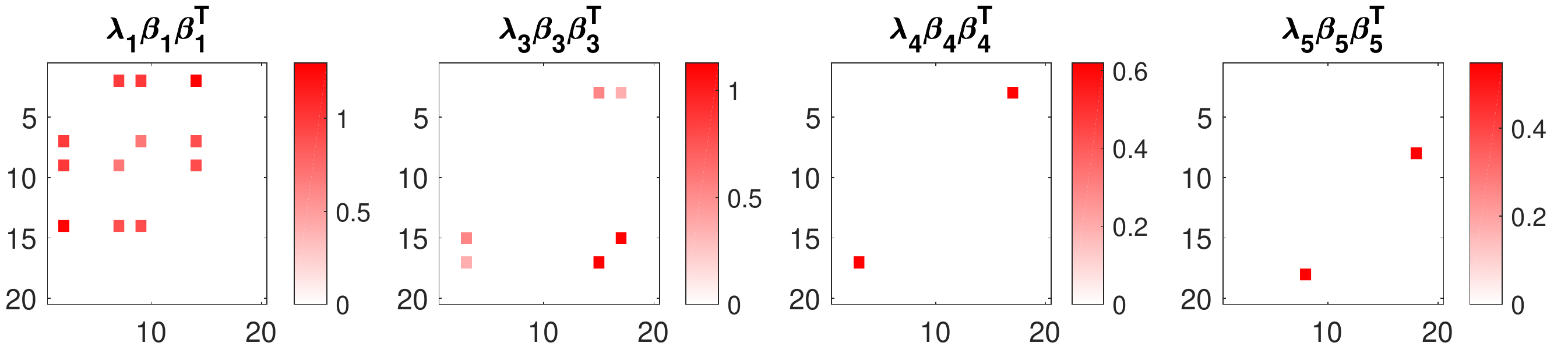} \\
\includegraphics[width=0.12\textwidth]{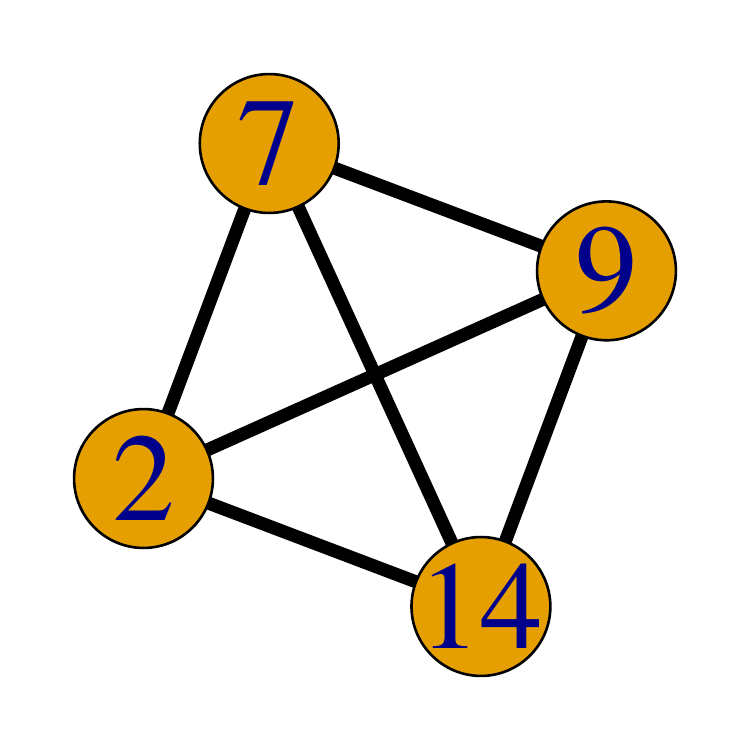} \includegraphics[width=0.12\textwidth]{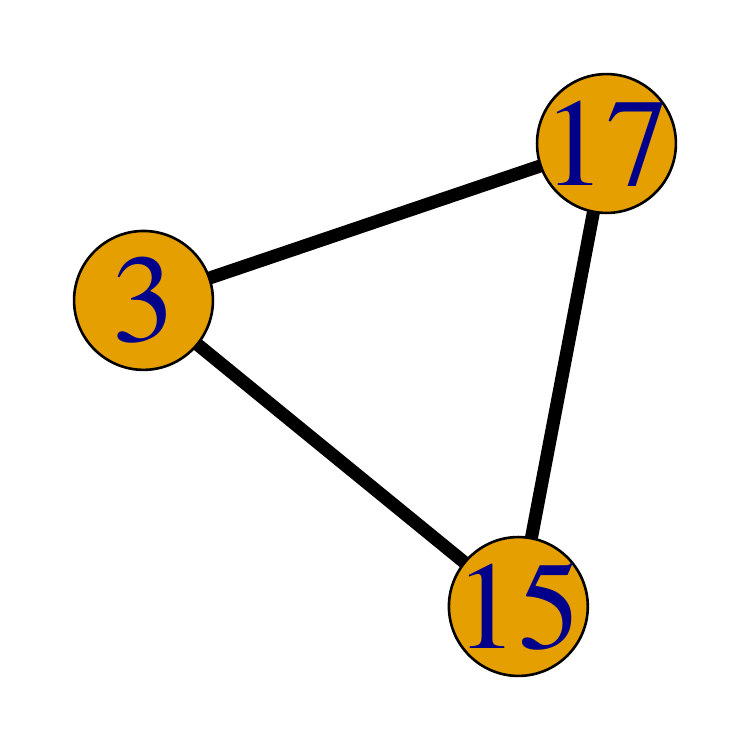} \includegraphics[width=0.11\textwidth]{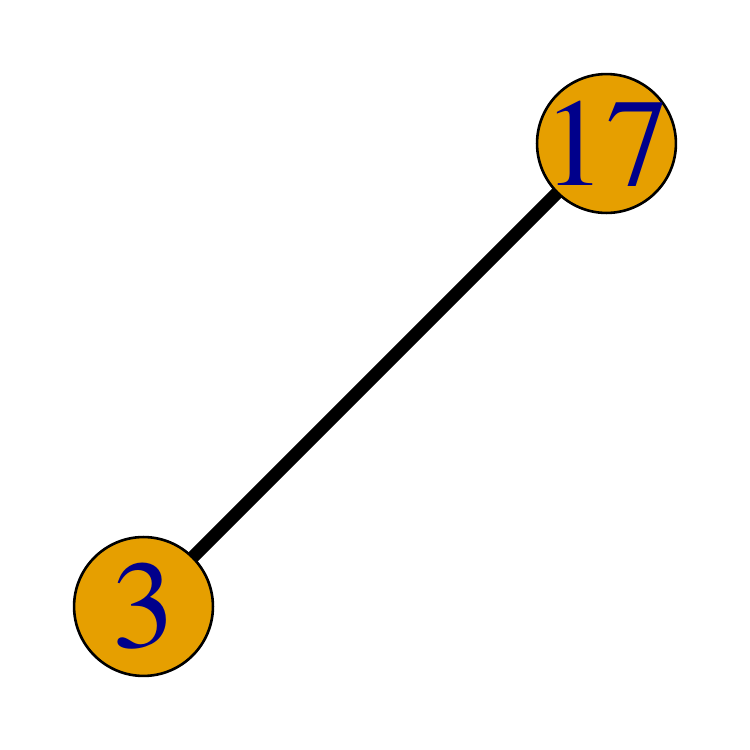} \includegraphics[width=0.11\textwidth]{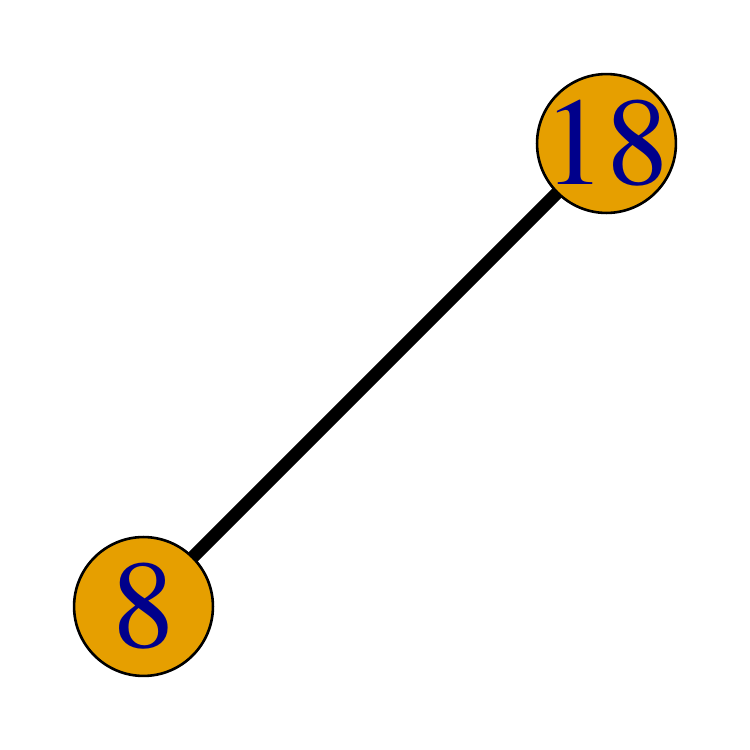} 
\caption{Estimated nonzero coefficient components $\{\lambda_{h}\boldsymbol{\beta}_{h}\boldsymbol{\beta}_{h}^{\top}\}$ from SBL (upper) and their selected subgraphs (lower) under high signal-to-noise ratio.  \label{fig:SBL_res_sim}}
\end{figure}

\begin{figure}[htb]
\centering
\includegraphics[width=0.5\textwidth]{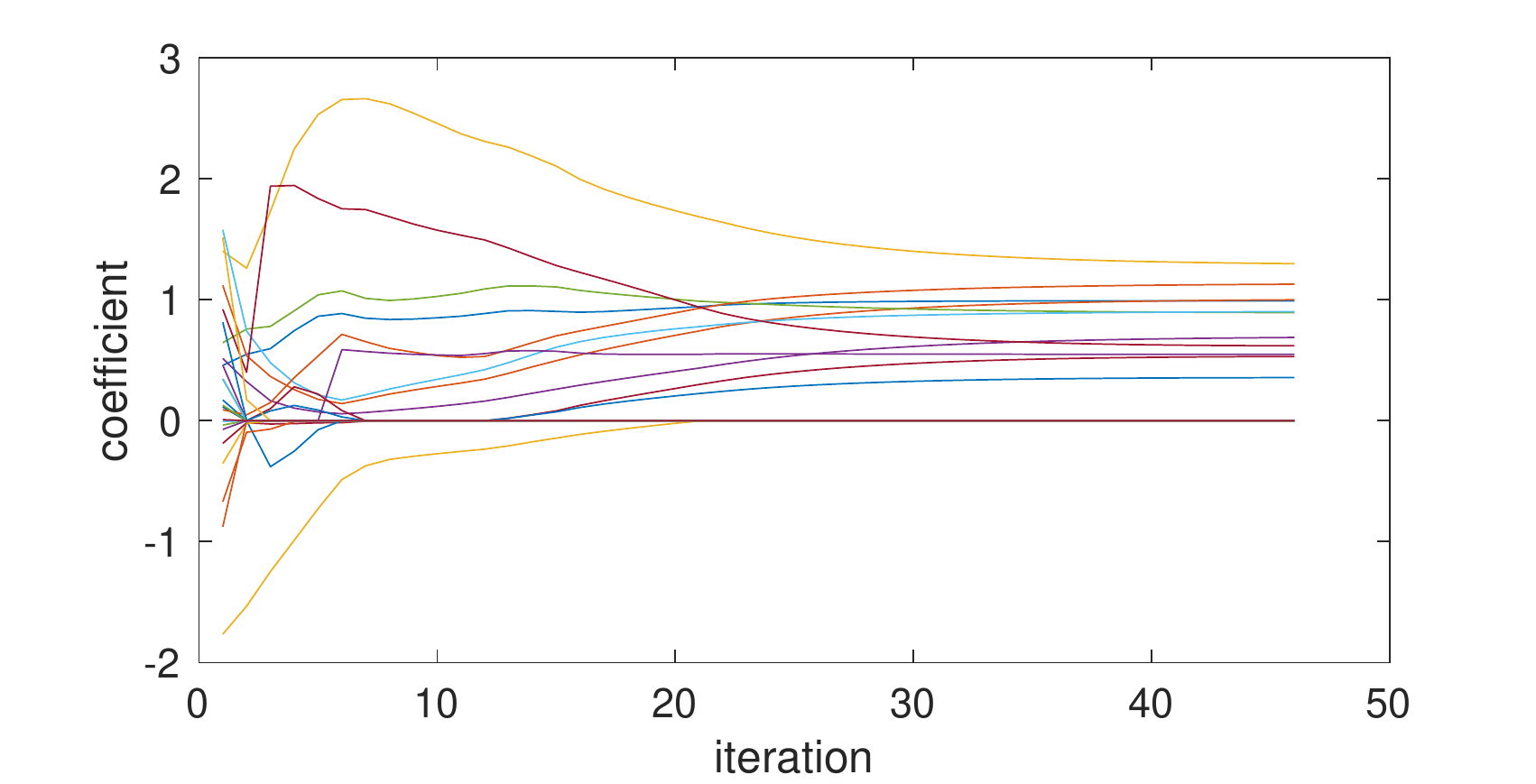} 
\caption{Profiles of estimated coefficients from SBL under high signal-to-noise ratio, showing how coefficient values $\{\lambda_{h}\beta_{hu}\beta_{hv}\}$ evolve over iterations for the estimated nonzero coefficients and 20 randomly selected zero coefficients in Figure \ref{fig:SBL_res_sim}.  \label{fig:SBL_coef_path}}
\end{figure}

The estimated coefficient components for $\{\lambda_{h}\boldsymbol{\beta}_{h}\boldsymbol{\beta}_{h}^{\top}\}_{h=1}^{5}$ from SBL as well as the selected subgraphs are displayed in Figure \ref{fig:SBL_res_sim}, where 4 out of 5 components are nonempty. Figure \ref{fig:SBL_res_sim} shows that our model recovers all the true signal subgraphs -- a single edge, a triangle and a 4-node clique, though the component $\lambda_{4}\boldsymbol{\beta}_{4}\boldsymbol{\beta}_{4}^{\top}$ repeatedly selects an edge in the true triangle signal. Figure \ref{fig:SBL_coef_path} displays the evolution of the estimated nonzero coefficients $\{\lambda_{h}\beta_{hu}\beta_{hv}\}$ and 20 randomly selected zero coefficients in Figure \ref{fig:SBL_res_sim} over iterations, which shows that the sequences of component coefficients converge as the objective function converges. In practice, we can always check such profiles of evolution for component coefficients and select a proper tolerance $\epsilon$ in Algorithm \ref{CD-SBL} to guarantee the convergence of solution sequences.

We use 10 initializations to run Algorithm \ref{CD-SBL} in this case, as the best local minimum found does not change when increasing to 20 initializations. The total runtime is 32.2 seconds. But since the numerical experiments were conducted in a machine with one Intel Core i5 2.7 GHz processor and 8 GB of RAM, there are substantial margins to reduce the computational time if parallel computing were employed in a multi-core machine.

\begin{figure}[!hbt]
\centering
\includegraphics[width=0.14\textwidth]{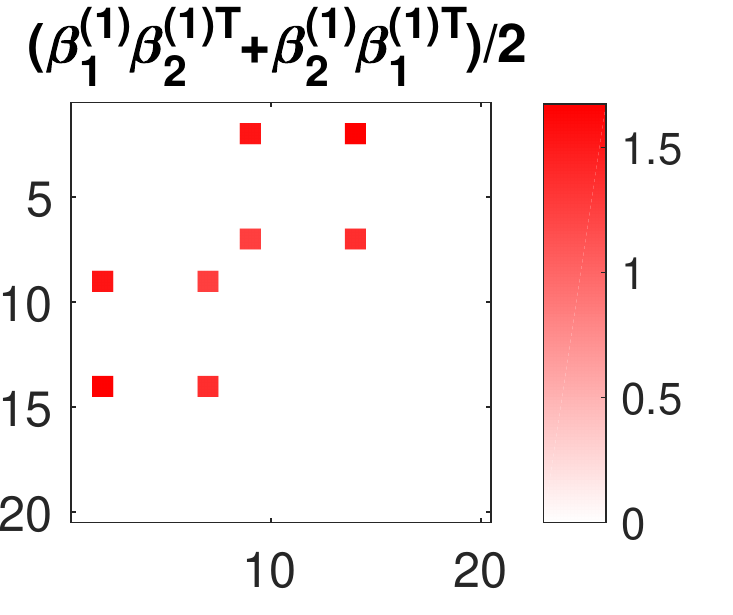}\includegraphics[width=0.11\textwidth]{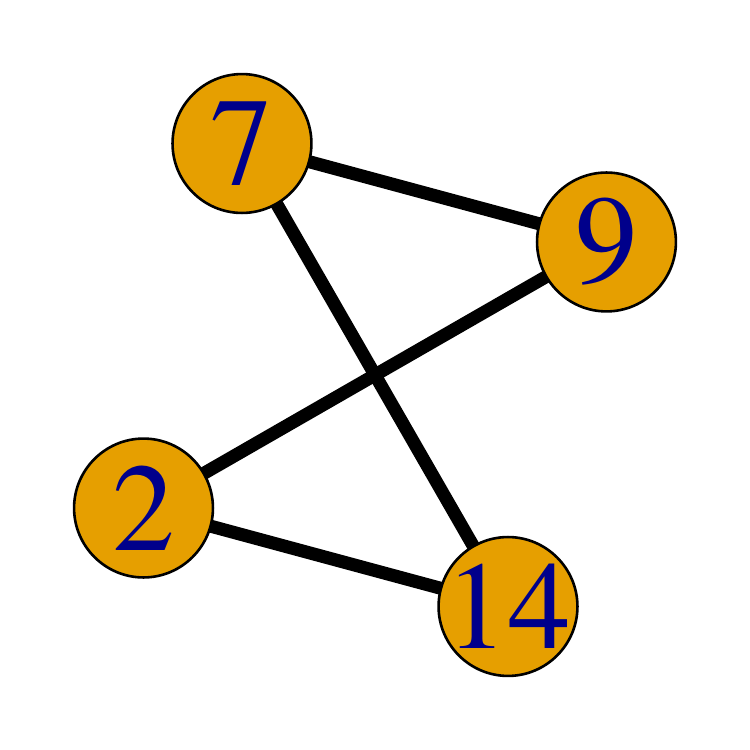}\includegraphics[width=0.14\textwidth]{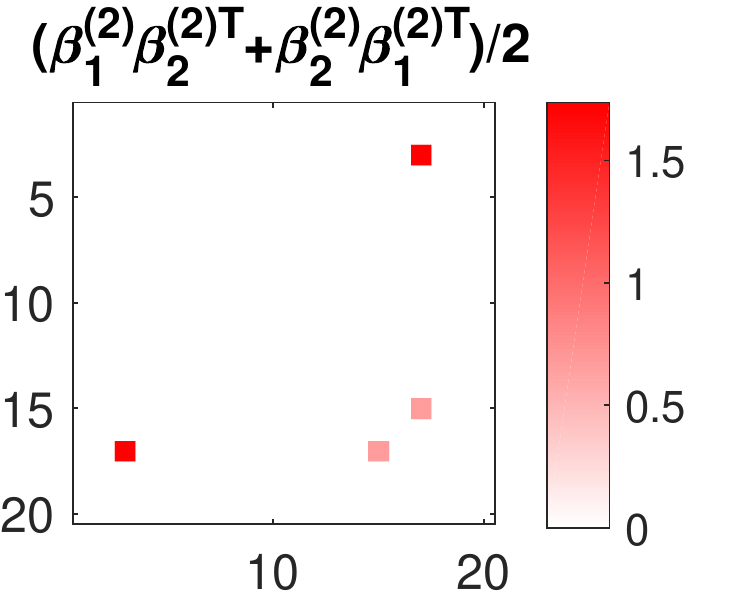}\includegraphics[width=0.11\textwidth]{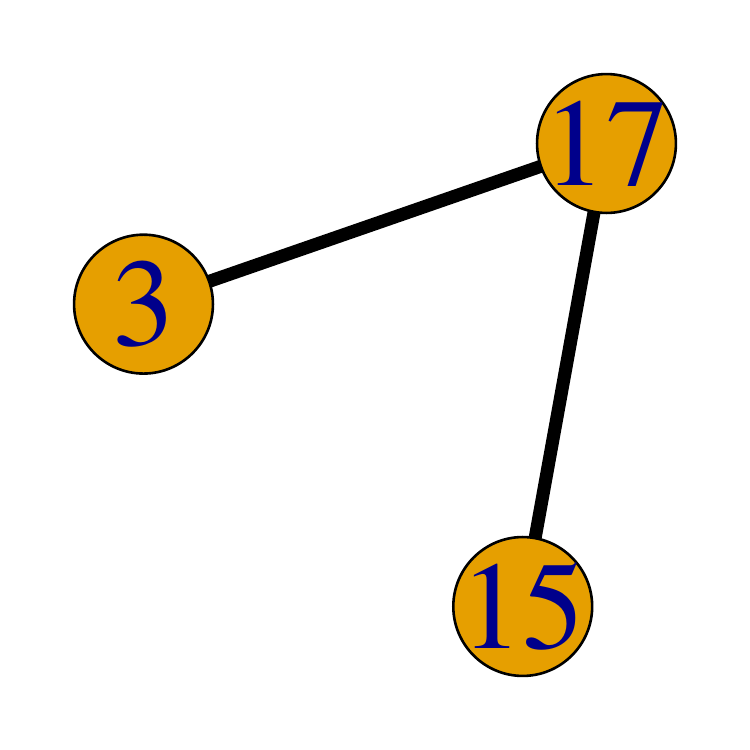}  
\caption{Estimated nonzero coefficient components $\{ ( \boldsymbol{\beta}_{1}^{(k)}\boldsymbol{\beta}_{2}^{(k)\top} + \boldsymbol{\beta}_{2}^{(k)}\boldsymbol{\beta}_{1}^{(k)\top} )/2 \}$ from naive TR and their selected subgraphs under high signal-to-noise ratio. \label{fig:TR_res_sim}}
\end{figure}

The naive TR is applied in this case under the same convergence criterion and initializations as in SBL. The estimated coefficient components $\{ ( \hat{\boldsymbol{\beta}}_{1}^{(k)}\hat{\boldsymbol{\beta}}_{2}^{(k)\top} + \hat{\boldsymbol{\beta}}_{2}^{(k)}\hat{\boldsymbol{\beta}}_{1}^{(k)\top} )/2 \}_{k=1}^{5}$ as well as the selected subgraphs are displayed in Figure \ref{fig:TR_res_sim}, where 2 out of 5 components are nonempty. Figure \ref{fig:TR_res_sim} shows that the naive TR model partially recovers the 4-node clique and the triangle signal, though misses the single-edge signal.

The procedure described above is repeated 100 times, where each time we generate a synthetic dataset based on \eqref{eq:simu_dgp} and \eqref{eq:simu_y}, and record the out-of-sample MSE (at the optimal $L_1$ penalty factor for lasso, naive TR and SBL), the true positive rate (TPR) representing the proportion of true signal edges that are correctly identified, and the false positive rate (FPR) representing the proportion of non-signal edges that are falsely identified, for lasso, TN-PCA, LRS, naive TR and SBL. Table \ref{tab:res_low_nsr} displays the mean and standard deviation (sd) of the MSE, TPR and FPR for the five methods in the high signal-to-noise ratio scenario. Although LRS has the lowest average MSE in Table \ref{tab:res_low_nsr},  its TPR and FPR are both 1, indicating that LRS selects all the edges in the network in each simulation. SBL has a bit higher average FPR than that of lasso and the highest TPR on average excluding LRS.

\begin{table}[htb]
\caption{Mean and sd of the MSE, TPR and FPR across 100 simulations under high signal-to-noise ratio.}
\label{tab:res_low_nsr}
\centering
\begin{tabular}{lccc}
\toprule 
 & MSE & TPR & FPR\tabularnewline
\midrule
lasso & 10.98$\pm$4.40 & 0.837$\pm$0.138 & \textbf{0.002$\pm$0.005}\tabularnewline
TN-PCA & 10.04$\pm$4.66 & 0.449$\pm$0.499 & 0.449$\pm$0.499\tabularnewline
LRS & \textbf{6.71$\pm$2.86} & \textbf{1.000$\pm$0.000} & 1.000$\pm$0.000 \\
naive TR & 15.94$\pm$6.93 & 0.696$\pm$0.122 & 0.024$\pm$0.027 \\
SBL & 10.08$\pm$4.51 & \textbf{0.848$\pm$0.169} & 0.005$\pm$0.007\tabularnewline
\bottomrule
\end{tabular}
\end{table}

\subsubsection{Low signal-to-noise ratio}

In this case, the noise level $\sigma=100\%$ of the standard deviation of the conditional mean $E(y_i \mid W_i)$ in the generating process \eqref{eq:simu_y}.

\begin{figure}[htb]
\centering
\includegraphics[width=.25\textwidth]{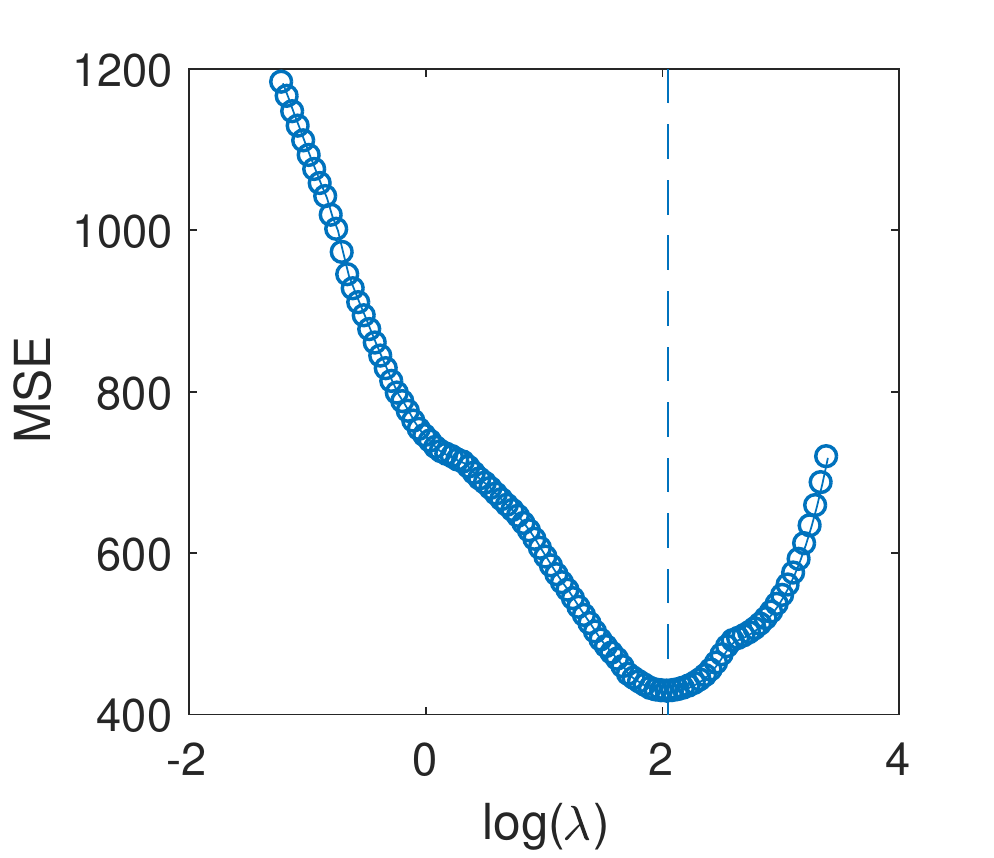}\includegraphics[width=0.25\textwidth, trim={0cm 0.1cm 0.6cm 0.8cm}, clip]{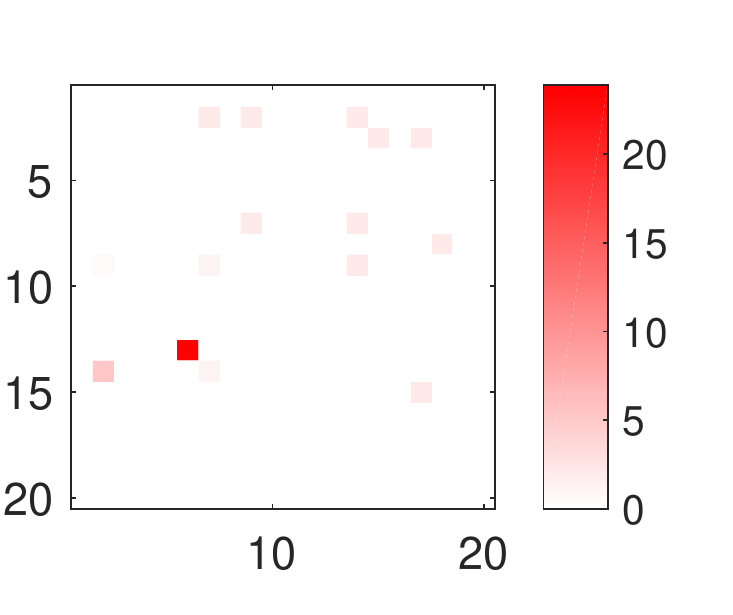}
\caption{Left: out-of-sample MSE from lasso under low signal-to-noise ratio. Right: estimated coefficients from lasso (lower-triangular) where the $L_1$ penalty factor is set corresponding to the vertical line on the left plot; the true coefficients for each edge of the network are shown in the upper triangle. \label{fig:sim2_lasso}}
\end{figure}

\begin{figure}
    \centering
    \includegraphics[width=.25\textwidth]{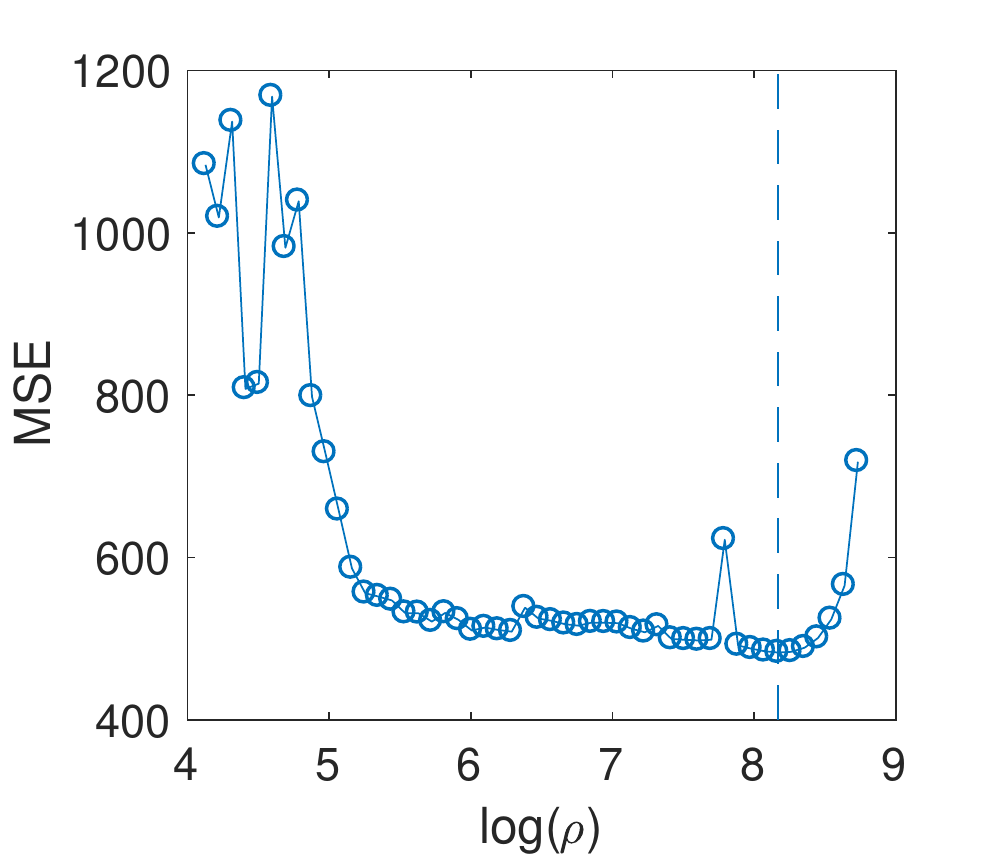}\includegraphics[width=.25\textwidth]{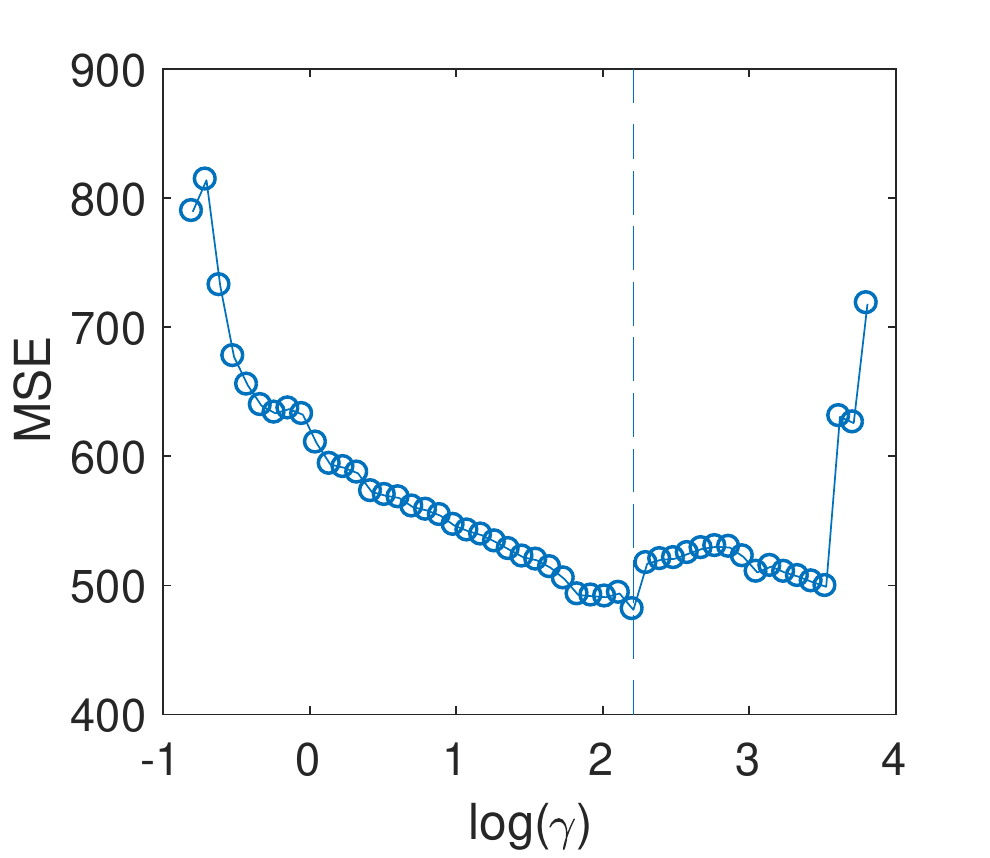}
    \caption{Out-of-sample MSE from naive TR (left) and SBL (right) under low signal-to-noise ratio. The vertical line in either plot indicates the selected value of the $L_1$ penalty factor in coefficient estimation.  \label{fig:MSE2_SBL_TR}}
    \vspace{0pt}
\end{figure}

Figure \ref{fig:sim2_lasso} and \ref{fig:MSE2_SBL_TR} display the MSE on test data versus the $L_1$ penalty factor for lasso, naive TR and SBL respectively. We set the optimal $L_1$ penalty factor for each model at the value that produces the minimum out-of-sample MSE as indicated in Figure \ref{fig:sim2_lasso} and \ref{fig:MSE2_SBL_TR}.

The estimated coefficients from lasso are displayed in the lower-triangular matrix in the left plot of Figure \ref{fig:sim2_lasso}, which shows that lasso misses many true signal edges and selects a false edge with very large coefficient.

The MSE on test data from LRS is 1271.5 in this case and that from TN-PCA is 1249.1, much higher than the minimum MSE 482.8 for naive TR, 481.1 for SBL and 427.5 for lasso. The solution for coefficient matrix $B$ from LRS is a dense matrix with full rank. The linear regression on the network PC scores from TN-PCA shows that none of the 20 components are significant in this case.

\begin{figure}[!hbt]
\centering
\includegraphics[width=0.14\textwidth]{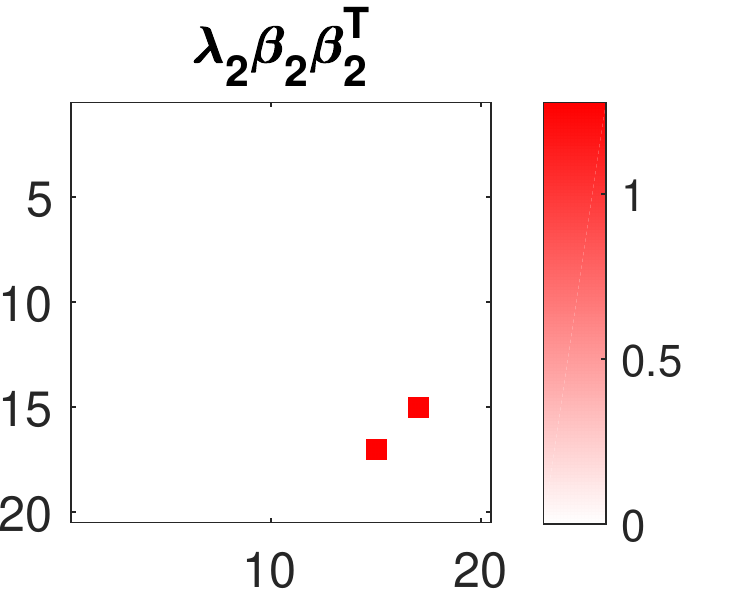}\includegraphics[width=0.10\textwidth]{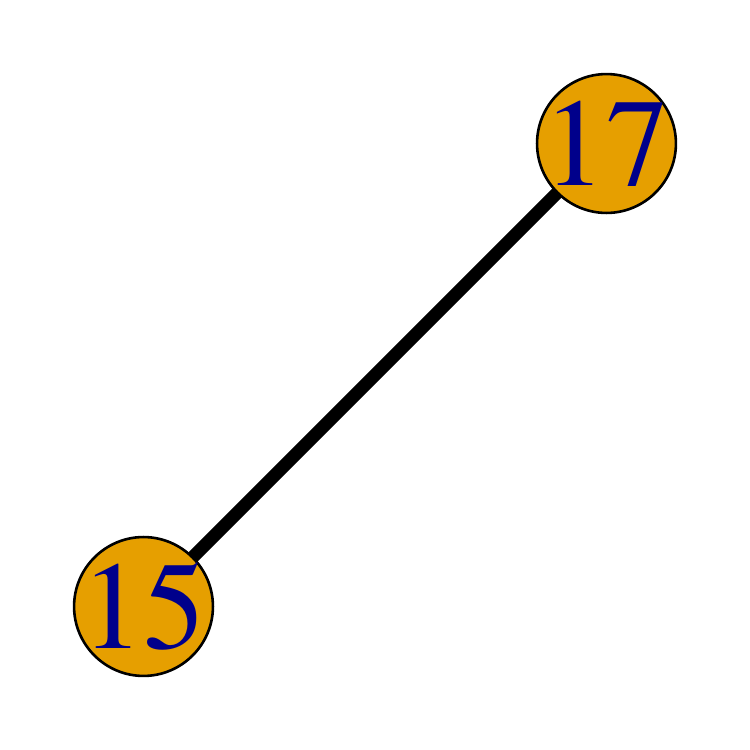}\includegraphics[width=0.14\textwidth]{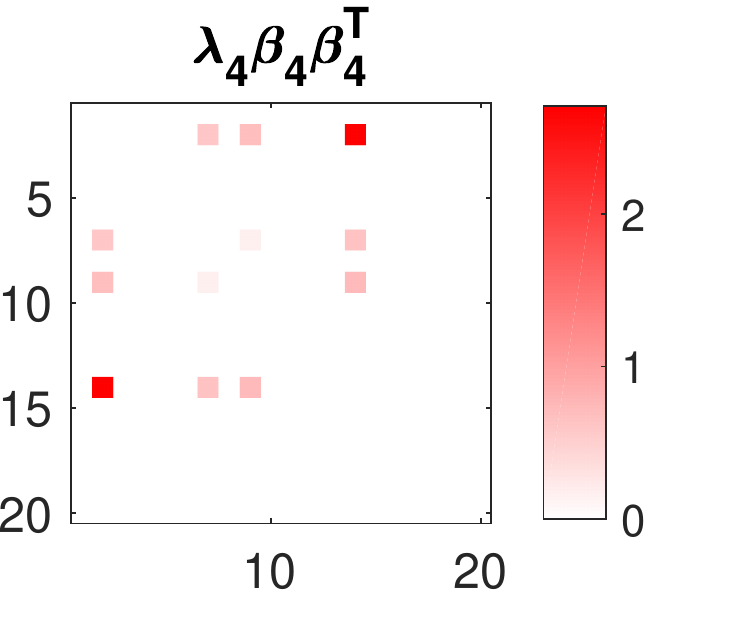}\includegraphics[width=0.12\textwidth]{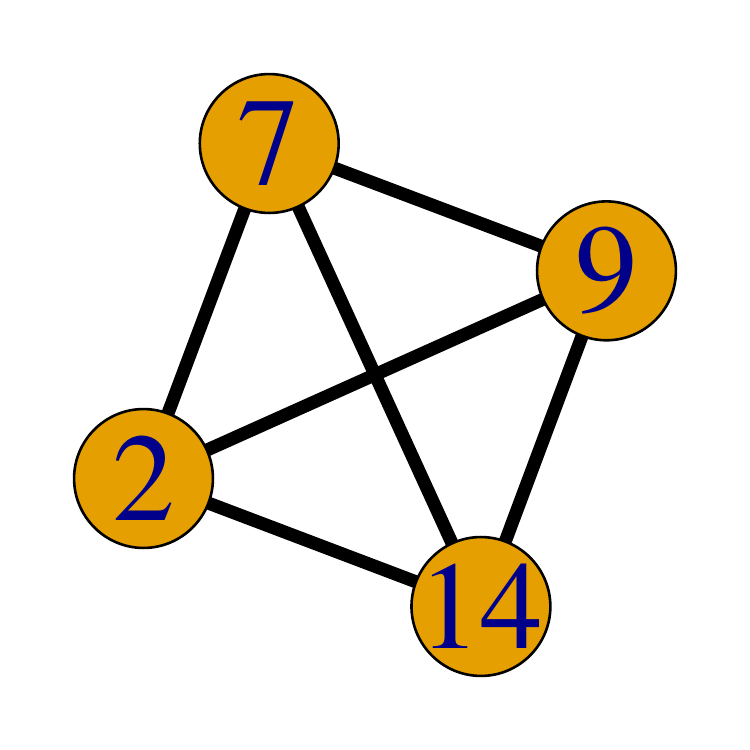}  
\caption{Estimated nonzero coefficient components $\{\lambda_{h}\boldsymbol{\beta}_{h}\boldsymbol{\beta}_{h}^{\top}\}$ from SBL and their selected subgraphs under low signal-to-noise ratio.  \label{fig:SBL_res_sim2}}
\end{figure}

\begin{figure}[htb]
\centering
\includegraphics[width=0.5\textwidth]{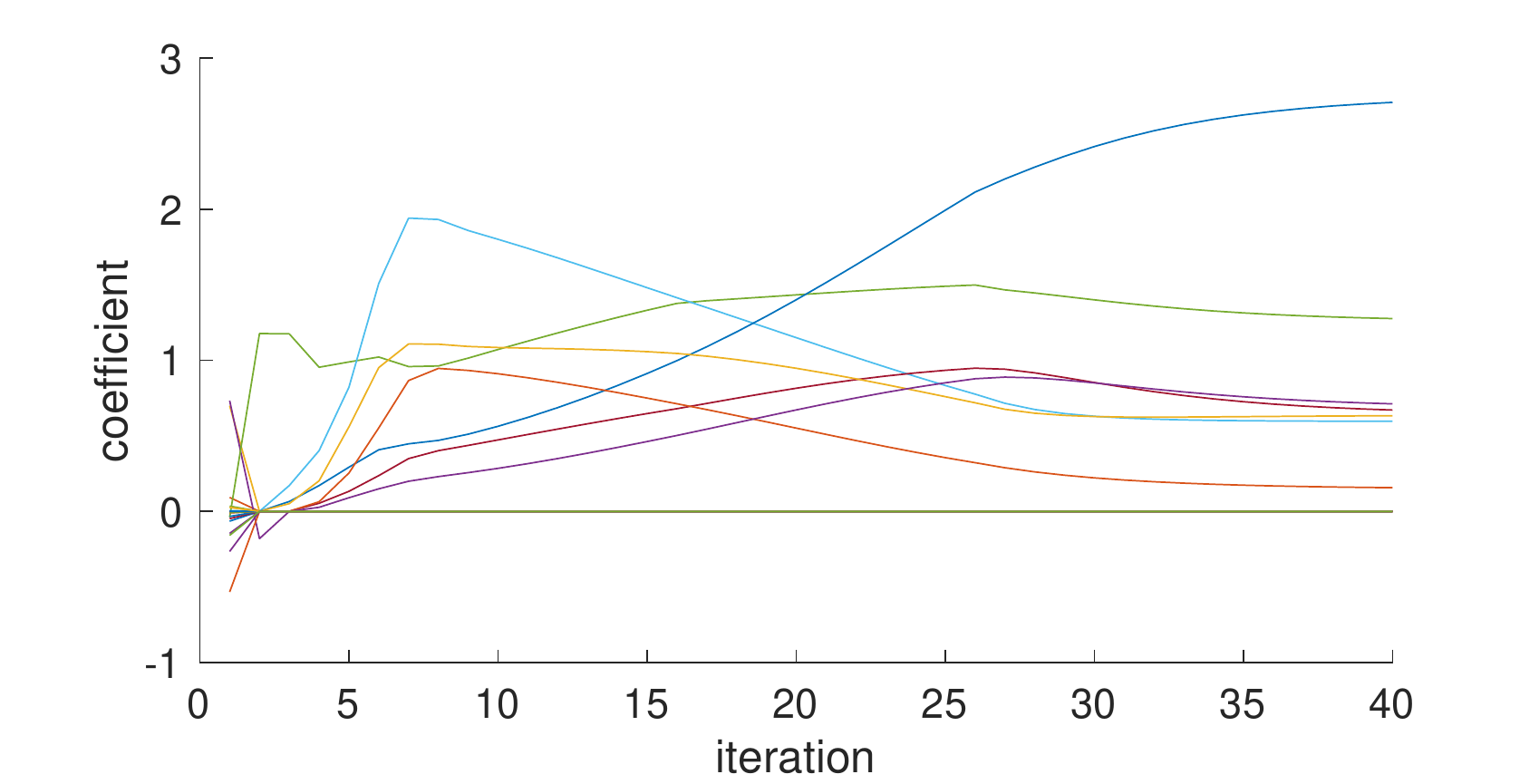} 
\caption{Profiles of estimated coefficients from SBL under low signal-to-noise ratio, showing how coefficient values $\{\lambda_{h}\beta_{hu}\beta_{hv}\}$ evolve over iterations for the estimated nonzero coefficients and 20 randomly selected zero coefficients in Figure \ref{fig:SBL_res_sim2}.  \label{fig:SBL_coef_path2}}
\end{figure}

SBL selects two nonzero coefficient components $\{\lambda_{h}\boldsymbol{\beta}_{h}\boldsymbol{\beta}_{h}^{\top}\}$ out of 5 in this case, which are displayed in Figure \ref{fig:SBL_res_sim2} along with the selected subgraphs. Figure \ref{fig:SBL_res_sim2} shows that our model perfectly recovers one true signal subgraph -- the 4-node clique, though partially recovers the triangle signal by identifying one edge and misses the single-edge signal. The evolution profiles of the estimated nonzero coefficients and 20 randomly selected zero coefficients in Figure \ref{fig:SBL_res_sim2} are displayed in Figure \ref{fig:SBL_coef_path2}, which indicates the convergence of the coefficients. The total runtime under 10 initializations is 18 seconds in this case.

\begin{figure}[!hbt]
\centering
\includegraphics[width=0.22\textwidth]{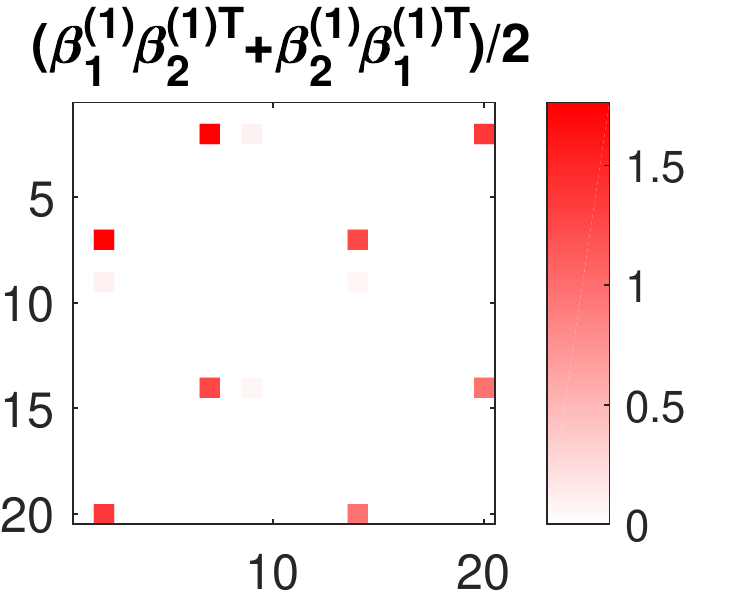}\includegraphics[width=0.15\textwidth]{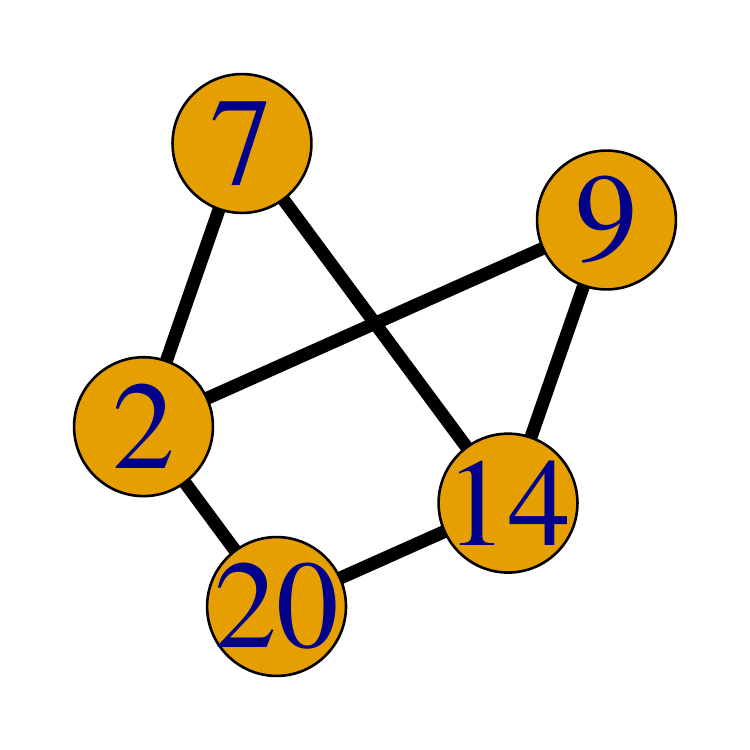}
\caption{Estimated nonzero coefficient component $( \boldsymbol{\beta}_{1}^{(1)}\boldsymbol{\beta}_{2}^{(1)\top} + \boldsymbol{\beta}_{2}^{(1)}\boldsymbol{\beta}_{1}^{(1)\top} )/2 $ from naive TR and its corresponding subgraph under low signal-to-noise ratio.  \label{fig:TR_res_sim2}}
\end{figure}

The naive TR is applied under the same convergence criterion and initializations as in SBL, where 1 out of 5 components is nonempty as displayed in Figure \ref{fig:TR_res_sim2}, which  shows that the naive TR method partially recovers the 4-node clique while selecting 2 false edges.

The procedure described above is again repeated 100 times and Table \ref{tab:res_high_nsr} displays the mean and sd of the out-of-sample MSE, TPR and FPR for the five methods in the low signal-to-noise ratio scenario. Table \ref{tab:res_high_nsr} shows that SBL has the lowest out-of-sample MSE on average. Although naive TR obtains a bit higher TPR on average than SBL in this case, it has much higher average FPR than that of lasso and SBL.

\begin{table}[htb]
\caption{Mean and sd of the MSE, TPR and FPR across 100 simulations under low signal-to-noise ratio.}
\label{tab:res_high_nsr}
\centering
\begin{tabular}{lccc}
\toprule 
 & MSE & TPR & FPR\tabularnewline
\midrule
lasso & 448.3$\pm$195.3 & 0.445$\pm$0.141 & \textbf{0.025$\pm$0.037}\tabularnewline
TN-PCA & 624.0$\pm$287.8 & 0.060$\pm$0.239 & 0.060$\pm$0.238\tabularnewline
LRS & 636.7$\pm$258.3 & \textbf{1.000$\pm$0.000} & 1.000$\pm$0.000 \\
naive TR & 394.5$\pm$157.1 &  \textbf{0.572$\pm$0.181} & 0.176$\pm$0.238 \\
SBL & \textbf{393.7$\pm$159.2} & 0.539$\pm$0.210 & 0.029$\pm$0.038\tabularnewline
\bottomrule
\end{tabular}
\end{table}

\subsection{Sensitivity to $K$}
\label{sensitivity K}

In the experiments above, the rank $K$ is set at 5 in SBL, which is an upper bound for the true rank of the generating process \eqref{eq:simu_y}, as recommended in Section \ref{remarks}. To assess the sensitivity of SBL's performance to the choice of $K$ in practice, we rerun SBL with $K=6$ and $K=7$ for the experiments in both high and low signal-to-noise ratio (SNR) scenarios. The mean and sd of the out-of-sample MSE, TPR and FPR are displayed in Table \ref{tab:sense_K}. Compared to Table \ref{tab:res_low_nsr} and \ref{tab:res_high_nsr} in either case, the average MSEs, TPRs and FPRs are very similar among different choices for $K$ in SBL, implying that Algorithm \ref{CD-SBL} is robust to the chosen upper bound for the rank.

\begin{table}[htb]
\caption{Mean and sd of the MSE, TPR and FPR for SBL with different choices of $K$ in high and low signal-to-noise ratio (SNR).}
\label{tab:sense_K}
\begin{tabular}{c|cccc}
\toprule 
 &  & MSE & TPR & FPR\tabularnewline
\midrule 
high & $K=6$ & 10.21$\pm$4.62 & 0.856$\pm$0.182 & 0.004$\pm$0.011\tabularnewline
SNR & $K=7$ & 10.15$\pm$4.61 & 0.858$\pm$0.172 & 0.005$\pm$0.009\tabularnewline
\midrule
low & $K=6$ & 394.5$\pm$158.0 & 0.570$\pm$0.224 & 0.020$\pm$0.021 \tabularnewline
SNR & $K=7$ & 395.4$\pm$158.8  & 0.548$\pm$0.208 & 0.020$\pm$0.024 \tabularnewline
\bottomrule
\end{tabular}\end{table}

\section{Application}
\label{application}

We applied our method to the Human Connectome Project (HCP) dataset \cite{van2012human}, exploring the association between the brain connectome and two cognitive abilities, auditory language comprehension ability and oral reading ability. The dataset contains sMRI and dMRI data for 1065 subjects and for each subject, a weighted brain network of fiber counts among 68 regions was constructed by a state-of-the-art dMRI processing pipeline \cite{zhang2018mapping}.

\subsection{Picture Vocabulary Data}
\label{picvocab}

The HCP dataset contains age-adjusted scale scores of the subjects in a picture vocabulary (PV) test where respondents are presented with an audio recording of a word and four photographic images on the computer screen and are asked to select the picture that most closely matches the meaning of the word. 

\begin{figure}[htb]
\centering
\includegraphics[width=0.5\textwidth]{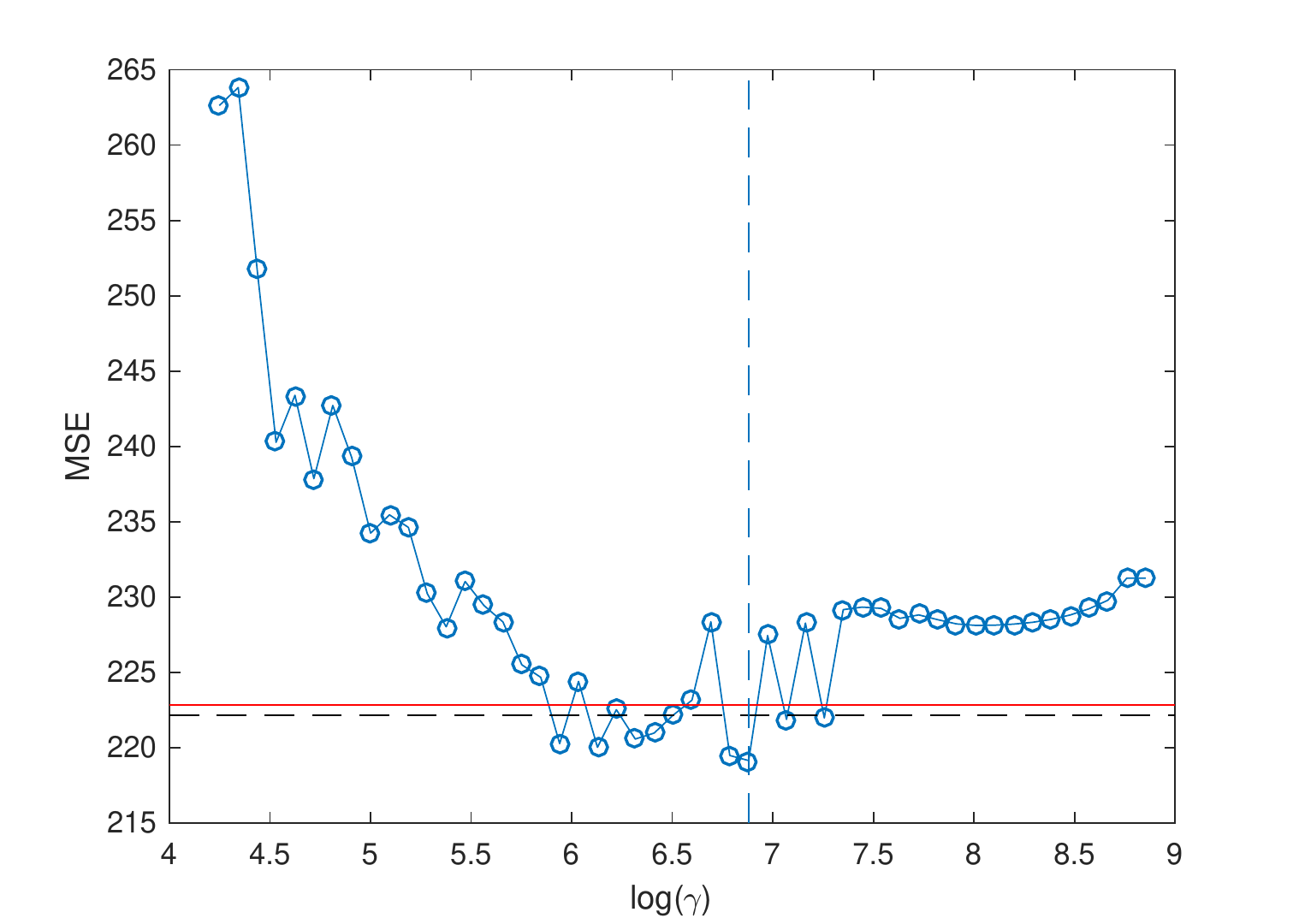}
\caption{Out-of-sample MSE of SBL on picture vocabulary data. The dashed vertical line indicates the selected value of $\gamma$ in inference; the red horizontal line indicates the minimum MSE of lasso; the black horizontal line indicates the MSE of TN-PCA.  \label{fig:picvoc_SBL_MSE}}
\end{figure}

We first compare the predictive performance for the PV scores among lasso, TN-PCA and SBL. The dataset is partitioned into a training set of 565 subjects and a test set of 500 subjects. We set $K=10$ for SBL. Five initializations are enough for Algorithm \ref{CD-SBL} to produce robust estimates for this dataset. The MSEs of PV scores on test data from SBL under different values of the $L_1$ penalty factor $\gamma$ are shown in Figure \ref{fig:picvoc_SBL_MSE}. The optimal $\gamma$ is set at the value that produces the smallest MSE, which is smaller than the minimum MSE of lasso, indicating better predictive performance. We set the rank $K=68$ in TN-PCA, which explains approximately 93\% of the variation in the brain networks. The out-of-sample MSE of TN-PCA is 222.1, which is higher than the minimum MSE of SBL as indicated in Figure \ref{fig:picvoc_SBL_MSE}. The linear regression of the PV scores on the low-dimensional embeddings of the brain networks shows that none of the 68 components are significant at the 5\% significance level.

\begin{figure}[htb]
\centering
\includegraphics[height=4.5cm]{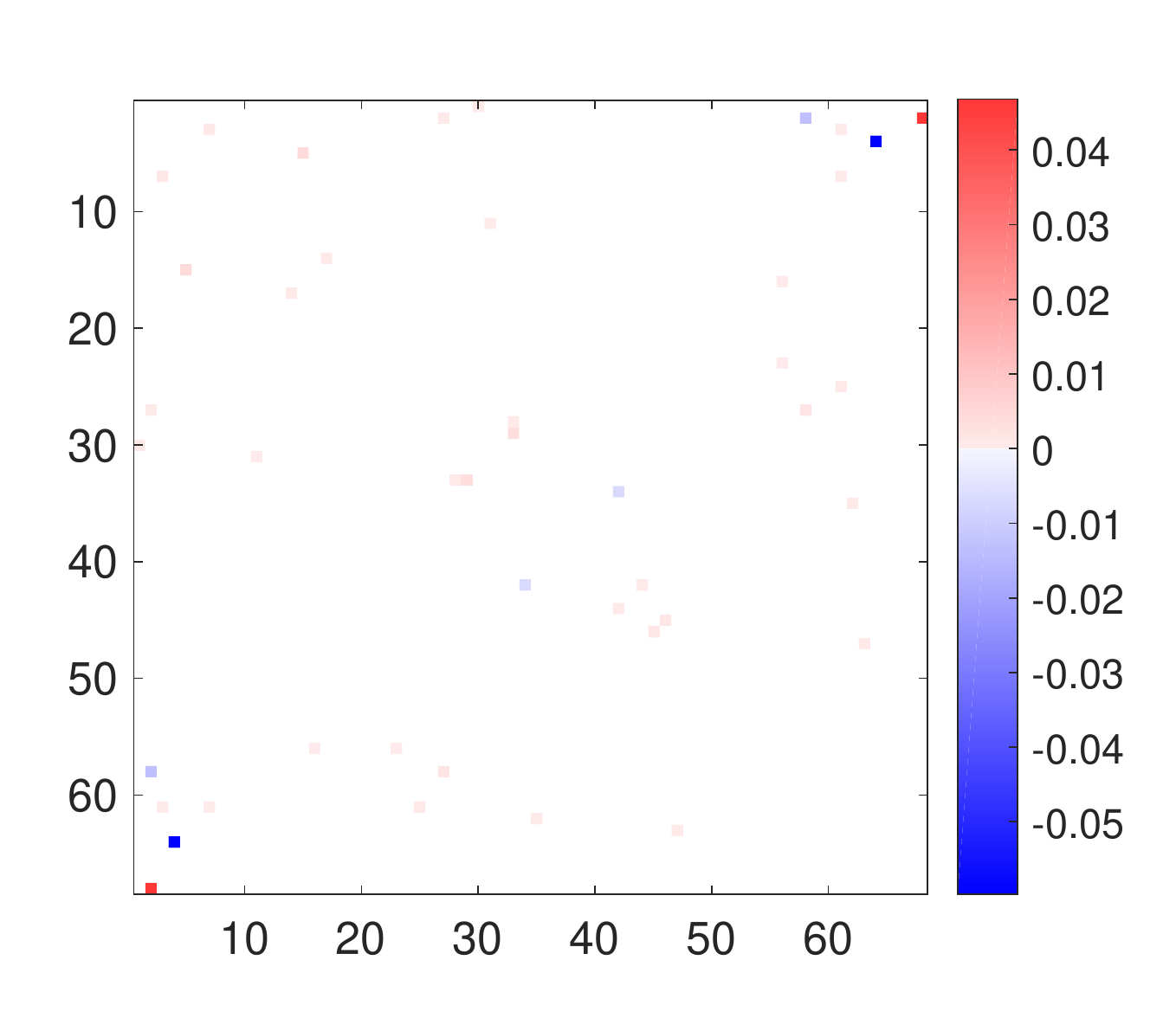}\includegraphics[height=4.5cm, trim={4cm 0 4cm 1cm}, clip]{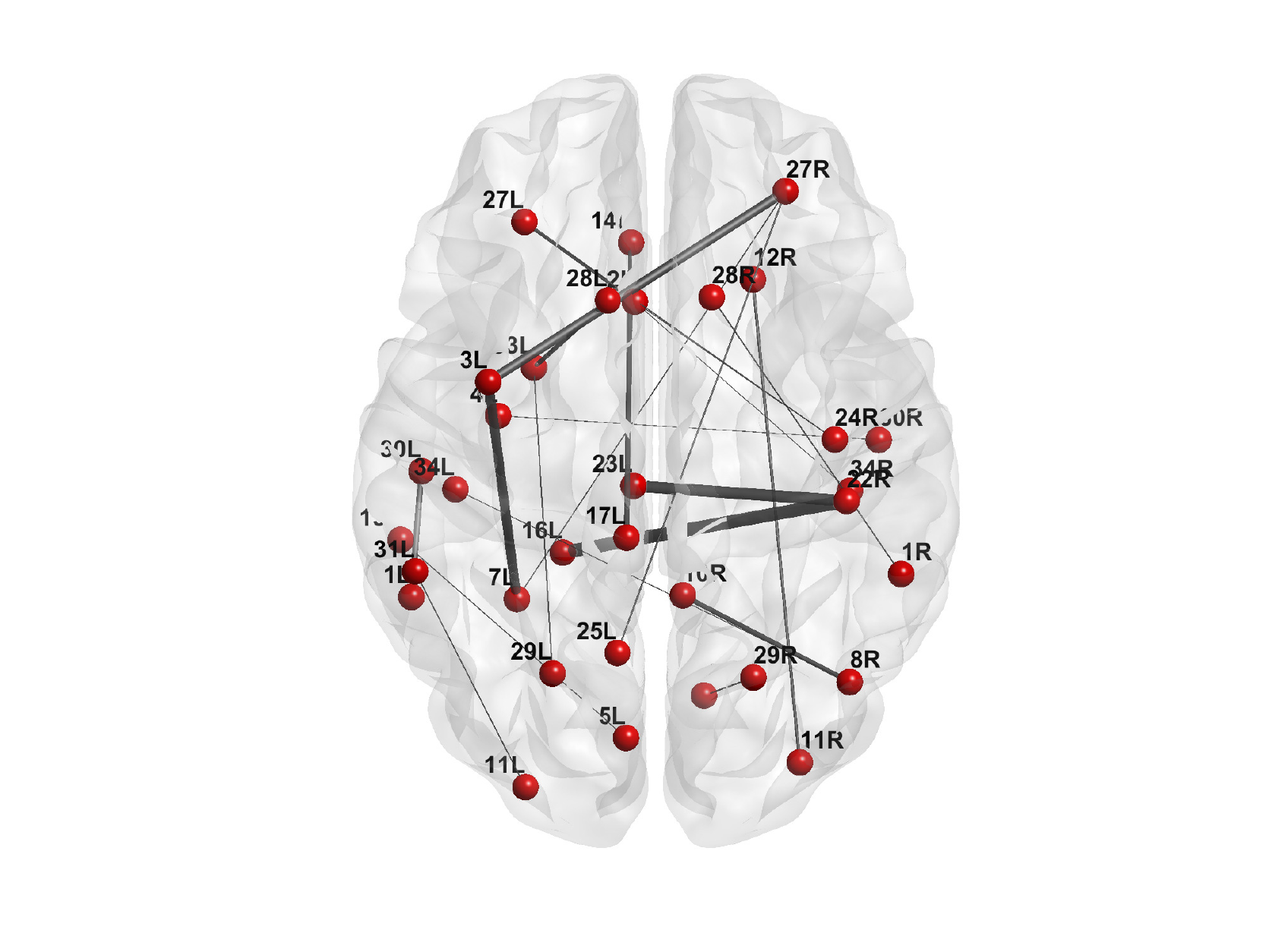}
\caption{Estimated coefficients from lasso in matrix form (left) and the structural connections in the brain corresponding to nonzero coefficients (right). The thickness of each edge is proportional to the average fiber count between the pair of regions. \label{fig:picvoc_lasso}}
\end{figure}

The estimated coefficients from lasso and the structural connections in the brain corresponding to the nonzero coefficients are displayed in Figure \ref{fig:picvoc_lasso}. As can be seen, these identified connections lack meaningful structure and are difficult to justify neurologically.

\begin{figure}[hbt]
\includegraphics[width=0.5\textwidth]{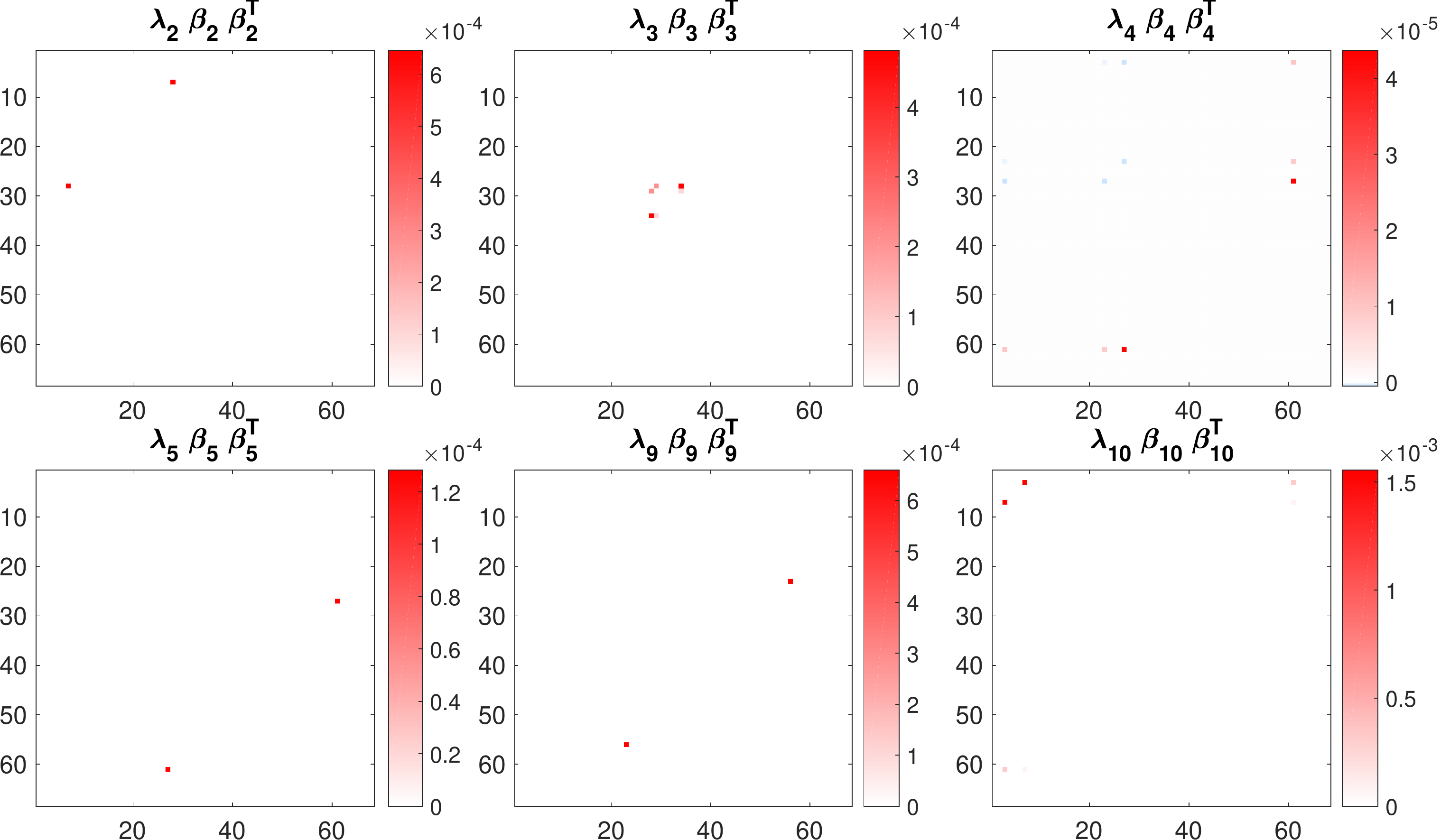}\\
\includegraphics[width=0.16\textwidth, trim={3cm 0 3cm 0}, clip]{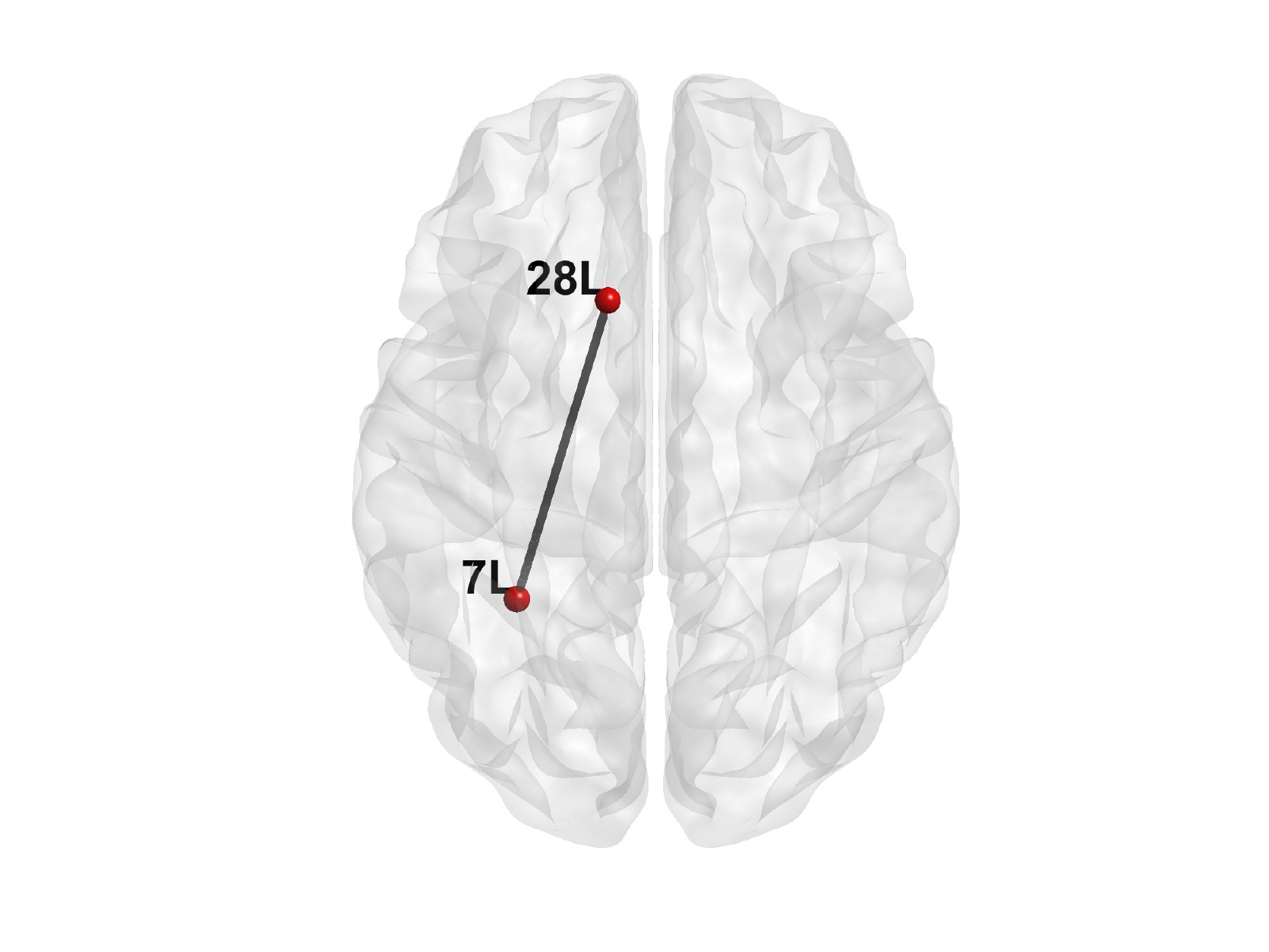}\includegraphics[width=0.16\textwidth, trim={3cm 0 3cm 0}, clip]{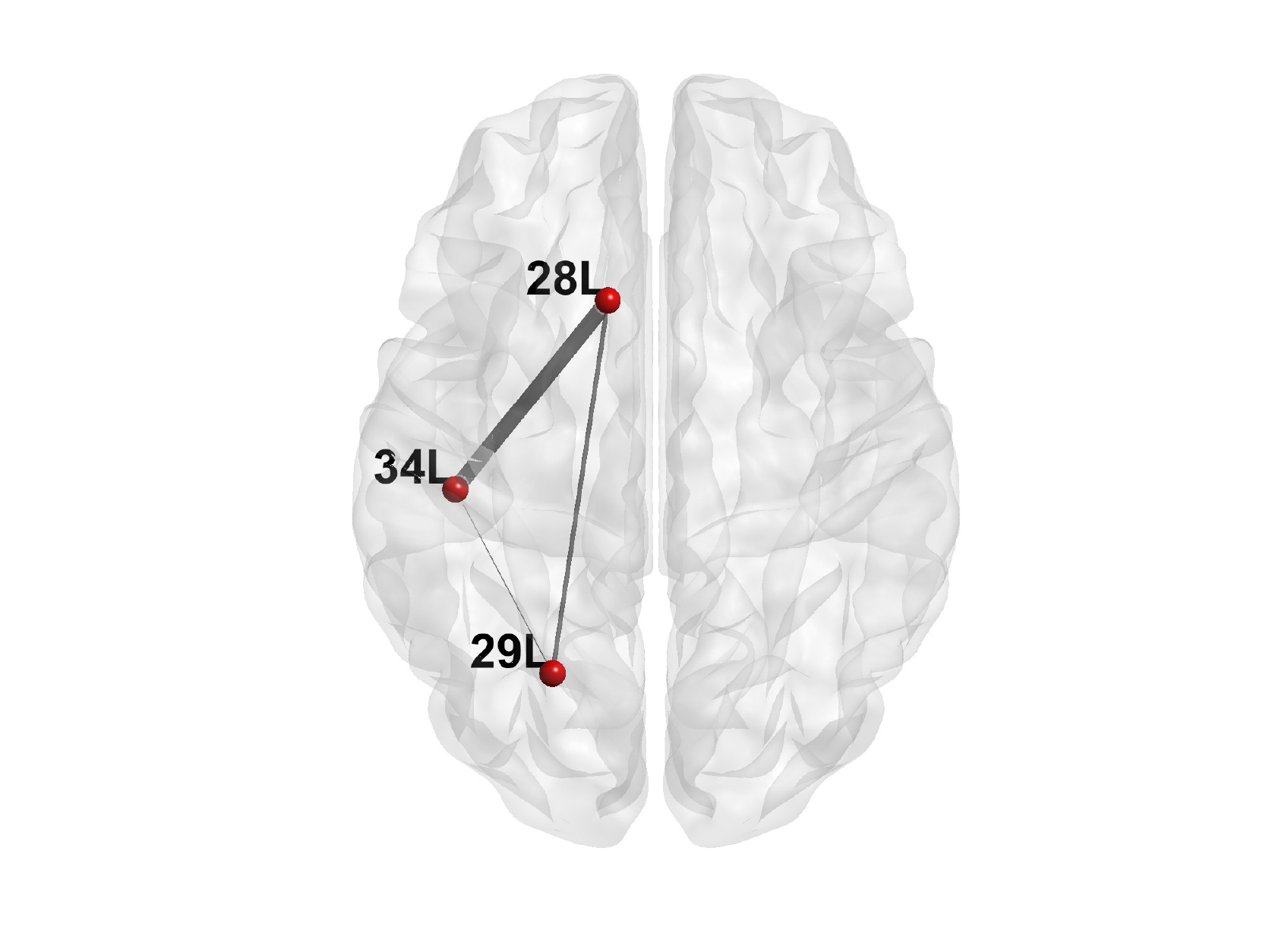}\includegraphics[width=0.16\textwidth, trim={3cm 0 3cm 0}, clip]{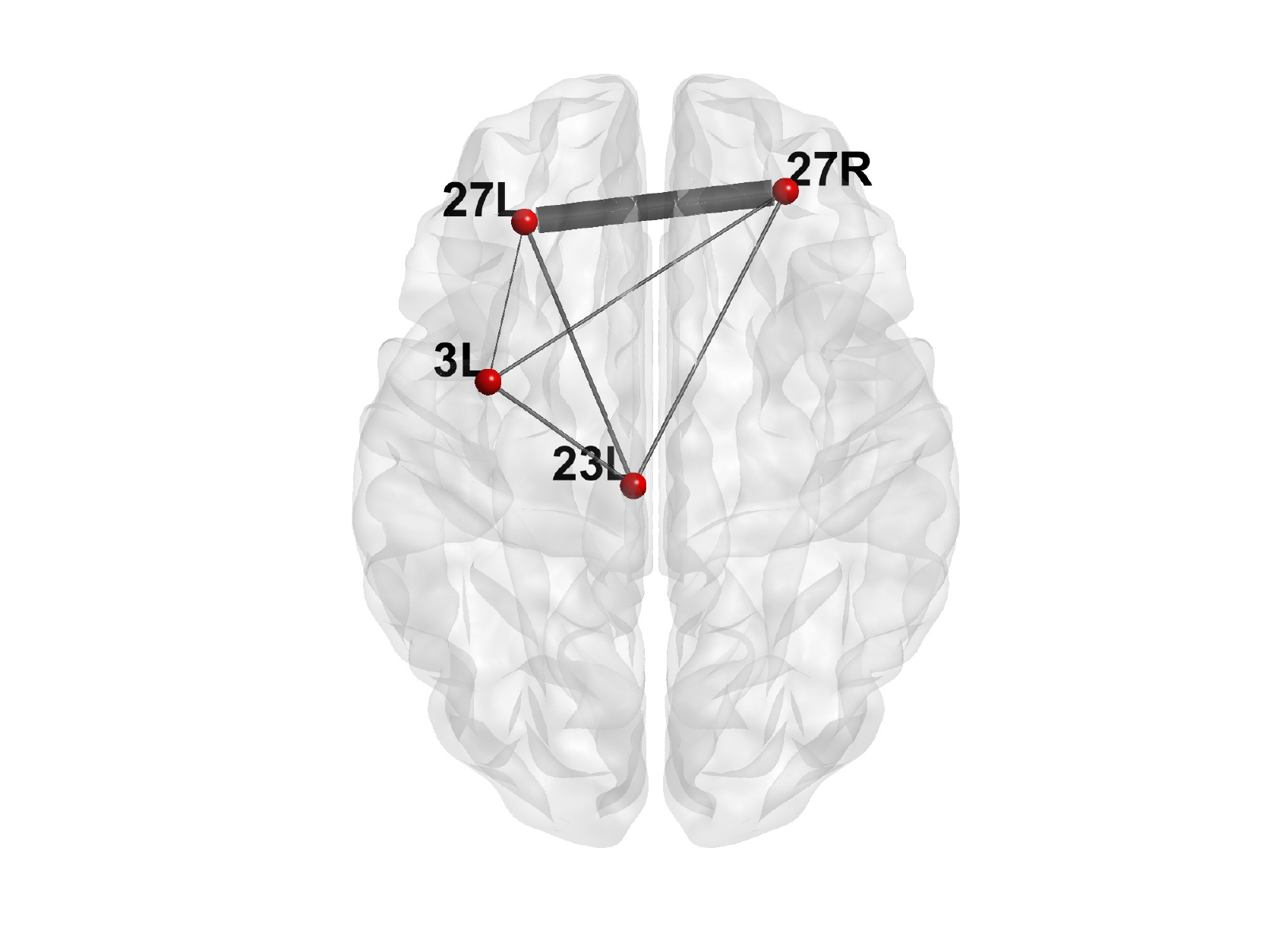} \\
\includegraphics[width=0.16\textwidth, trim={3cm 0 3cm 0}, clip]{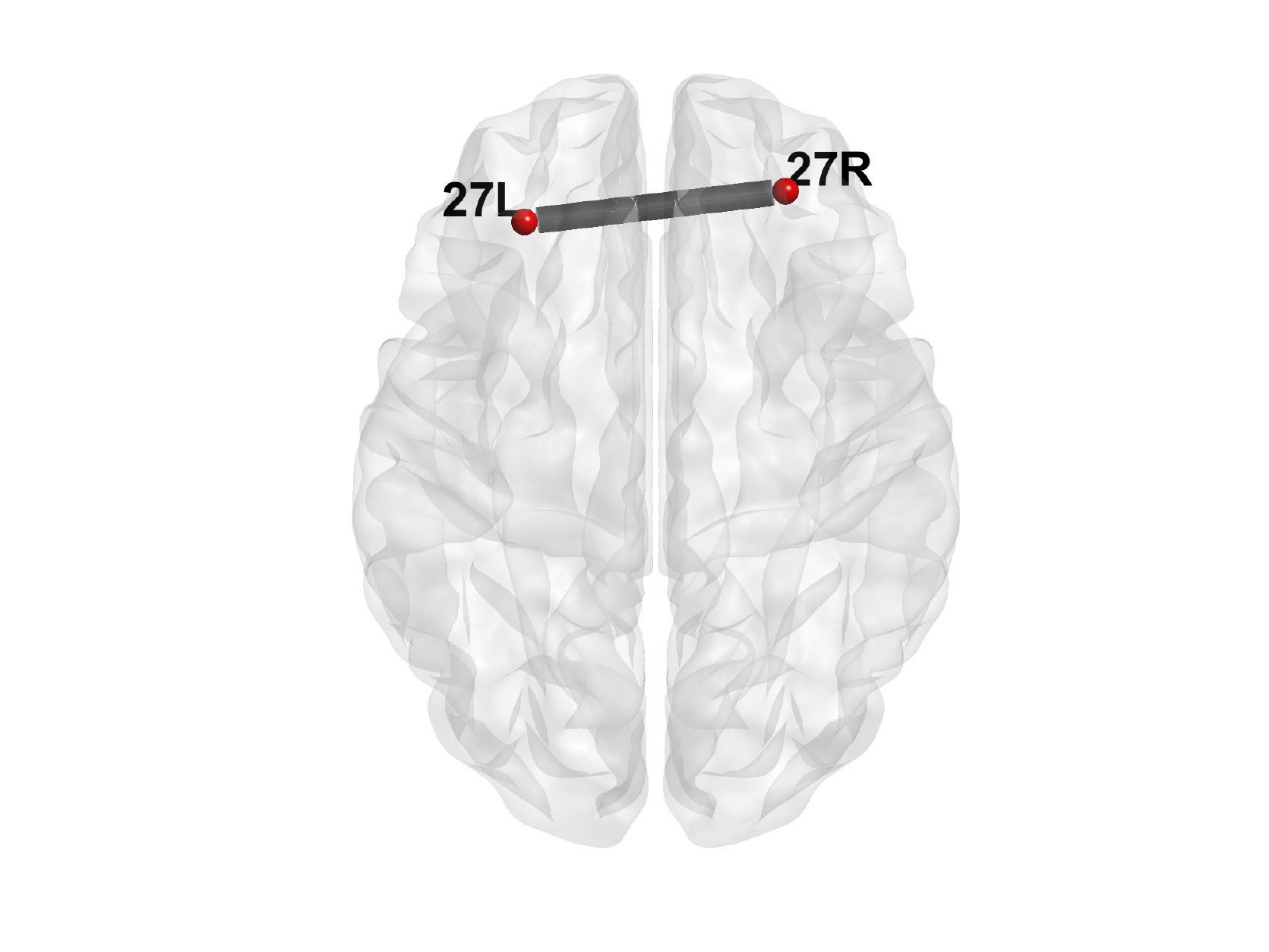}\includegraphics[width=0.16\textwidth, trim={3cm 0 3cm 0}, clip]{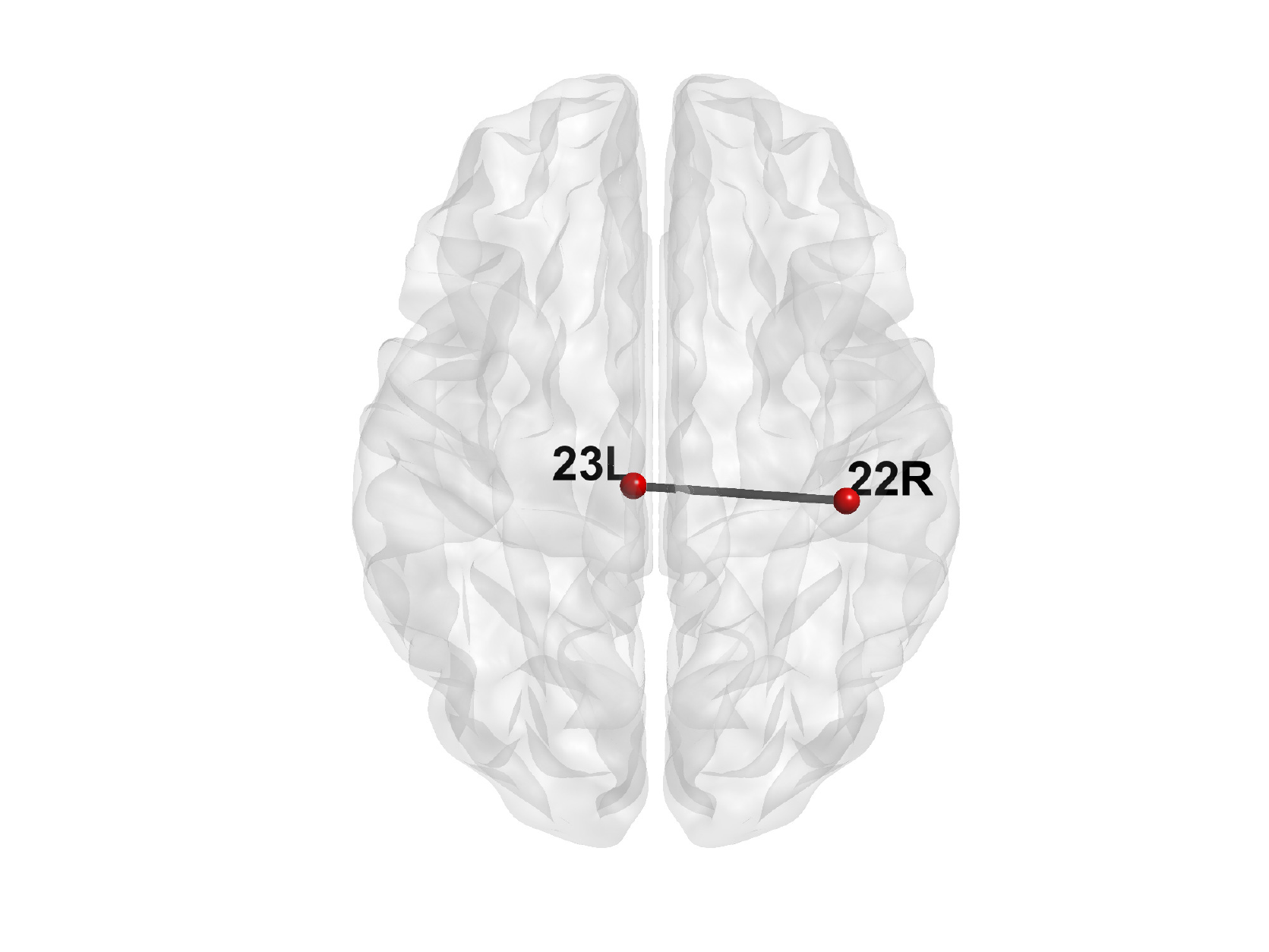}\includegraphics[width=0.16\textwidth, trim={3cm 0 3cm 0}, clip]{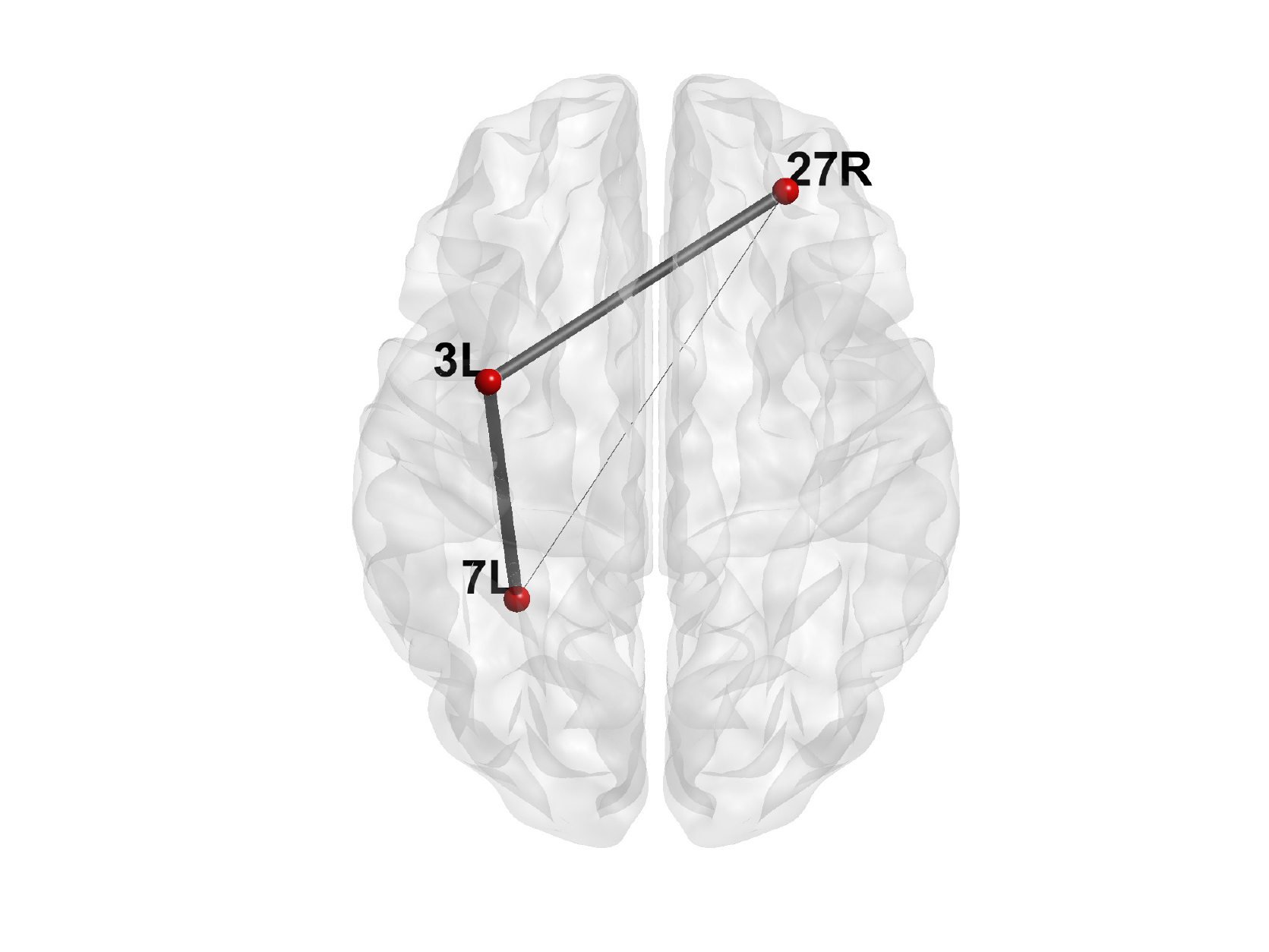}
\caption{Estimated nonzero component matrices $\{\lambda_{h}\boldsymbol{\beta}_{h}\boldsymbol{\beta}_{h}^{\top}\}$ for picture vocabulary data (upper) and their selected subgraphs in the brain (lower). The thickness of each edge is proportional to the average fiber count between the pair of brain regions.\label{fig:picvoc_sbl}}
\end{figure}

For $L_1$-penalized symmetric bilinear regression, only 6 out of 10 coefficient component matrices $\{\lambda_{h}\boldsymbol{\beta}_{h}\boldsymbol{\beta}_{h}^{\top}\}_{h=1}^{K}$ have nonzero entries, implying $K=10$ is large enough to capture all the signal subgraphs for this dataset. The estimated nonzero component matrices and their corresponding structural connections in the brain are displayed in Figure \ref{fig:picvoc_sbl}, which shows that SBL locates multiple simple subgraphs in the brain that may form some anatomical circuits in linguistic processing of sound to meaning. Three subgraphs in Figure \ref{fig:picvoc_sbl} only contain a single connection verifying the flexibility of the model. We also observe that some brain regions repeatedly appear in the subgraphs in Figure \ref{fig:picvoc_sbl}, which may indicate important roles of these regions in auditory comprehension. For example, $27L$, $27R$ (left and right superior frontal gyrus), $7L$ (left inferior parietal gyrus) and $29L$ (left superior temporal gyrus) are among activated regions when shifting from listening to meaningless pseudo sentences to listening to meaningful sentences \cite{saur2008ventral, dronkers2011neural}. Figure \ref{fig:picvoc_sbl} also shows that most estimated coefficients of the strengths of these signal connections are positive, implying that stronger neural connections among these regions are expected to lead to higher auditory comprehension ability. These identified anatomical sub-networks in the brain are consistent with the notion that auditory language processing is a complex process, which is the product of the coordinated activities of several brain regions. 

\subsection{Reading Recognition Data}

The HCP dataset also contains the age-adjusted scale scores of the subjects in an oral reading recognition (RR) test where participants were scored on reading and pronouncing letters and words. We apply our method to find sub-networks in the brain connectome relevant to oral reading ability.
Following the same procedure of partitioning data as in Section \ref{picvocab}, we compare the predictive performance for the RR scores among lasso, TN-PCA and SBL. The minimum out-of-sample MSE of SBL is 201.8, which is smaller than that of lasso, 205.9. Although TN-PCA obtains the smallest MSE, 194.7, in this case, the resulting 16 significant components select all the connections in the brain network.

\begin{figure}[hbt]
\centering
\includegraphics[width=0.11\textwidth, trim={4cm 0 4cm 0}, clip]{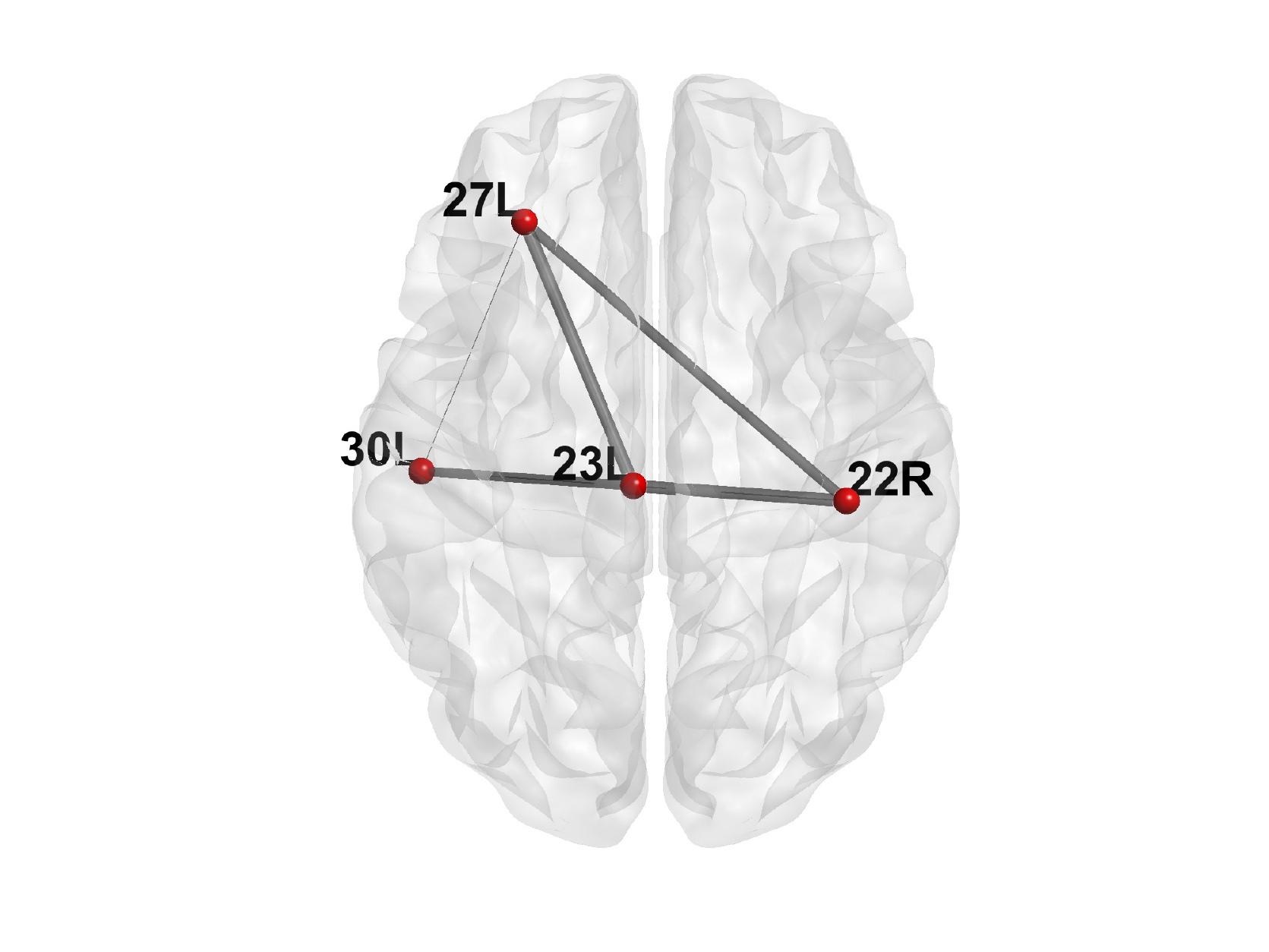}
\includegraphics[width=0.11\textwidth, trim={4cm 0 4cm 0}, clip]{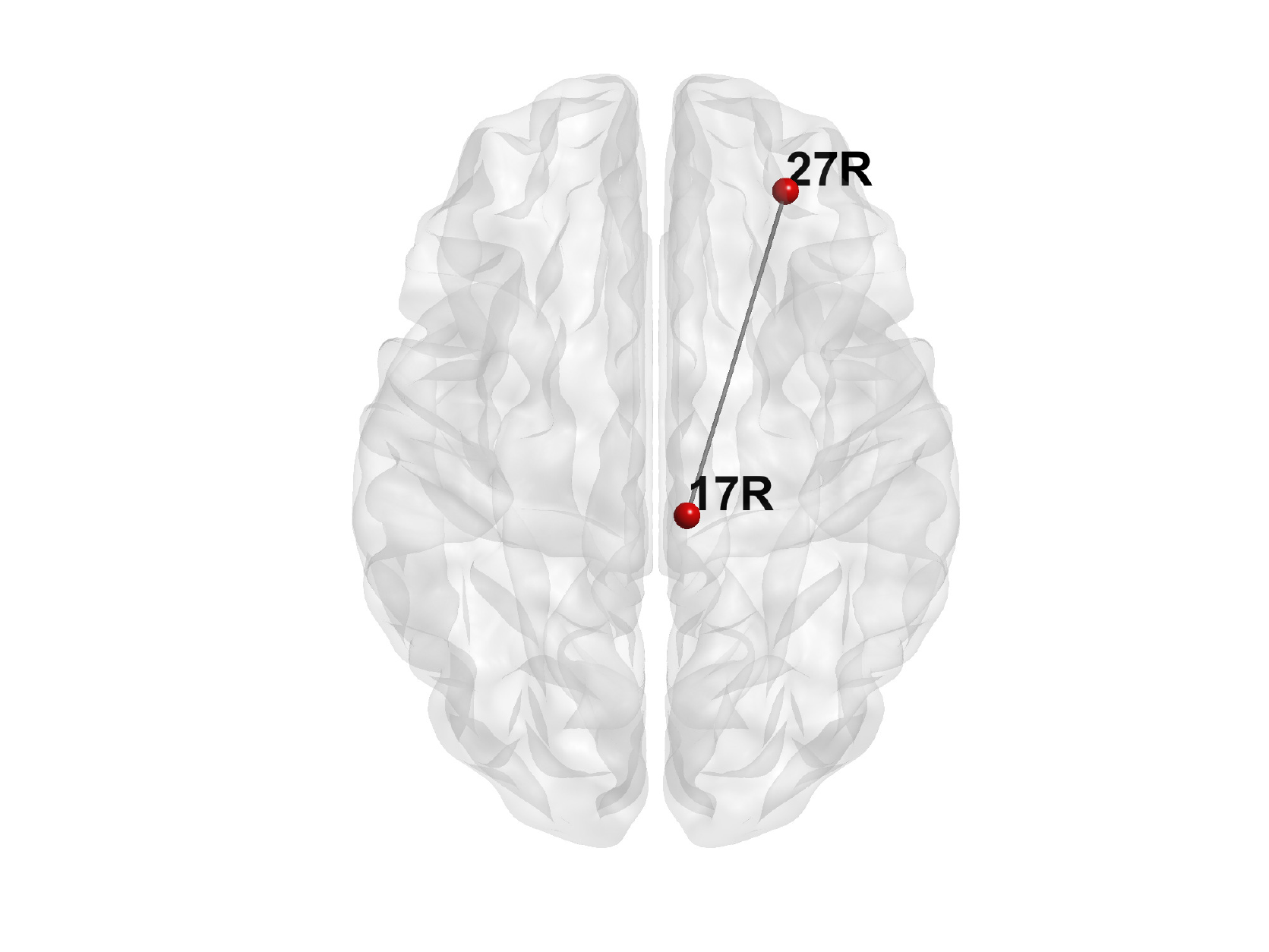}
\includegraphics[width=0.11\textwidth, trim={4cm 0 4cm 0}, clip]{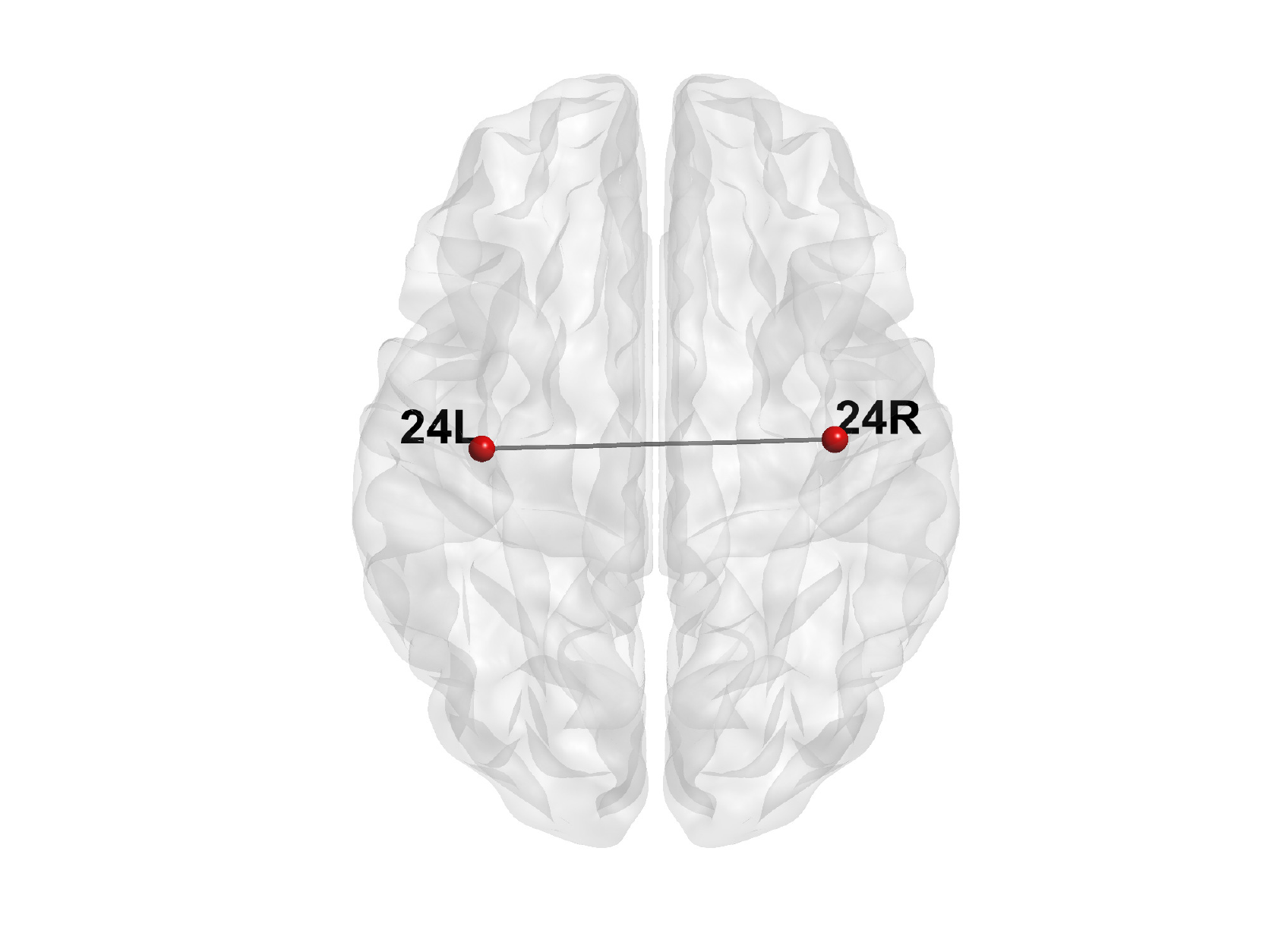}
\includegraphics[width=0.11\textwidth, trim={4cm 0 4cm 0}, clip]{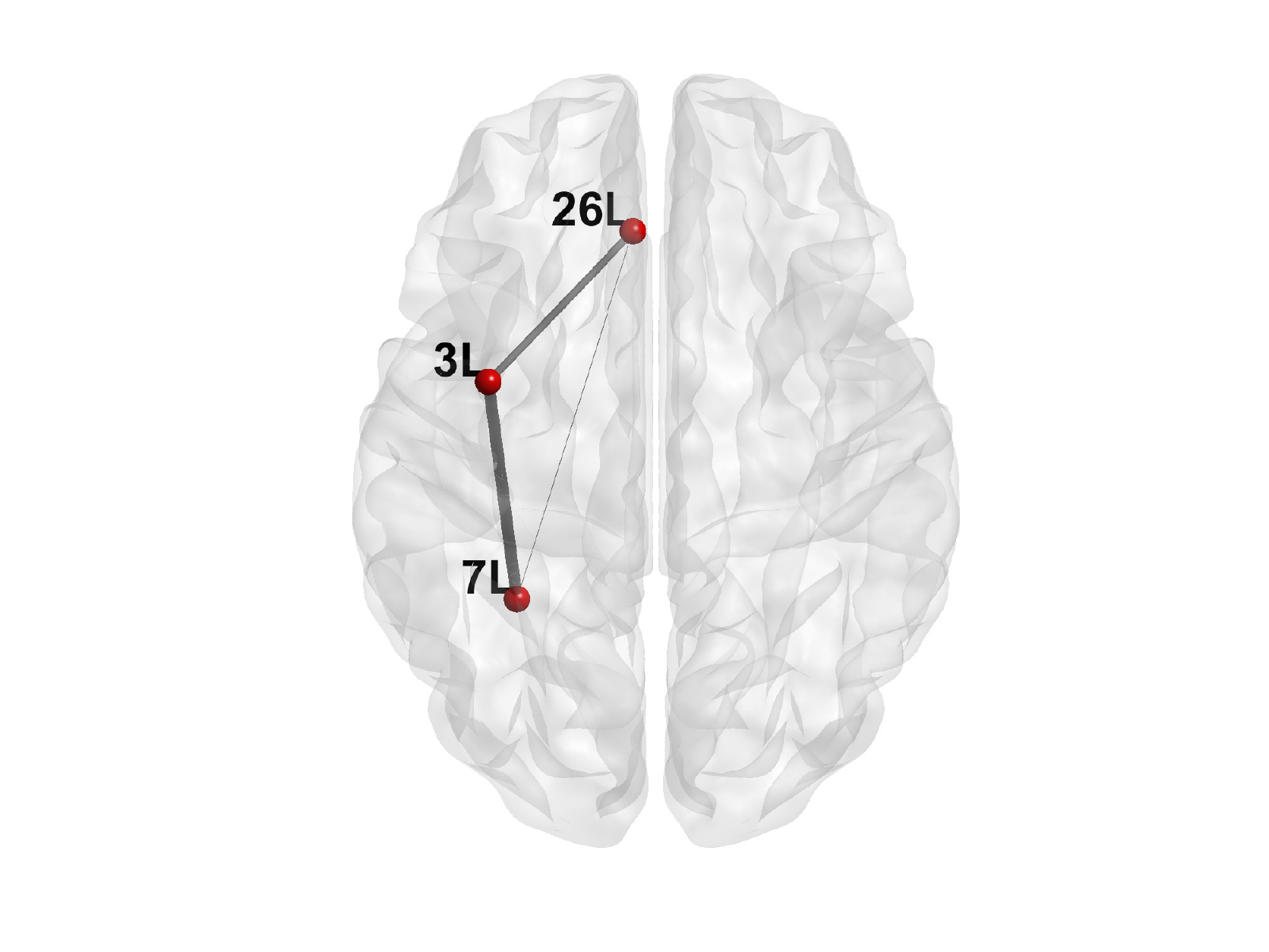} \\
\includegraphics[width=0.15\textwidth, trim={3cm 0 3cm 0}, clip]{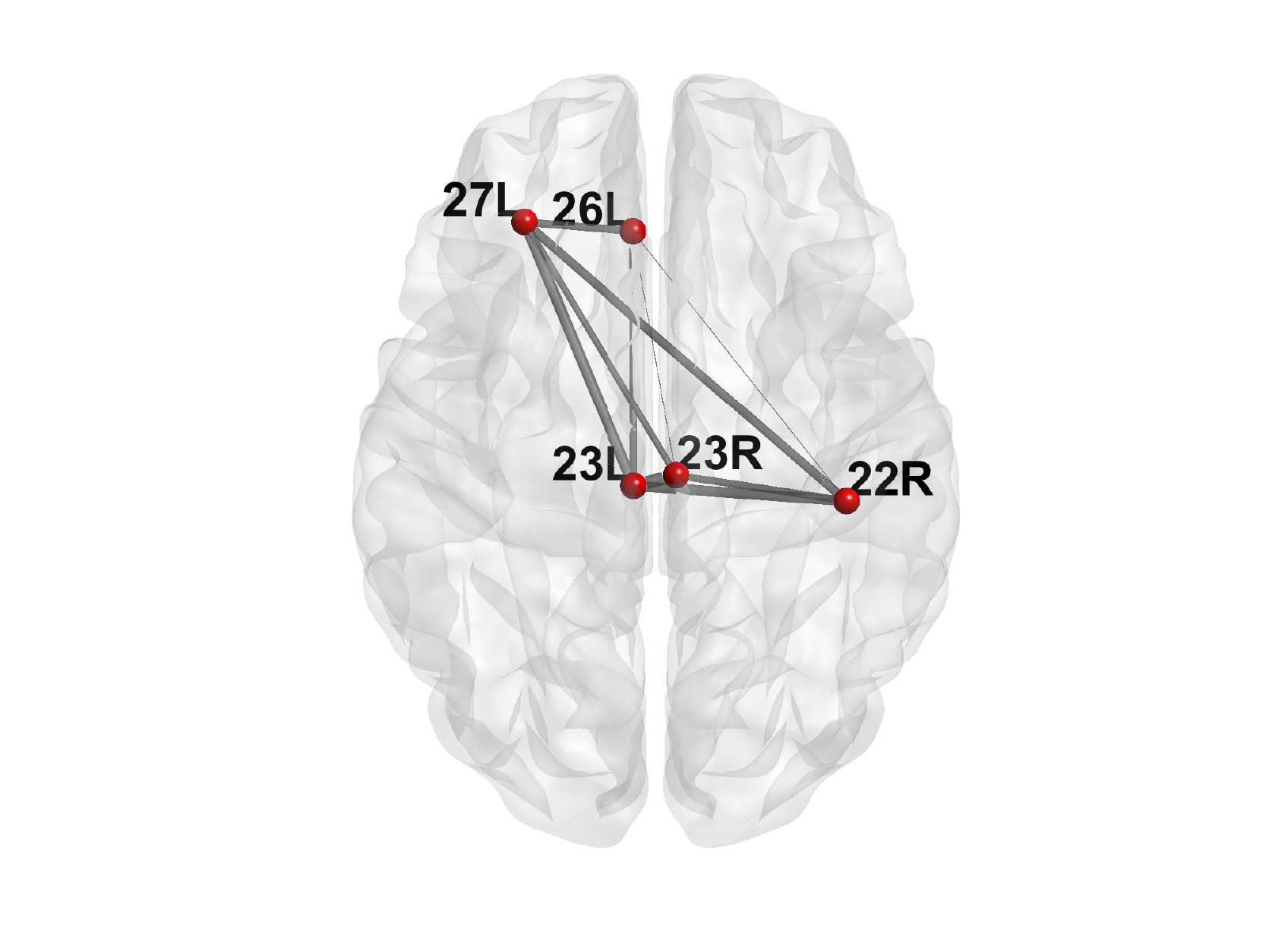}
\includegraphics[width=0.15\textwidth, trim={3cm 0 3cm 0}, clip]{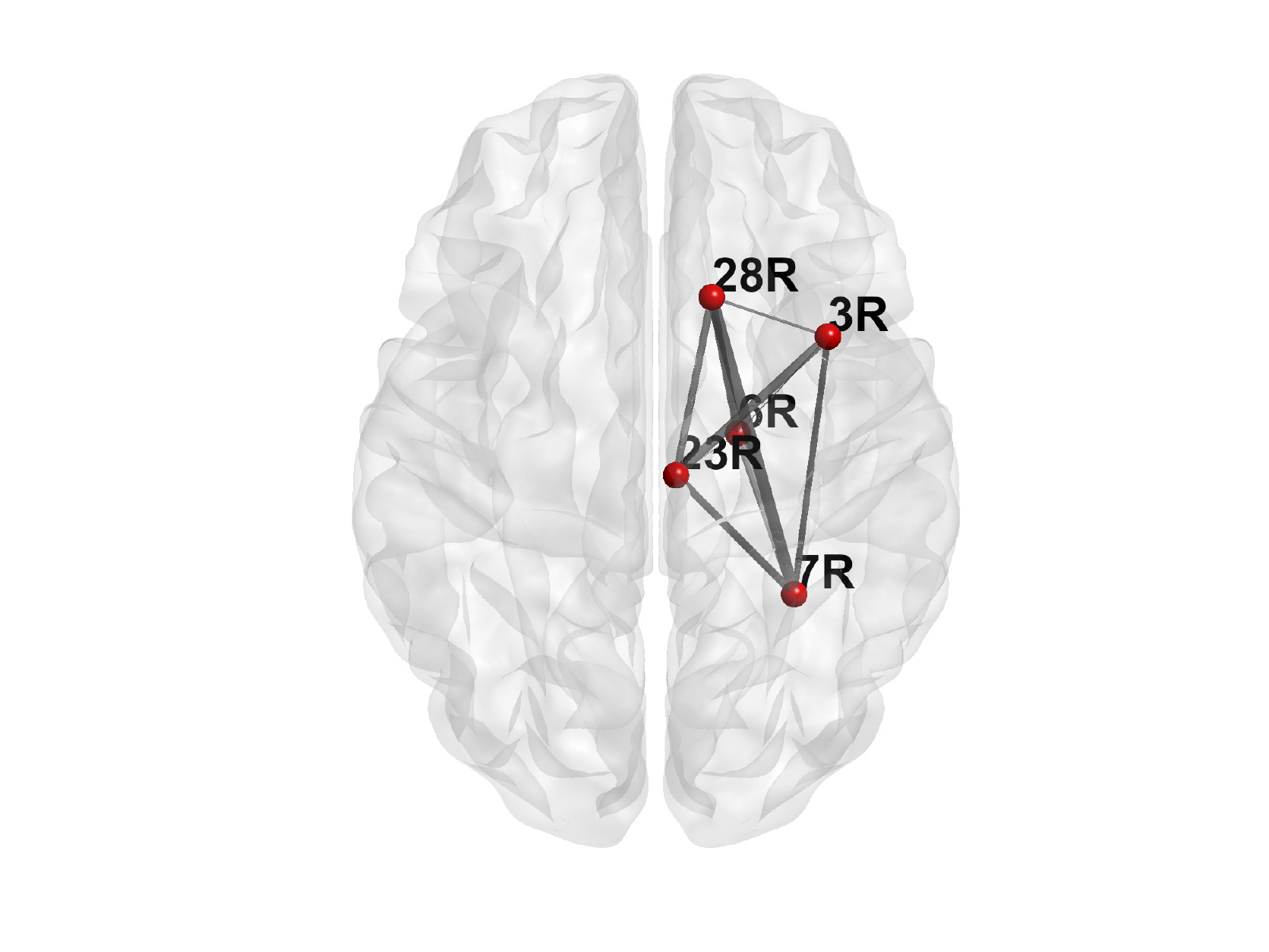}
\includegraphics[width=0.15\textwidth, trim={3cm 0 3cm 0}, clip]{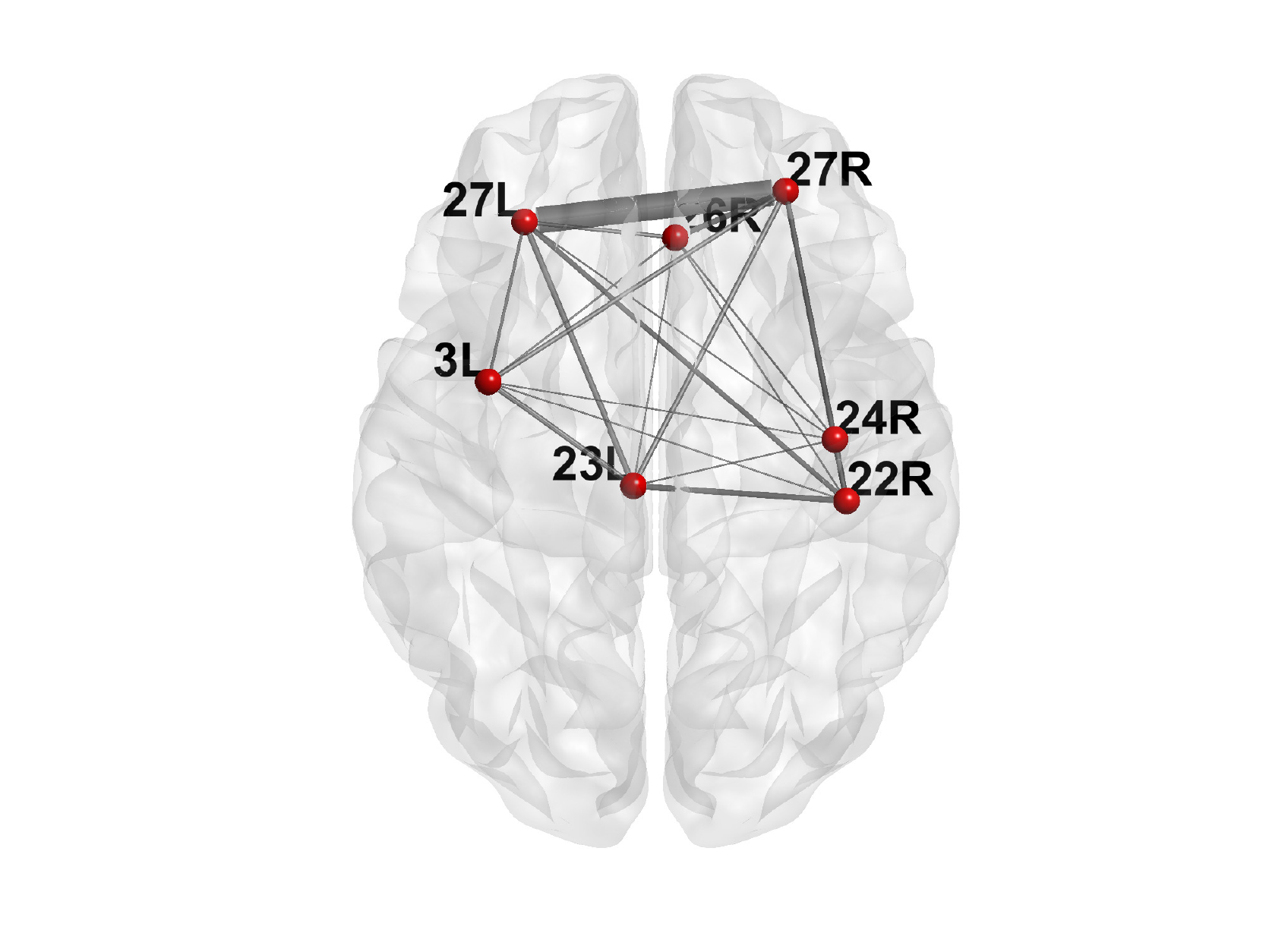}

\caption{The selected subgraphs in the brain relevant to oral reading ability. The thickness of each edge is proportional to the average fiber count between the pair of brain regions.\label{fig:readeng_sbl}}
\end{figure}

In this case, SBL selects 7 non-empty components $\{\lambda_{h}\boldsymbol{\beta}_{h}\boldsymbol{\beta}_{h}^{\top}\}$ out of 10 with penalty factor $\gamma$ set at the optimal value. The subgraphs of brain connectome corresponding to these nonzero components are displayed in Figure \ref{fig:readeng_sbl}. We notice that a triangle subgraph repeatedly appears in these subgraphs,  consisting of three regions: $27L$ (left superior frontal),  $23L$ (left precentral) and $22R$ (right posterior cingulate). This triangle subgraph may form a core anatomical circuit in the phonological reading pathway. These regions agree with the findings in neuroscience that the superior frontal gyrus is associated with word reading \cite{cloutman2011neuroanatomical}, left precentral gyrus is involved in phonological output \cite{safi2016recruitment} and the posterior cingulate cortex is associated with language comprehension \cite{smallwood2013default}.

\section{Conclusion}
\label{conclude}

In summary, the symmetric bilinear model is a useful tool in analyzing the relationship between an outcome and a network-predictor, which produces much more interpretable results than unstructured regression does, while maintaining competitive predictive performance. 
We develop an effective coordinate descent algorithm for $L_1$-penalized symmetric bilinear regression which outputs a set of small outcome-relevant subgraphs. Our method contributes to an insightful understanding of the sub-structure of networks that is relevant to the response and has wide applications in various fields such as neuroscience, internet mapping and social networks. Although we have focused on a continuous response, the methods are straightforward to adapt to classification problems and count responses by a simple modification of the goodness-of-fit component of the loss function.

\section*{Acknowledgment}

We would like to thank support for this project from Army Research Institute (ARI grant W911NF-16-1-0544).

\ifCLASSOPTIONcaptionsoff
  \newpage
\fi

\bibliographystyle{IEEEtran}
\bibliography{IEEEabrv,sdr_ref}

\begin{IEEEbiography}[{\includegraphics[width=1in,height=1.25in,clip,keepaspectratio]{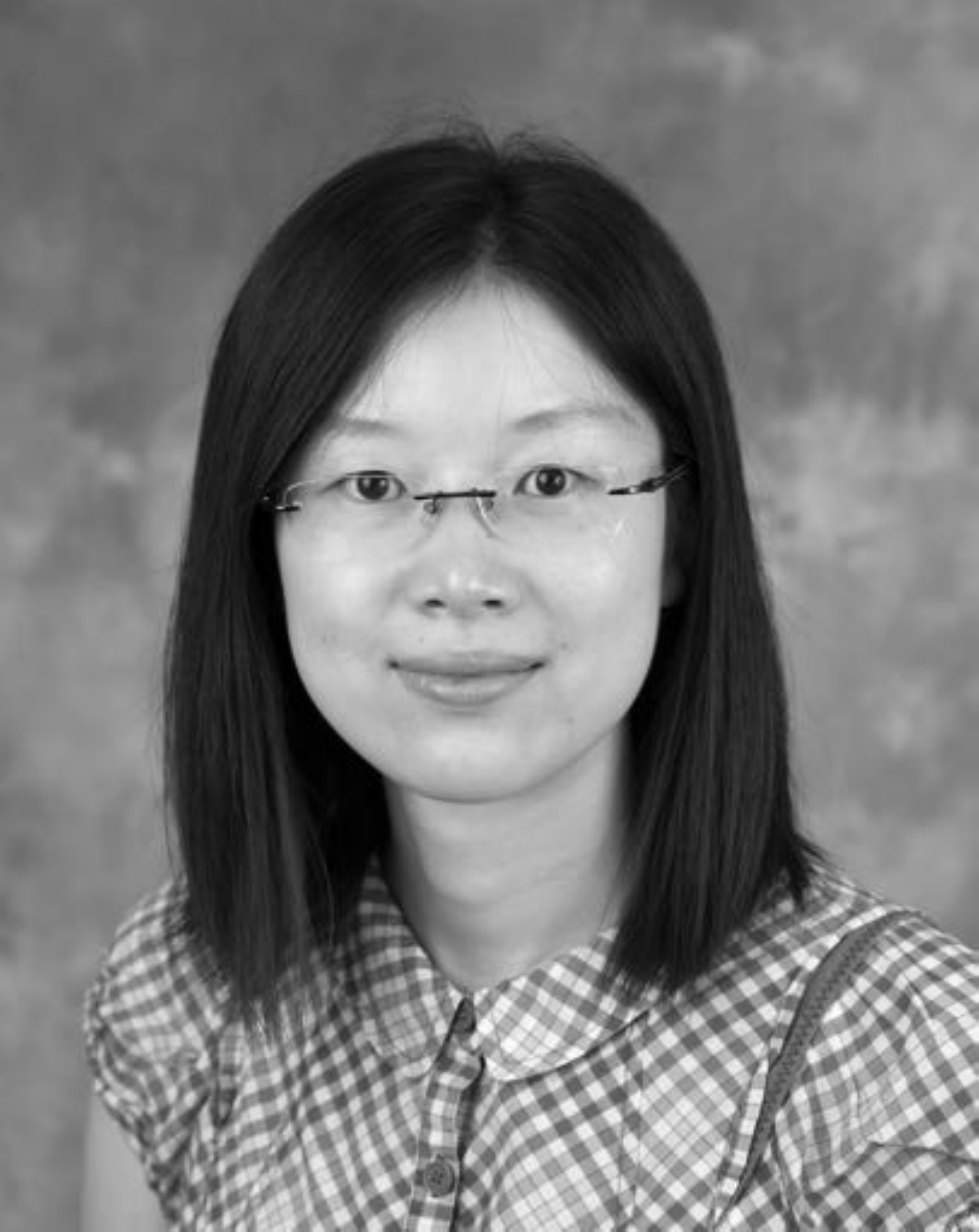}}]{Lu Wang}
received her Ph.D. degree in Statistics from Duke University in May 2018. She is currently an assistant professor in the Department of Statistics at the Central South University in China. Her research interests include network analysis, Bayesian modeling and high dimensional optimization.
\end{IEEEbiography}

\begin{IEEEbiography}[{\includegraphics[width=1in,height=1.25in,clip,keepaspectratio]{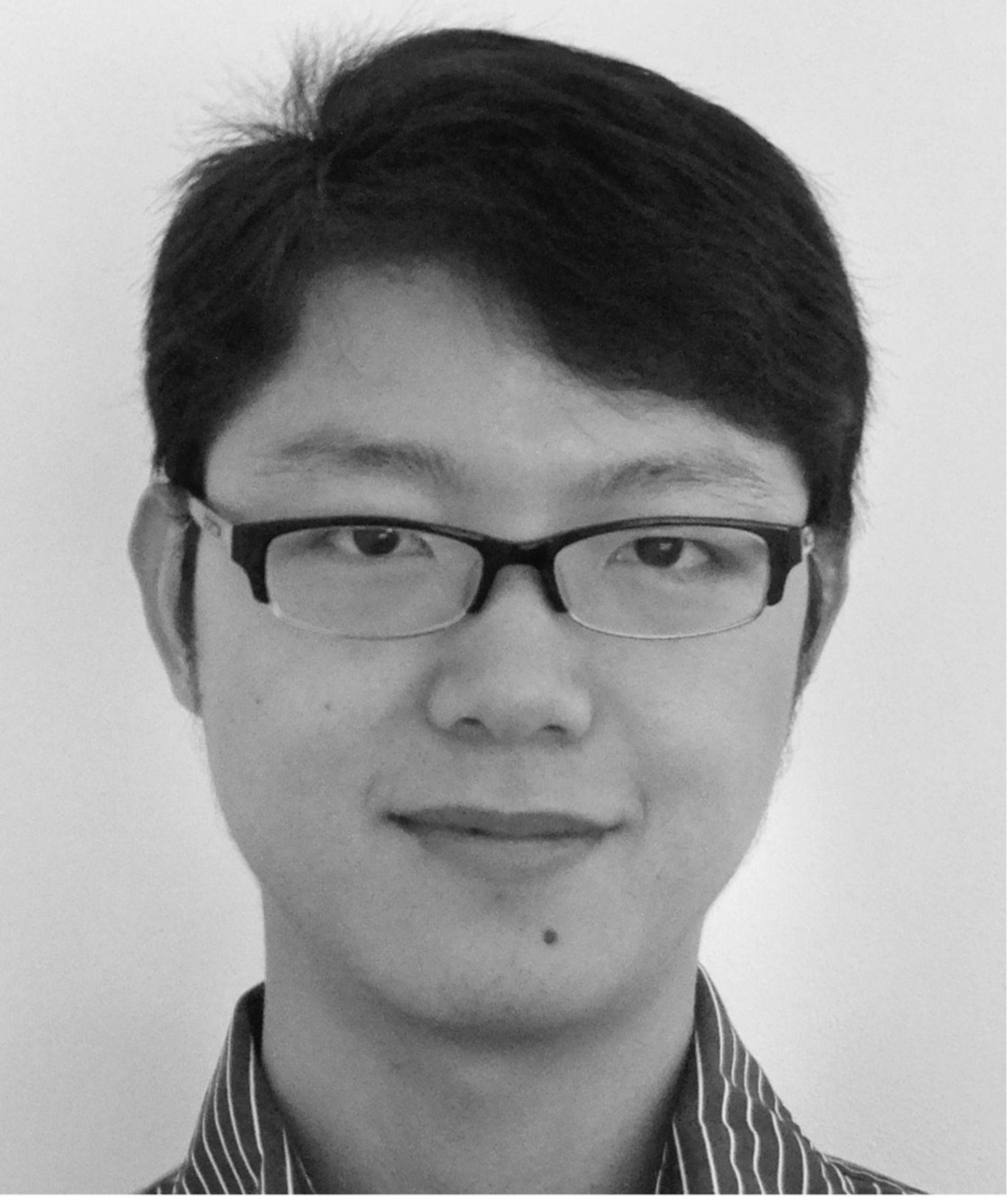}}]{Zhengwu Zhang} received his Ph.D. degree in Statistics from Florida State University in May 2015. He is currently an assistant professor in the Department of Biostatistics and Computational Biology at the University of Rochester. His research interests include statistical image analysis, statistical shape analysis,  Bayesian statistics, network analysis, and computational neuroscience.
\end{IEEEbiography}

\begin{IEEEbiography}[{\includegraphics[width=1in,height=1.25in,clip,keepaspectratio]{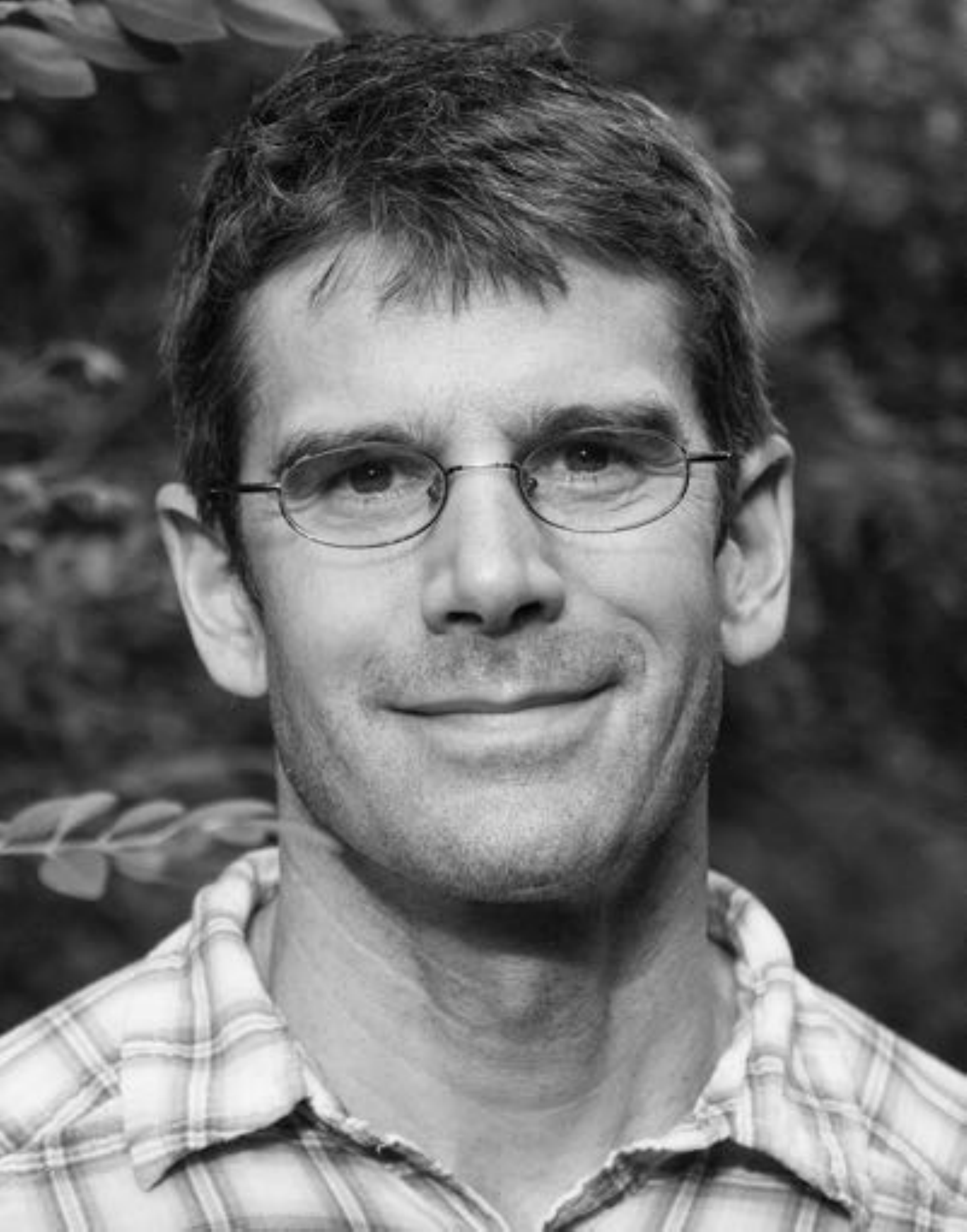}}]{David Dunson}
is Arts and Sciences Distinguished Professor of Statistical Science, Mathematics and ECE at Duke University.  He has made broad contributions in Bayesian statistical and signal processing methodology for complex and high-dimensional data, with a particular emphasis on nonparametric Bayesian approaches, dimensional reduction, and object data analysis.  His methodology work is often directly motivated by and applied to data from scientific studies, with a particular focus on environmental health, genomics and neuroscience.  He is a Fellow of the American Statistical Association, Institute of Mathematical Statistics, and International Society for Bayesian Analysis.  He won the 2010 COPSS President’s Award given annually to one top statistician internationally age 40 or under.  
\end{IEEEbiography}

\end{document}